\documentclass[epsf,amsfonts,12pt]{article}
\usepackage{graphicx}
\usepackage{stackengine,lipsum,boxhandler}
\stackMath
\usepackage{epstopdf}
\usepackage{epsfig}

\DeclareGraphicsExtensions{.png,.jpg,.pdf,.mps,.pdf} 
\usepackage{caption}
\usepackage{subcaption}
\usepackage{float}
\usepackage[bottom,hang,flushmargin]{footmisc}
\usepackage{amsmath, multirow}
\usepackage{amsmath,bm}
\usepackage{titling}
\usepackage{adjustbox}
\usepackage{afterpage}
\usepackage{lscape}
\usepackage{pdflscape}
\usepackage{booktabs,threeparttable}
\usepackage{tabularx,ragged2e,booktabs}
\usepackage{dirtytalk}
\usepackage{caption}
\usepackage{lipsum}
\usepackage[a4paper,left=2.4cm,right=2.4cm,top=2.4cm,bottom=2.4cm]{geometry}
\usepackage[11pt]{moresize}
\usepackage{adjustbox}

\usepackage{array, booktabs, makecell}

\usepackage{geometry}
\usepackage[hidelinks]{hyperref}
\usepackage{tikz}
\usepackage{enumerate}
\usetikzlibrary{arrows,calc}
\usetikzlibrary{shapes}
\usetikzlibrary{positioning,shadows}
\usetikzlibrary{decorations.markings}
\def\gleq{{\kern5pt\raise1.40ex\hbox{$>$}
           \kern-1.7ex\hbox{$=$}
           \kern-2.0ex\lower1.67ex\hbox{$<$}
           \kern5pt}}

\newcommand{\beq}{\begin{equation}}
\newcommand{\eeq}{\end{equation}}
\newcommand{\beqa}{\begin{eqnarray}}
\newcommand{\eeqa}{\end{eqnarray}}
\newcommand{\beqan}{\begin{eqnarray*}}
\newcommand{\eeqan}{\end{eqnarray*}}

\newtheorem{querytheorem}{Query}

\usepackage{amsmath}
\usepackage{graphicx}
\usepackage{eurosym}
\usepackage{tikz}
\usepackage{subcaption}
\usepackage{float}
\usepackage{longtable,booktabs}
\usepackage[noblocks]{authblk}
\usepackage[ddmmyyyy]{datetime}

\usepackage{titlesec}
\usepackage{alltt}
\usepackage{upquote}

\titleformat*{\section}{\large\bfseries}
\titleformat*{\subsection}{\large\bfseries}
\titleformat*{\subsubsection}{\normalsize\bfseries}
\titleformat*{\paragraph}{\large\bfseries}
\titleformat*{\subparagraph}{\large\bfseries}

\newcommand*{\Scale}[2][4]{\scalebox{#1}{$#2$}}%

\usepackage{xcolor}
\definecolor{gold}{RGB}{193, 165, 57} 
\preauthor{\vskip 1.0em \large}
\DeclareRobustCommand{\authoring}{
\begin{center}
\textbf{Andrew Mountford} \vskip 0.5em
{\normalsize Royal Holloway, University of London and CReAM}
}
\author{\authoring}
\postauthor{\end{center} \vskip 0.75em}

\begin{document}
\title{\Large \textbf{How Large are Monetary Policy Effects? \\ Identifying the Shocks in a VAR also Identifies the Constants}\thanks{Many thanks to Jonathan Wadsworth for his very helpful comments.  Address for correspondence: A.Mountford,
Department of Economics, Royal Holloway, University of London, Egham, TW20 OEX, U.K.  Tel: +44 1784 443906 email: A.Mountford@rhul.ac.uk}}
\date{\textbf{November 2026}}
\maketitle
\begin{abstract}
\noindent
\baselineskip=11pt {\small \rm{Restrictions on the contemporaneous effects matrix used to identify fundamental shocks in a structural VAR, also determine the mapping from the structural constant terms to the reduced form  constant terms. In some models one will have priors about these structural constant terms and these should therefore be included in a Bayesian estimation procedure.  We illustrate the potential significance of this using the analysis of monetary policy by Baumeister and Hamilton (2018) as a motivating example. We show that imposing priors over the structural constant terms can lead to a more intuitive estimated monetary policy rule and a much larger estimated role for monetary policy in describing the evolution of the data. The analysis also allows for a clear visual presentation of the structural model, highlighting its multiple contending elements. For example the observed inflation series is estimated to be the result of positive demand and commodity price effects offset by negative monetary policy and  risk spread effects.}  (157  words)   \bigskip  \vspace{0.1in} \newline
\noindent \vspace{0.1in} Keywords: Vector Autoregressions, Historical Decompositions,  Monetary Policy}
\end{abstract}
\thispagestyle{empty}

\pagebreak

\setcounter{page}{1}

\section{Introduction}   
\label{sec:Intro}

The constant terms in a structural vector autoregression (VAR) have not received the same attention in the literature as its impulse response functions. Yet the structural constant terms are subject to the same identification transformations as the structural shock processes and can play a large role in the structural model underpinning the VAR. Furthermore, economists may have priors over the contribution of these structural constant terms and these should be incorporated into a Bayesian estimation procedure. We illustrate the potential significance of this using Baumeister and Hamilton's (2018) analysis of monetary policy as a motivating example. We show that the simple addition of intuitive priors over the constants of the structural model can significantly increase the estimated role of monetary policy on the evolution of the macroeconomy.

This paper makes contributions to the literature in two ways; to the identification of VARs in general and to the specific literature on the effects of monetary policy. Baumeister and Hamilton's (2015) seminal analysis derived a Bayesian estimation framework which allowed the incorporation of beliefs and incomplete information about the structural parameters.  They then applied and extended this methodology to the analysis of monetary policy in Baumeister and Hamilton (2018)  and to oil price shocks in Baumeister and Hamilton (2019). The contribution of this paper is to show the force of this approach when it is applied to the structural constant terms. In doing so this paper also builds on the recent work of Bergholt et al. (2025), who showed that the reduced form constant terms in a VAR are often estimated imprecisely, and that this in turn leads to an increased variance in the estimated contribution of the identified shocks.  They showed how a single-unit-root prior can solve this problem by shrinking the uncertainty around the estimated deterministic component.\footnote{ \ \ \ Brignone and Piffer (2026) show how this modeling of the deterministic terms can aid prediction.}  In this paper, we focus on the role of the structural constant, extending the approach of Mountford and Wadsworth (2025), by imposing priors over the structural constants. We how these priors can affect posterior parameter distributions and enable them to become more consistent with economic intuition.\footnote{ \ \ \ Mountford and Wadsworth (2025) analyze the labor market effects of immigration where the structural constant associated with the growth of immigration share in the labor market negatively affects the growth of native wages.}  

An additional benefit of focusing on the structural constants is that it allows the multiple contending structural forces acting on the data to be presented clearly. By including the contributions of the structural constants into the historical decompositions, the competing influence of each structural equation is depicted, both in terms of the mean and the variation around it.  For example, in Figure \ref{fig:BHHD_6var_1} below, we estimate the observed long run level of inflation to be the result of positive demand and commodity price effects and negative monetary policy and credit risk effects, with commodity prices and risk effects playing prominent roles in its variation. 

The effects of monetary policy is a fundamental topic in macroeconomics with intrinsic policy implications. There is consequently a large literature estimating these effects using a variety of approaches. This paper contributes to the structural VAR strand of this literature. This attempts to identify the effects of monetary policy using time series methods without imposing strong parametric assumptions.\footnote{ \ \ \ We do not attempt to summarize here the large related literature deriving restrictions on VARs from DSGE models, see Baumeister and Hamilton (2018), Canova (2007) and Canova and Ferroni (2022) for a discussion of this topic.}   A seminal paper in this field is Leeper, Sims and Zha (1996) which found, using US data, that shocks to monetary policy were responsible for only a small proportion of the variation in output and prices, and that most of the variation in the policy interest rate itself can be attributed to systematic reactions to other macroeconomic shocks. This was an intuitive result that is consistent with how one might expect a monetary authority to behave. Of course, what constitutes a strong assumption is open to question, and this has been explored.  Uhlig (2005) using apparently minimal identifying sign restrictions, also found only a small role for monetary policy in the variation of output and prices.  However other studies, using alternative plausible sign restrictions, have found significant effects of monetary policy, such as Canova and De Nicol\`o (2002) for G7 economies and Mumtaz and Surico (2009) for the UK economy. 

Baumeister and Hamilton (2015) showed that the methodology used in the previous sign restriction literature imposed stronger restrictions on the parameters than had previously been believed and developed a new method for imposing informative priors on the structural parameters of a VAR.  In their analysis of  monetary policy, Baumeister and Hamilton (2018)  found, in common with Leeper, Sims and Zha (1996) and  Uhlig (2005), that monetary policy shocks had only a small influence on the evolution of income, inflation and interest rates, with demand and supply shocks responsible for most of the variation in the data.  

In this paper we show that the identification in Baumeister and Hamilton (2018) implies a counter-intuitive monetary policy rule where interest rates are insignificantly different from zero in the environment where output is at its potential and inflation is zero. We show that by imposing a prior such that the constant in the monetary policy rule is positive, the estimated monetary policy rule becomes more intuitive and results in monetary policy having a greater estimated role in the evolution of the macroeconomy.   In our three variable VAR we find in common with the analysis of Bergholt et al. (2025), that the inflation surge after 2020 was the result of demand and supply factors.  In our six variable VAR this surge was primarily explained by a specific supply shock, the commodity price inflation shock,  but with other shocks, including that of monetary policy, also being present. This is consistent with the analysis of Brignone and Piffer (2025) for the UK economy.

The paper is organized as follows. In section \ref{sec:Accounting} we describe the relationship between the structural constants and the reduced form constants and the contribution of structural constants to the historical decompositions of the data.  Section \ref{sec:Structural_priors} then describes how we implement priors over the structural constants which are the focus of the paper.   Section \ref{sec:BH3VAR} applies this method to the three variable three equation  model of Baumeister and Hamilton (2018) and section \ref{sec:BH6VAR} extends this to a six variable VAR. Sections  \ref{sec:robust}  and \ref{sec:howlarge} discuss the robustness of the results and the size of the estimated monetary policy effects. 

\section{Accounting for the constant terms} 
\label{sec:Accounting}

A structural VAR with $n$ variables and $p$ lags is described by equation (\ref{eq:Struct_VAR}),
\small
\begin{equation}
Ay_t= C + B_1 y_{t-1} + B_2 y_{t-2} + \dots + B_p y_{t-p} + u_t 
\label{eq:Struct_VAR}
 \end{equation}
 \normalsize
where $A$, is an $(n \times n)$ matrix of contemporaneous effects, $C$, is a $(n \times 1)$ vector of constants,  the matrices of the parameters for lag $i$, $B_i$, are dimension $(n \times n)$, and $u_t$, is the vector of structural shocks on the VAR. The $(n \times n)$ variance covariance matrix for these shocks, $E[u_t u^{\prime}_{t}]=D$, is assumed to be diagonal. The diagonal structure implies the shocks are  `fundamental' in the sense of having a zero covariance with each other. 

The reduced form of the VAR in equation (\ref{eq:Struct_VAR}) is given by multiplying both sides by the inverse of the $A$ matrix, so that  
\small
\begin{equation}
\begin{aligned}
 y_t &= A^{-1} C + A^{-1} B_1 y_{t-1} +A^{-1} B_2 y_{t-2} + \dots + A^{-1} B_p y_{t-p} + A^{-1} u_{t} \\
     &= \kappa+ \Phi_1 y_{t-1} + \Phi_2 y_{t-2} + \dots + \Phi_p y_{t-p} + \epsilon_{t}
\label{eq:VAR_rf}
\end{aligned}
\end{equation}
\normalsize
\noindent The reduced form coefficient matrices and error terms in equation (\ref{eq:VAR_rf}) -  $\kappa, \Phi_j$ and $\epsilon_{t}$, -   are their counterparts in the structural equation multiplied by $A^{-1}$. 

Macroeconomic research has focused on identifying the structural shocks which make up $u_t$, by restrictions on  the matrix $A$. Given $A$, the effects of the structural shocks on the time series can be described using impulse response functions and historical decompositions. 

Following Hamilton (1994) impulse response functions can be calculated by rewriting equation (\ref{eq:VAR_rf}) as a VAR of 1 lag,  
\small
\begin{equation}
\underset{(np\times 1)}{Y_t}= \underset{(np\times 1)}{K} +\underset{(np\times np)}{F} \underset{(np\times 1)}{Y_{t-1}}+  \underset{(np \times 1)}{E_{t} } 
\label{eq:RF_VAR1}
\end{equation}
\normalsize 
where
\small
\begin{equation*}
\begin{aligned}
Y_t= \left[
\begin{array}{c}
\underset{(n\times 1)}{y_t} \\
\vdots \\
\underset{(n\times 1)}{y_{t-p+1}}
\end{array}
\right] &
K=\left[
\begin{array}{c}
\kappa_{n \times 1} \\
0_{(n(p-1) \times 1)}
\end{array}
\right] &
F= \left[
\begin{array}{ccc}
\Phi_1 & \dots & \Phi_p \\
\multicolumn{2}{c}{I_{n(p-1)}} & 0_{n(p-1),n}
\end{array}
\right] & \ \ 
E_{t} = \left[
\begin{array}{c}
\epsilon_{t}\\
 0_{n(p-1),1}
\end{array}
\right]
\end{aligned}
\end{equation*}
\normalsize 

\noindent The iteration of equation (\ref{eq:RF_VAR1}) backward from period $t+s$, implies that the observation $Y_{t+s}$ in period $\tiny{t+s}$ can be decomposed into three contributions: the initial conditions, the constant terms and the innovations in the previous s periods.  i.e.
\small
\begin{equation}
Y_{t+s} = \underbrace{F^s Y_{t}}_{\text{initial conditions}}  + \underbrace{F^{s-1}K + F^{s-2}K +\dots +K}_{\text{constant terms}}  + \underbrace{F^{s-1}E_{t+1} + F^{s-2}E_{t+2}+ \dots + E_{t+s}}_{\text{innovations}} 
\label{eq:decomp}
\end{equation}
\normalsize
where the variables of interest are the first $n$ rows of this system. The contribution of the structural shocks at time period $t$, $u_{t}$, $s$ periods after the shock, denoted $H_s$, is thus given by 
\begin{equation}
H_s= F_n^{s-1} A^{-1} u_{t} 
\label{eq:IRF}
\end{equation}
where $F^{s-1}_n$ denoted the top left $n \times n$ sub matrix of $F^{s-1}$. The impulse response functions are $H_s$  calculated for all periods $s$, for  $s=1,T^{\Scale[0.65]{\mbox{max}}}$, where $T^{\Scale[0.65]{\mbox{max}}}$ is a time horizon chosen to be sufficient to determine the impulse response function's shape.  

Equation (\ref{eq:decomp}) shows that the $F$ matrix determines the mapping of both the structural shocks and the structural constant terms into the observed times series. One can therefore apply the same analysis of equation (\ref{eq:IRF})  for the contribution of the structural constant term to $Y_{t+s}$, denoted $J_s$,  so that 
\begin{equation}
J_s= \sum^{s-1}_{l=0} F_n^i A^{-1} C
\label{eq:J_IRF}
\end{equation}

Furthermore the mean of the vector time series, denoted $\mu$,  is given by 

\begin{equation}
\begin{aligned}
\mu = &[I_n - \Phi_1 -\Phi_2 -\dots \Phi_p]^{-1}A^{-1}C 
\end{aligned}
\label{eq:VAR_mean}
\tag{HAM1}
\end{equation}
\normalsize

As the $F$ and $\Phi_i$ matrices are functions of matrix $A$, then matrix $A$  plays an important role in determining how the mean of each variable in the VAR is explained by each structural constant term. 

In some models one will have priors about some or all of the structural constant terms. Including such priors in the estimation process will therefore affect the estimated $A$ matrix and so will influence the estimated contribution of each structural constant. In doing this it will also effect the contribution of each structural shock as this also depends on $A$.  In the next section we describe how we implement the priors over the structural constants and in the following section we demonstrate the potential significance of including such priors using the example of Baumeister and Hamilton's (2018) analysis of monetary policy.

\section{\Scale[0.95]{\mbox{Implementing priors over the structural constants}}}
\label{sec:Structural_priors}

To implement priors over the structural constants we follow Baumeister and Hamilton (2015) and apply the 'mixed estimation' method described in Theil (1971). Baumeister and Hamilton (2015) use Bayesian methods to estimate the structural VAR of equation (\ref{eq:Struct_VAR}). They employ natural conjugate priors for the autoregressive and variance parameters independently across equations and  conditional on the parameters of the $A$ matrix. Specifically, conditional on $A$  and $D$,  the prior for the $np+1$, $B$ coefficients of each equation i in equation (\ref{eq:Struct_VAR}) - denoted $b_i \equiv [B_1(i,:) \hdots B_p(i,:)  \ C(i)]$ -  are independently distributed $\mathcal{N}(m_i,d_{ii}M_i)$ across equations. Each diagonal element, $d_{i}$, of matrix $D$, is independently inverse gamma distributed.

Baumeister and Hamilton (2015) show how, given the matrix $A$, the mean  and variance of the posterior for the coefficients of the structural VAR can be estimated equation by equation, using an OLS regression where the dependent variable in each equation is $ [a_i^{\prime}y_1 \hdots a_T^{\prime}y_Y  \ m_i^{\prime}P_i] $ where $P_i$ is the Cholesky factors of $M^{-1}_i$, and where the dependent variables are $ [x_i \hdots x_{T-1} \ P_i] $. Additional prior information, which can be represented by linear combinations on the parameters of one equation,  can be implemented following Theil (1971) by further augmenting the data matrices. For example, following Baumeister and Hamilton (2015), consider the case where there are $h_i$ pieces of prior information for the parameters in equation $i$ which  can be represented by the equation, $r_i=R_ib_i +\nu_i$, where $\nu_i$ is normally distributed error term $\mathcal{N}(0,V_i)$. This prior information can be included in the estimation by augmenting the data matrices so that the dependent and independent variables in the OLS regression, denoted, $\widetilde{Y}$ on $\widetilde{X}$, become,
\small
\begin{equation}
\begin{aligned}
\underset{\Scale[0.7]{(1 \times (T + k +h_i))}}{\widetilde{Y}}=&[\underset{\Scale[0.7]{(1 \times n)(n \times 1)}}{a_i^{\prime}y_1} \hdots \underset{\Scale[0.7]{(1 \times n)(n \times 1)}}{a_T^{\prime}y_T} \underset{\Scale[0.7]{(1 \times k)(k \times k)}}{m_i^{\prime}P_i}  \underset{\Scale[0.7]{(1 \times h_i)(h_i \times h_i)}}{\mathbf{r_i^{\prime}P_{Vi}}}  ] \\
\underset{\Scale[0.7]{(k \times (T + k +h_i))}}{\widetilde{X}}=&[\underset{\Scale[0.7]{(k \times 1)}}{x_i} \hdots \underset{\Scale[0.7]{(k \times 1)}}{x_{T-1}} \underset{\Scale[0.7]{(k \times k)}}{P_i}  \underset{\Scale[0.7]{(k \times h_i)(h_i \times h_i)}}{\mathbf{R_i^{\prime}P_{Vi}}}  ] \\
\end{aligned}
\label{eq:BH_Ytilde}
\tag{BH(2015)}
\end{equation}
\normalsize 
where  $P_{Vi}$ is the Cholesky factor of  $V^{-1}_i$. Thus a prior that the structural constant in equation $i$ should be close to 0.1 is a single constraint,($h_i=1$), with $r_i=0.1$, $R_i$ a $\Scale[0.85]{(np+1 \times 1)}$ vector with a value of 1 in the position of the constant term and 0 elsewhere and where the uncertainty about this prior is represented by the variance term, $V_i$.  

We are also imposing the prior that the mean of the reduced form VAR is close to the sample means of the data series via a single-unit-root prior as recommended by Bergholt et al. (2025). This single-unit-root prior is implemented by further augmenting the data matrices of both the independent variables and the dependent variables at all lags by $\mu_i/\delta$, where $\mu_j$ is the sample mean of variable $j$ and where $\delta$ represents the precision of this prior, where we set $\delta=0.01$. Thus, via equation (\ref{eq:VAR_mean}),  a prior over one structural constant will also effect the other structural constants.

\section{\Scale[0.95]{\mbox{An example of identifying the monetary policy rule}}}
\label{sec:BH3VAR}

We demonstrate the implications of including priors over the structural constants using Baumeister and Hamilton's (2018) analysis of monetary policy as a motivating example. In this  section we analyse the three variable structural VAR in the output gap, $y_t$, inflation, $\pi_t$ and the nominal interest rate, $r_t$.  In the following section we estimate a six variable VAR with three additional variables.  We estimate this model using an updated sample from 1980(Q1) to 2025(Q2) and show how that, without additional priors,  a large proportion of the posterior distribution for the constant of the monetary policy rule is counter intuitively negative. We then show that imposing a prior on this constant can produce results both more consistent with economic theory and with a larger role for monetary policy in explaining the evolution of macroeconomic variables.  

\newpage

We first describe the Baumeister and Hamilton (2018) model emphasizing the economic interpretation of the constant terms. Their structural VAR is written below in equation (\ref{eq:BH1}), where $x_{t-1}$ is a matrix containing the lagged values of $y_t$, $\pi_t$ and $r_t$.  \vspace{0.2in} 
\small
\begin{equation}
\begin{aligned}
\left[ 
\begin{array}{ccc}
1 & -a_{12} &0 \\
1 & -a_{22} &-a_{23} \\
-a_{31} & -a_{32} &1 \\
\end{array}
\right] \left[
\begin{array}{c}
y_{t} \\
\pi_{t} \\
r_{t} 
\end{array}
\right]= 
\left[
\begin{array}{c}
c^s \\
c^d \\
c^m
\end{array}
\right] + \underset{\Scale[0.9]{(3 \times np)}}{\mathbf b_{}} \underset{\Scale[0.9]{(np \times 1)}}{\mathbf x_{t-1}} + 
 \left[
\begin{array}{c}
u^s_{t} \\
u^d_{t} \\
u^m_{t} 
\end{array}
\right]
\end{aligned}
\label{eq:BH1}
\tag{BH1}
\end{equation}
\normalsize 

The interpretation of the first equation is a supply relationship where contemporaneous supply responds non-negatively to inflation, as in the Phillips curve. Thus the parameter $a_{12}$ in equation (\ref{eq:BH1}) is restricted to be non-negative in the prior. Interest rates are assumed to affect this relationship only with a lag, a common assumption in the macroeconomics literature, and so $r_t$ is multiplied by a zero in this equation. The constant in this equation, $c^s$, therefore has the interpretation as a determinant of supply which is unrelated to inflation or the interest rate i.e. an exogenous determinant of supply. Algebraically, if inflation and the interest rate were at zero for $p$ periods, then the output gap would only depend on $c^s$ and the parameters of its own lags. The intuition of the Phillips Curve is that an economy that is below its steady state inflation rate for an extended period of time will be below its potential output level and hence intuitively we would expect  $c^s$ to be negative. 

The second equation is interpreted as a demand equation where contemporaneous demand responds non-positively to the interest rate. Thus the parameter $ a_{23}$ in equation (\ref{eq:BH1}) is restricted to be non-positive in the prior. Inflation could plausibly affect demand positively or negatively and so $a_{22}$'s sign is not restricted in the prior.   The constant in this equation, $c^d$,  has the interpretation as a determinant of demand which is unrelated to inflation or the interest rate i.e. an exogenous determinant of demand.  A  positive constant, $c^d$, implies that the forces of aggregate demand cause the economy to `overheat' at zero interest rates and inflation,  which is intuitive. 

The third equation is interpreted as the monetary policy rule of the Federal Reserve Board. It is assumed that the Federal Reserve will increase interest rates when inflation increases or output is high. Thus the parameters $a_{31}$ and $a_{32}$ in equation (\ref{eq:BH1}) are assumed to be non-negative in the prior.\footnote{\ \ \ Baumeister and Hamilton (2018) relate this equation to the Taylor rule of Taylor (1993) which is commonly used in the literature, see e.g. Lubik and Schorfheide (2004). They model the contemporaneous coefficients of this equation as the product of the contemporaneous influences on monetary policy $a_{31}$ and $a_{32}$ and a smoothing parameter $\rho$. Their model estimates all of these parameters with $\rho$ restricted to be between 0 and 1 so as not to change the signs of contemporaneous effects.} The constant in this equation, $c^m$  therefore has the interpretation as a determinant of the  interest rate unrelated to inflation or the output gap i.e. an exogenous determinant of the interest rate. A  positive constant, $c^m$, implies  that if the economy is operating at potential with zero inflation then the interest rate would be positive. This corresponds to a positive rate of time preference, which is intuitive.

Note that the long run level of each series depends on all the structural constants, $c^s$, $c^d$  and $c^m$ as well as the autoregressive coefficients, $b_i$, as described in equation (\ref{eq:VAR_mean}).

\subsection{Data}
\label{sec:BH3_data}

We use the updated version of the series in Baumeister and Hamilton (2018). For the three variable VAR, these are real GDP, real potential GDP, the PCE deflator and the Federal Funds Rate which we have taken from FRED website for the sample 1980(Q1)-2025(Q2).\footnote{\ \ \ https://fred.stlouisfed.org/.  } These are series GDPC1, GDPPOT, DPCERD3Q086SBEA and DFF, in FRED.   The output gap is the log of real gdp minus the log of real potential gdp, inflation is measured by the year on year difference of the log PCE deflator and the Federal Funds Rate is not transformed and uses quarterly averages.

\subsection{Estimation}
\label{sec:estimation}

Baumeister and Hamilton (2015)  show how the model can be estimated equation by equation using MCMC methods. We follow Baumeister and Hamilton (2018)  and use their “Minnesota prior” for the autoregressive parameters, asymmetric t distributions for the $A$ matrix parameters together with additional information for the parameters of the monetary policy equation. We take 10 million draws, discarding the first 5 million draws as burn-in and saving 1 in 500 draws as the thinning ratio. This leaves a sample of 10,000 draws. The scale of the jumping distribution is adjusted  during the burn in and achieves acceptance ratios between 27 and 35 percent.

\subsection{Results for the three variable VAR}
\label{sec:BH3_results}

In this section we first discuss the impulse response functions.  These are the building blocks for the discussion of the historical decompositions which are the focus of our attention. The impulse response functions are displayed in Figure \ref{fig:IRF3}. For each structural shock they describe the responses of all variables in the VAR. This is done for each draw from the posterior distribution of the parameters which results in a distribution of impulse responses.  We have colored the responses to each shock differently so that we can use the same colors in the historical decomposition figures, thereby easily distinguishing the contribution of each structural equation to the evolution of the data.

\setcounter{figure}{0}

 \begin{figure}[t]
 
\caption{Impulse responses of the three variable VAR}
\vspace{0.25in}   
\centering\includegraphics[height=3.5in, keepaspectratio]{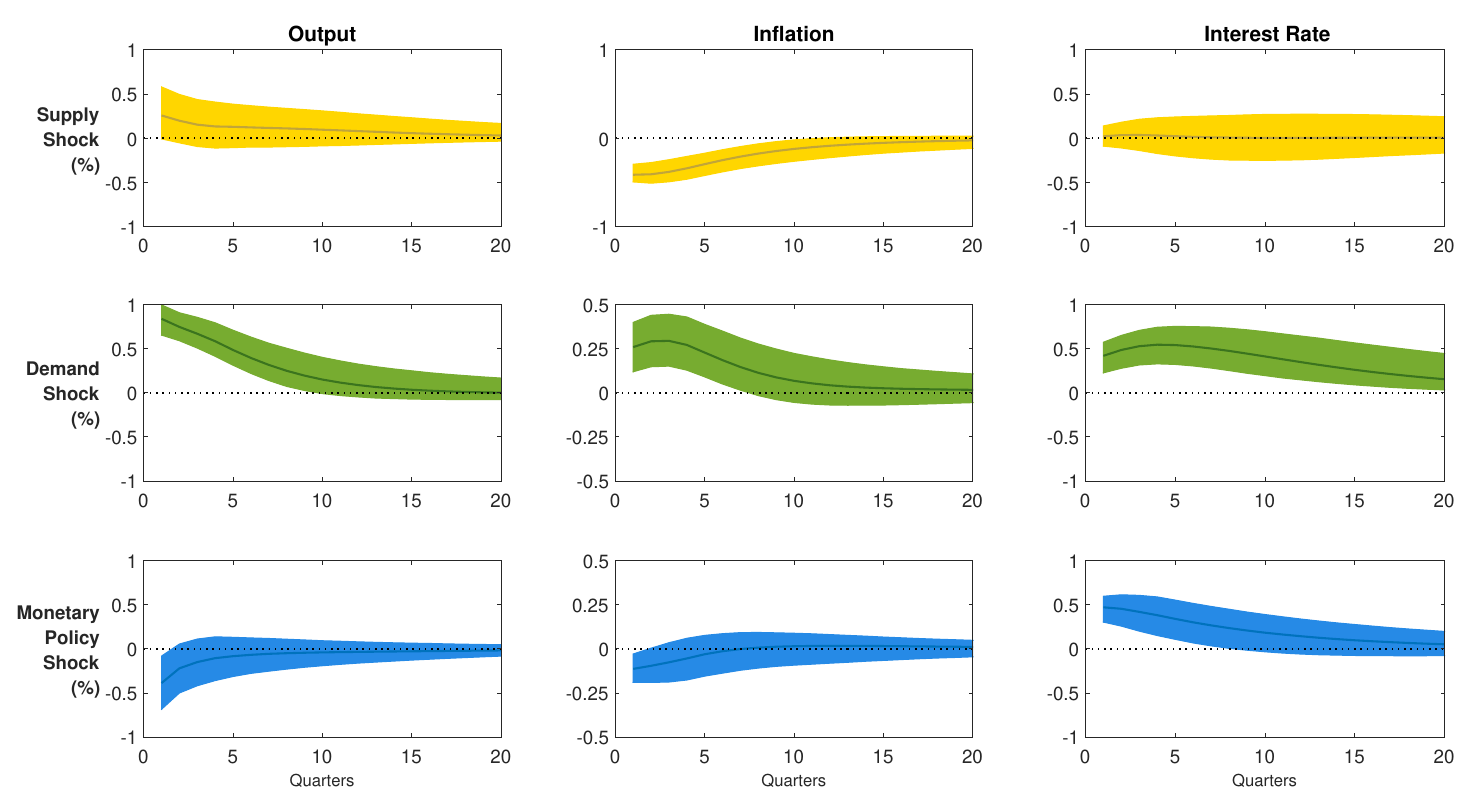}
\vspace{0.015in}  
\label{fig:IRF3}
\noindent \flushleft \footnotesize
Figure \ref{fig:IRF3} plots the impulse responses of the variables in the three variable VAR to the three fundamental shocks described in Section \ref{sec:BH3VAR}, for the case where there is a prior for the structural constant in the Taylor rule.  The impulses of the supply shocks are colored yellow, those of the demand shock colored green and those of the monetary policy shock colored blue.  The shaded region is the 95\% confidence band of the posterior distribution with the median colored with a darker line. 
 \vfill
\end{figure}

The first identified shock in Figure \ref{fig:IRF3}, colored yellow, is the supply shock. Its impulse responses show a positive effect on output alongside a negative effect on inflation. These impulses correspond to how we think a productivity shock might impact the economy. The economy is able to produce more for a given level of inflation, and is therefore a shift in the Phillips curve. The effect of this shock on interest rates is centered around zero, which is not surprising given the opposing forces operating on the Taylor rule. 

The second identified shock in Figure \ref{fig:IRF3}, colored green, is the demand shock. Its impulse responses are positive for all variables and correspond to how we think a demand shock might impact the economy. Increased demand raises output and also raises inflation via the Phillips curve. Interest rates also rise as both elements in the Taylor rule are increasing. 

The third identified shock in Figure \ref{fig:IRF3}, colored blue, is the monetary policy shock. Its impulse responses are also very intuitive. An unexpected increase in the interest rate reduces output  and inflation. Interest rates converge slowly back to its steady state level, with most of the posterior still higher than the steady state three years after impact.

\subsubsection*{Historical Decompositions}

Historical decompositions depict the contribution of each of these three structural shocks to the evolution of the three series  using a stacked bar graph.  We plot the median historical decomposition with the contribution of the demand shock colored green, the supply shock colored yellow and the monetary policy shock colored blue.  We also decompose the  constant terms  into their structural contributions, as per equation (\ref{eq:J_IRF}). We use the same colors for the constants, although with a slightly lighter shade to differentiate.  The actual data series are plotted in red and the contribution of the initial conditions is  colored light grey.   The initial conditions have a large contribution at the start of the sample. This is the inherited contribution of previous shocks, notably the higher than average inflation and interest rates in the years preceding the start of the sample. Consistent with equation (\ref{eq:decomp}), this contribution dies out over time.

\newpage
\thispagestyle{empty}
 \begin{figure}[H]
\caption{Historical decompositions of VAR without additional priors on the structural constants}
  \begin{subfigure}{.95\textwidth}
\vspace{0.1in}   \caption{Output gap }
\centering\includegraphics[width=0.95\linewidth, height= 6.0cm]{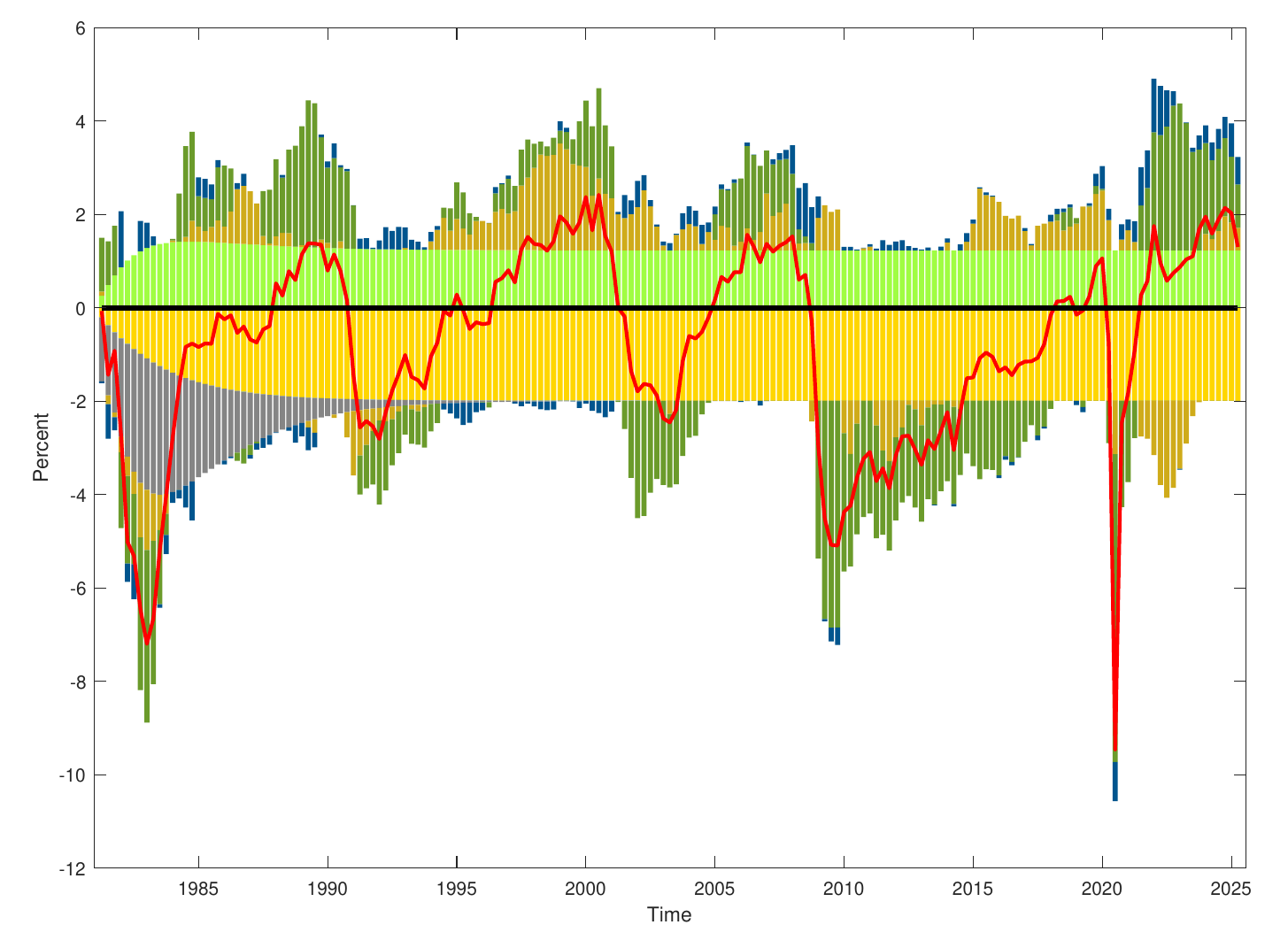}
\label{fig:outstack_NOPRIOR}
  \end{subfigure}
  \begin{subfigure}{.95\textwidth}
\vspace{0.01in}  \caption{Inflation}
\centering\includegraphics[width=0.95\linewidth, height= 6.0cm]{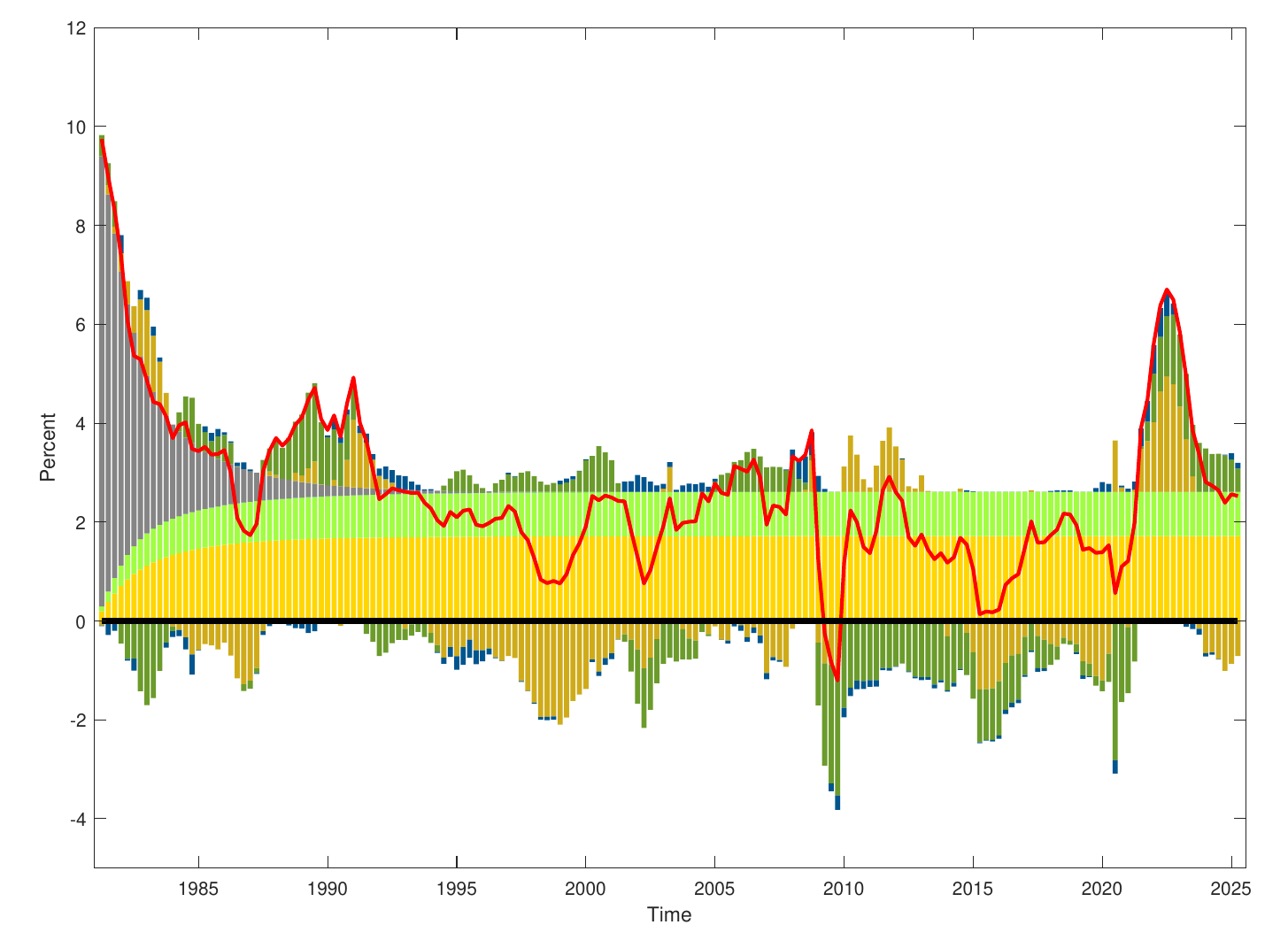}
 \label{fig:infstack_NOPRIOR}
  \end{subfigure}
  \begin{subfigure}{.95\textwidth}
\vspace{0.01in}       \caption{Interest Rate}
\centering\includegraphics[width=0.95\linewidth, height= 7.5cm]{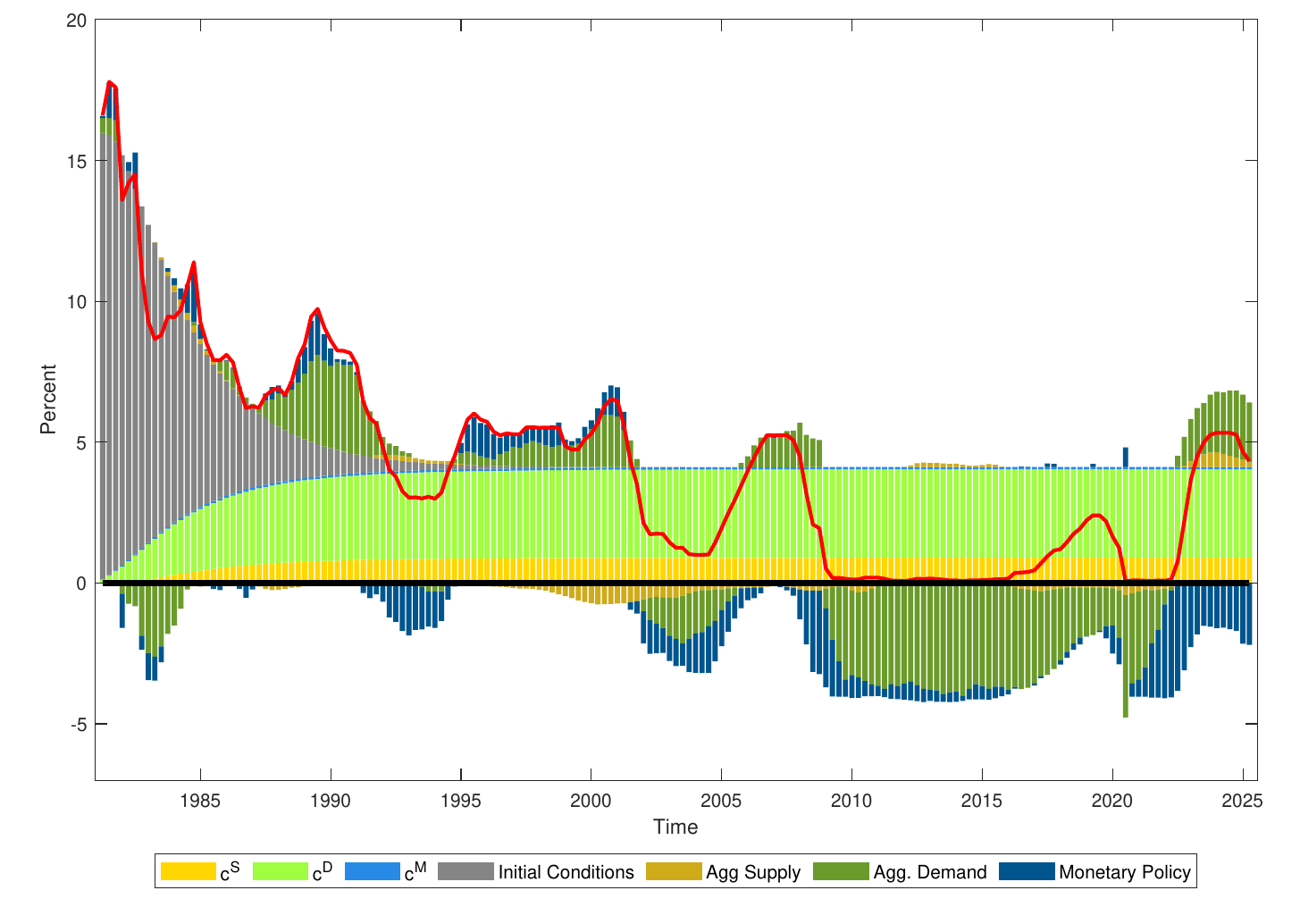}  
 \label{fig:intstack_NOPRIOR}
  \end{subfigure}
   \label{fig:BHHD}
\noindent \flushleft \footnotesize

Figure \ref{fig:BHHD} plots the median estimates of the historical decomposition of the output gap, inflation, and the interest rate for the case where there are no additional priors on  the structural constants. It also decomposes the reduced form VAR's constant terms into their structural components. The yellow bars are the shocks and structural constant from the supply equation, the green bars are those from the demand equation and the blue bars those from the monetary policy equation. The  constant terms are given a slightly lighter shade. In each panel the red line plots observed series and the effect of the initial conditions in gray. 
\end{figure}

\begin{figure}[H]
\caption{Historical decompositions of the VAR with prior on the structural constants}
  \begin{subfigure}{.99\textwidth}
\vspace{0.1in}   \caption{Output gap }
\centering\includegraphics[width=0.95\linewidth, height= 6.0cm]{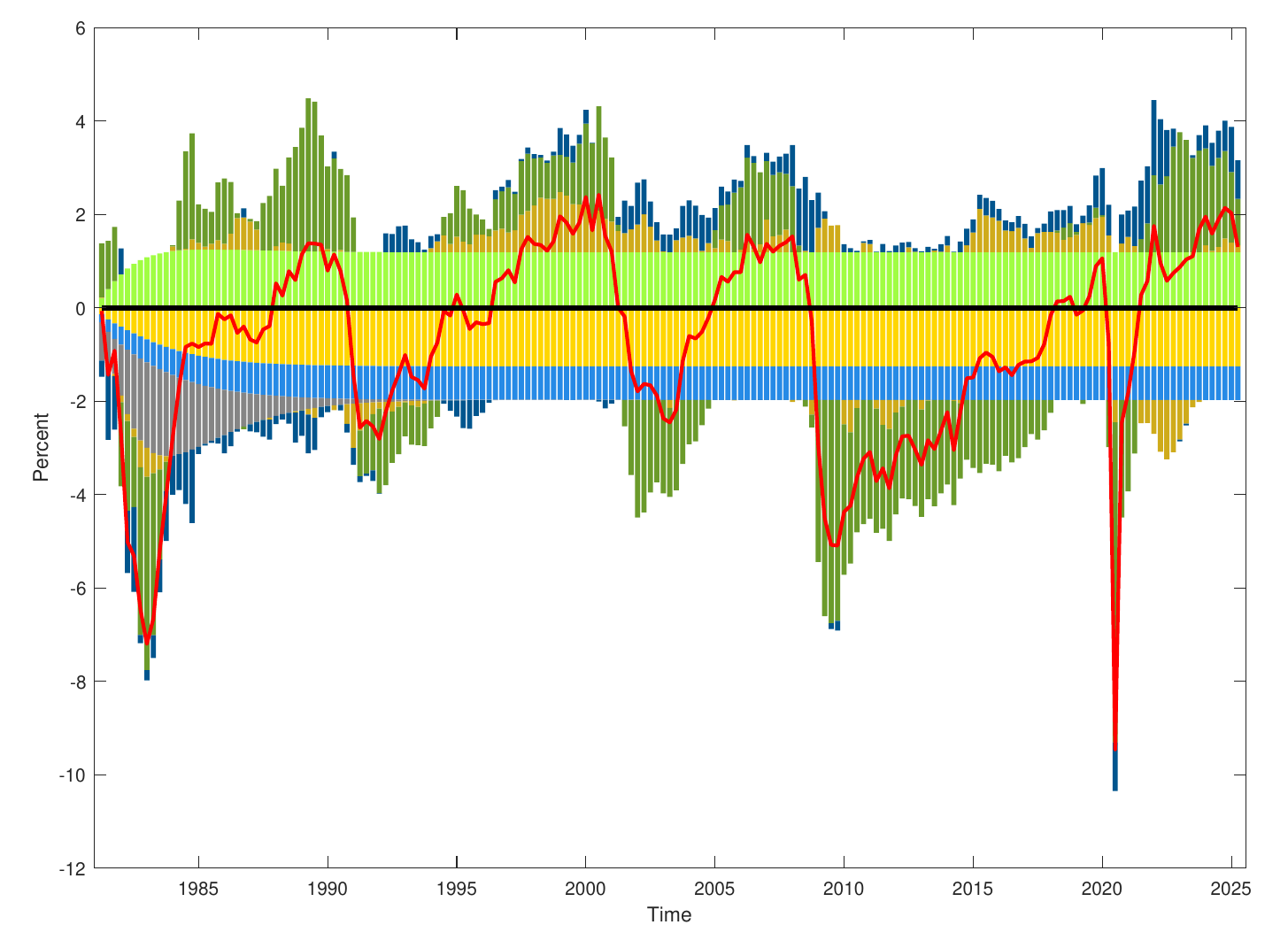}
\label{fig:Aggmigstack}
  \end{subfigure}
  \begin{subfigure}{.99\textwidth}
\vspace{0.1in}  \caption{Inflation}
\centering\includegraphics[width=0.95\linewidth, height= 6.0cm]{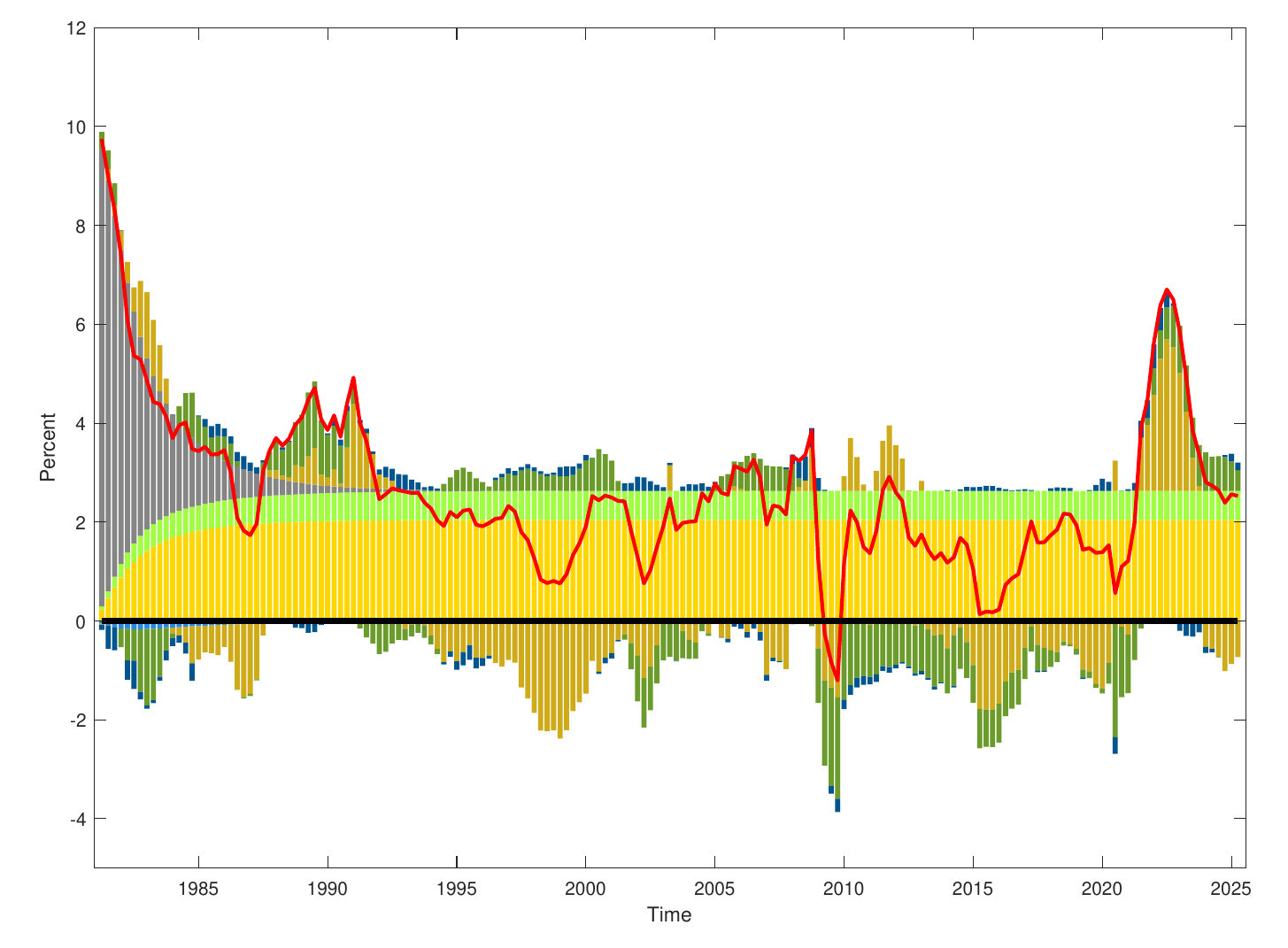}
 \label{fig:Agghoursstack}
  \end{subfigure}
  \begin{subfigure}{.99\textwidth}
\vspace{0.1in}       \caption{Interest Rate}
\centering\includegraphics[width=0.95\linewidth, height= 7.5cm]{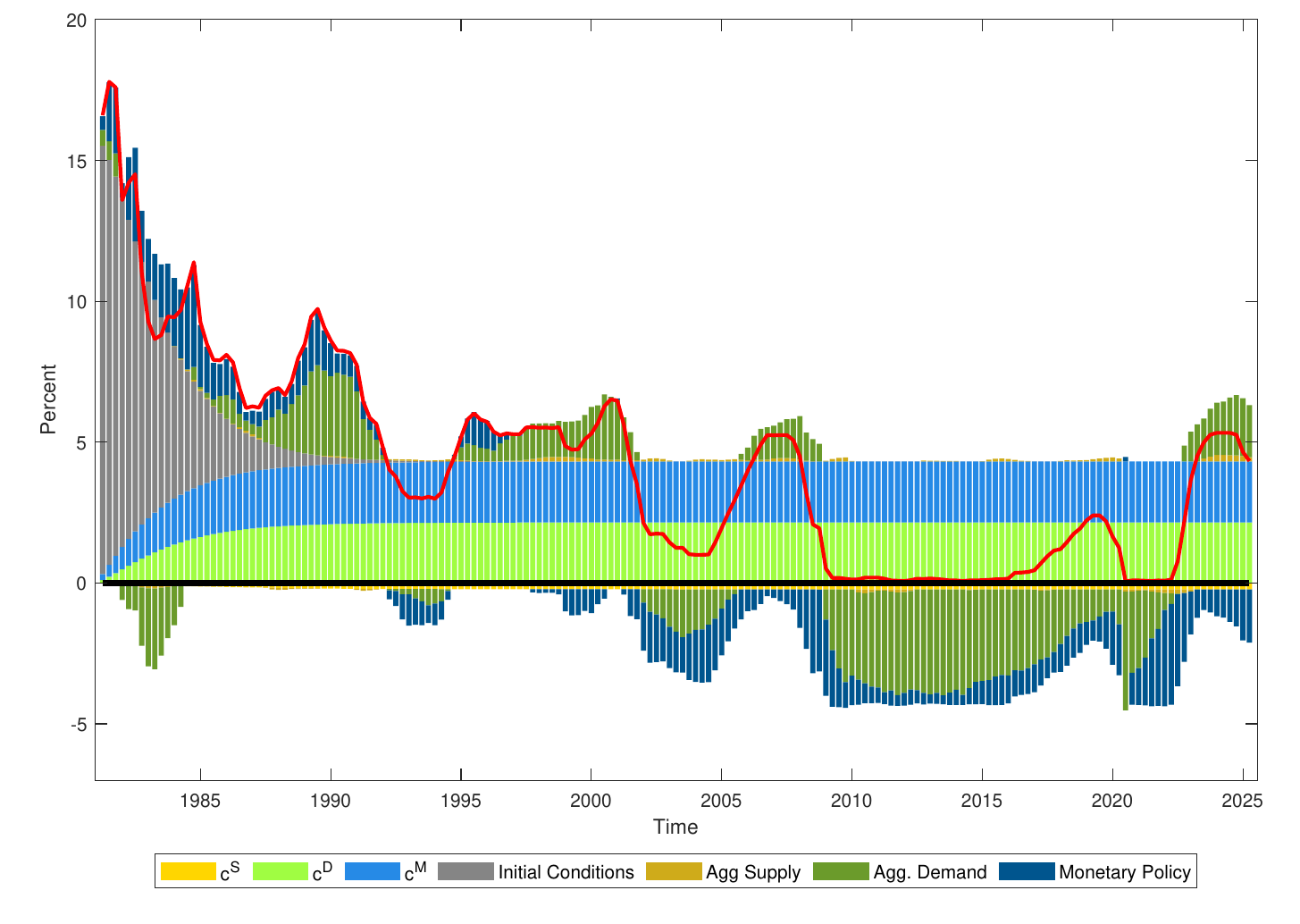}  
 \label{fig:AggWagesstackPEN}
  \end{subfigure}
   \label{fig:BHHD_prior}
\noindent \flushleft \footnotesize
Figure \ref{fig:BHHD_prior} plots the median estimates of the historical decomposition of the output gap, inflation, and the interest rate for the case where there is a prior for the structural constant of the Taylor Rule equation. It also decomposes the reduced form VAR's constant terms into their structural components. The yellow bars are the shocks and structural constant from the supply equation, the green bars are those from the demand equation and the blue bars those from the monetary policy equation. The  constant terms are given a slightly lighter shade. In each panel the red line plots observed series and the effect of the initial conditions in gray.       
 \vfill
\end{figure}

Figure \ref{fig:BHHD} displays the historical decompositions without an additional  prior over the structural constants and Figure \ref{fig:BHHD_prior} plots the case when there is such a prior on the structural constant for the Taylor Rule equation. The novel aspects of Figures \ref{fig:BHHD} and  Figure \ref{fig:BHHD_prior} compared to the previous literature is their inclusion of the decomposition of the contribution of the constant terms into their structural components as described in section \ref{sec:Accounting}. 

The contribution of the structural constants, $c^{s}$ and $c^d$, in Figure \ref{fig:BHHD} is quite intuitive. The contribution of $c^s$, the constant in the Phillips curve equation, represented by the yellow bars, is negative for the steady state output gap and positive for steady state inflation. This means that in the absence of the other equations, inflation would be positive but below the level needed for output to be at its potential. This is in line with textbook macroeconomic analysis based on the Phillips Curve e.g. Barro and Gordon (1983).  The contribution of  the constant in the demand equation, $c^d$, represented by the green bars, is positive on both steady state inflation and the output gap. This is consistent with the idea that the forces of aggregate demand cause the economy to run above potential and `overheat'.  For the variation, Figure \ref{fig:BHHD} also accords with our understanding of the sample period with respect to the contributions of demand and supply. There are large negative demand effects on output after the 2007-8 financial crisis and negative supply and positive demand effects after the covid epidemic. For inflation the post covid surge is accounted for by a  mixture of demand supply effects as in Bergholt et al. (2025). 

However the contribution of the constant in the monetary policy rule, $c^m$, represented by the blue bars, is almost zero.  This contribution is not intuitive. As noted above, one would expect an economy with a positive rate of time preferences to have a positive interest rate in the absence of excess demand and  inflation effects.  Instead the steady state interest rate is almost entirely determined by the constant in the demand equation,  $c^d$.  The contributions of the Taylor rule to both steady state inflation and the output gap is also very small, which runs counter to standard macroeconomic theory. For the variation in the data, monetary policy also plays very little role in the variation of output and inflation, and the variation in the interest rate is dominated by demand. 

\begin{figure}[!t]
\caption{Posterior distribution of the structural constants with and without additional prior }
  \begin{subfigure}{.49\textwidth}
\vspace{0.1in}   \caption{Without priors over structural constants}
\centering\includegraphics[height=2.5in, keepaspectratio]{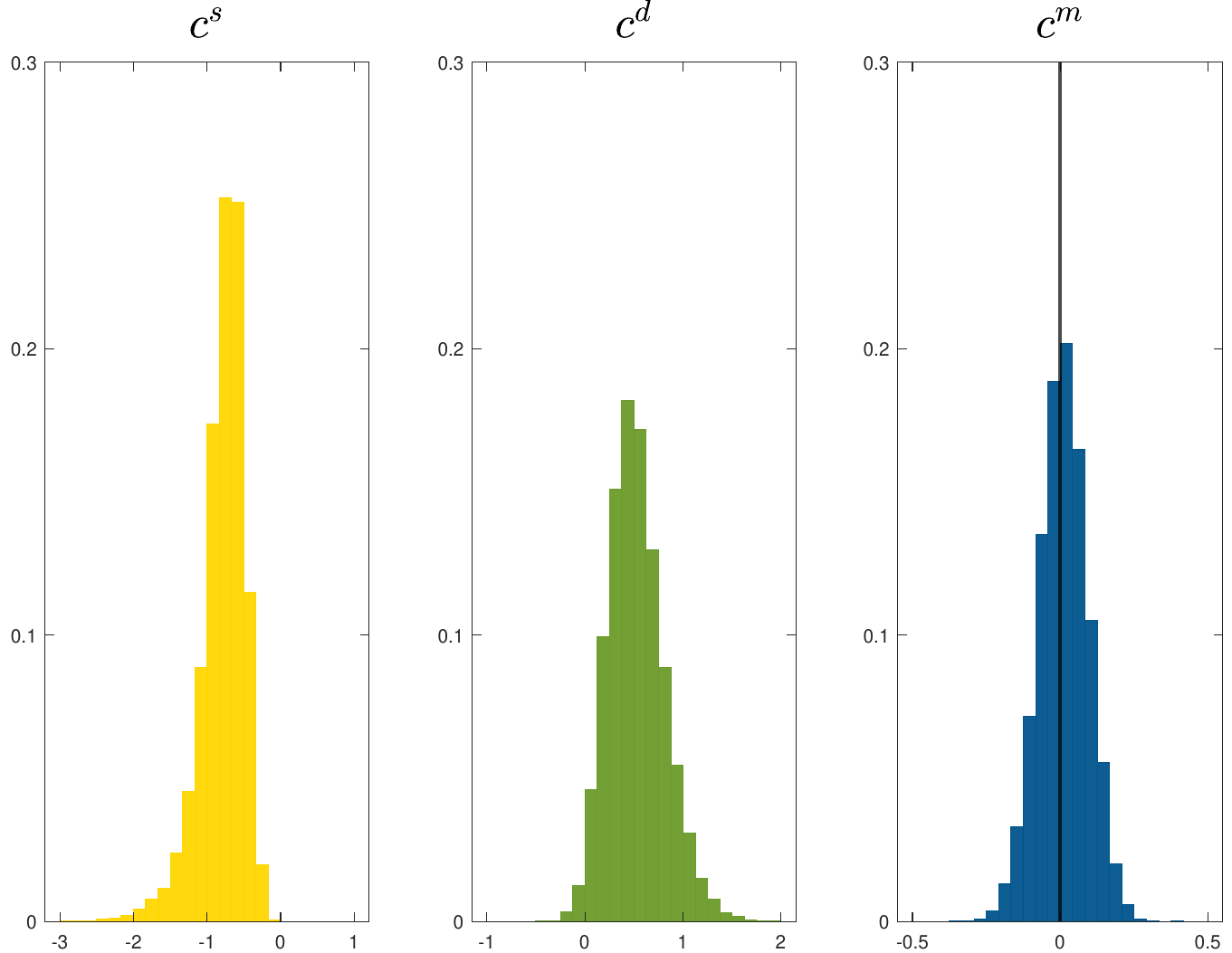}
     \label{fig:BHrep_CONST_ORIG}
  \end{subfigure}
    \begin{subfigure}{.49\textwidth}
  \vspace{0.1in}       \caption{With priors over structural constants}
\centering\includegraphics[height=2.5in, keepaspectratio]{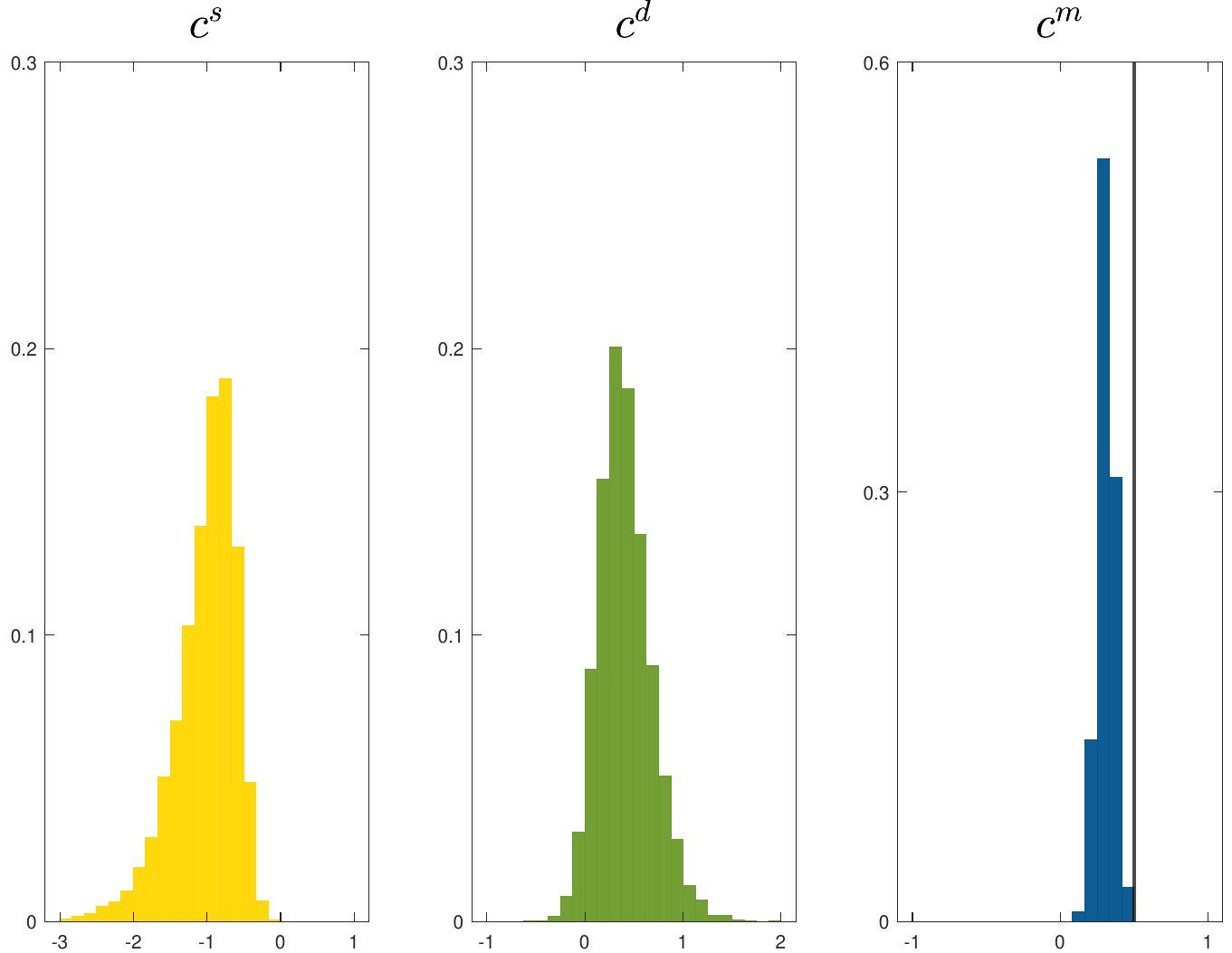}
    \label{fig:BHrep_CONST_PRIOR}
  \end{subfigure}
   \label{fig:BHrep_CONSTANTS}
\noindent \flushleft \footnotesize
Figure \ref{fig:BHrep_CONSTANTS} plots the posterior distributions of the structural constants. Panel a) plots the case  without additional priors over the structural constants.  Zero is represented by a vertical black line for $c^m$'s distribution.  Panel b) plots the case with an additional prior over $c^m$'s,  as described in section \ref{sec:Structural_priors}. The mean of the additional prior, $0.5$, is represented by a vertical black line.       
 \vfill
\end{figure}

The reason for the lack of influence of $c^m$ is shown in Figure \ref{fig:BHrep_CONSTANTS} which plots the posterior distributions of the structural constants for the models without and with an additional prior over the structural constant for the Taylor Rule.  When there is no additional prior, Figure \ref{fig:BHrep_CONST_ORIG} shows that the posterior distribution of $c^m$ has sizeable portions of the density above and below zero, with the mode being very close to zero.  This does not conform to our understanding of what a Taylor rule should be, and so these results motivate the estimation of the model with a prior over the structural constant in the Taylor rule equation.  

The prior we impose sets  the prior mean for $c^m$ equal to 0.5  and $V_i$ of $0.01$, using the notation of section \ref{sec:Structural_priors}. The effect of the prior on the posteriors of the structural constants is seen by comparing Figures \ref{fig:BHrep_CONST_ORIG} and \ref{fig:BHrep_CONST_PRIOR}. In Figure \ref{fig:BHrep_CONST_PRIOR} the posterior distribution for $c^m$ is shifted to the right with its mode in between that in Figure \ref{fig:BHrep_CONST_ORIG} and the prior mode of 0.5, as is standard in Bayesian analysis.

The effect of this prior on the historical decompositions is seen by comparing Figures \ref{fig:BHHD} and \ref{fig:BHHD_prior}. The most apparent difference is in the historical decomposition of the interest rate.  In Figure \ref{fig:BHHD_prior} the Taylor rule equation, colored blue,  significantly affects both the level and variation of the interest rate, with the latter being most striking during the 2007-2008 financial crisis and the covid epidemic. Nevertheless it is still the case that demand equation has a large influence on the level and variation of the interest rate, both in terms of variation and in its long run level. The supply equation again has very little effect on the interest rate as was the case in Figure \ref{fig:BHHD}. This is consistent with its impulse responses, shown in Figure \ref{fig:IRF3}, where the supply shock moved inflation and output in opposite directions. 

The historical decomposition for the output gap in Figure \ref{fig:BHHD_prior} shows the same dominant roles of the demand and supply equations, but importantly there is now a negative influence for the Taylor Rule equation both in its level and variation, being particularly evident in the rebound after the Covid epidemic. This negative role of the interest rate rule on output is consistent with macroeconomic textbook models.     

Finally, the historical decompositions for inflation in Figures \ref{fig:BHHD}  and \ref{fig:BHHD_prior} are very similar. Monetary policy plays very little role in determining inflation. This is counter-intuitive and may indicate that the three dimensional VAR is too small i.e. there may be omitted variable bias as in e.g. Sims (1992). We investigate this in section \ref{sec:BH6VAR} where we again follow Baumeister and Hamilton (2018) in estimating a larger six variable VAR with  commodity price inflation, the corporate risk premium and real wage growth being the additional variables. 

\newpage
\thispagestyle{empty}

\newpage

\section{\Scale[0.95]{\mbox{A six variable VAR}}}
\label{sec:BH6VAR}

In section \ref{sec:BH3VAR} we showed that the addition of a prior over the structural constant parameter in the Taylor Rule equation produced results with a greater role for monetary policy in determining interest rates. However it was still the case that monetary policy played very little role in explaining inflation. One possible reason for this is that the three variable VAR is too small and only imperfectly captures inflation dynamics. We therefore follow Baumeister and Hamilton (2018) and consider a larger VAR including additional data on commodity price inflation, real wage growth and the corporate risk spread, as we detail below.

\subsection{Data}
\label{sec:BH6_data}

The additional three variables used in the VAR are commodity price inflation, real wage growth and the corporate risk spread. They are all taken from publicly available series on the  FRED website. The commodity price index is the PPIACO series, which is `Producer Price Index by Commodity: All Commodities'. Commodity price inflation is the year on year difference of the log of this series. The real wage data is the AHETPI series, which is `Average Hourly Earnings of Production and Nonsupervisory Employees, Total Private, Dollars per Hour', which we deflate by the PCE deflator of section \ref{sec:BH3VAR} for consistency. Real wage growth is given by the year on year difference of the log of real wages.  The corporate risk spread data is the series BAA10YM, which is `Moody's Seasoned Baa Corporate Bond Yield Relative to Yield on 10-Year Treasury Constant Maturity'. This series is not transformed. 

\subsection{\Scale[0.95]{\mbox{Identifying and interpreting the six variable VAR }}}
\label{sec:BH6_identify}

\begin{figure}[!t]
\caption{Posterior distribution of structural constants of six variable VAR }
  \vspace{0.1in}      
\centering\includegraphics[width=6.5in, height=2.5in]{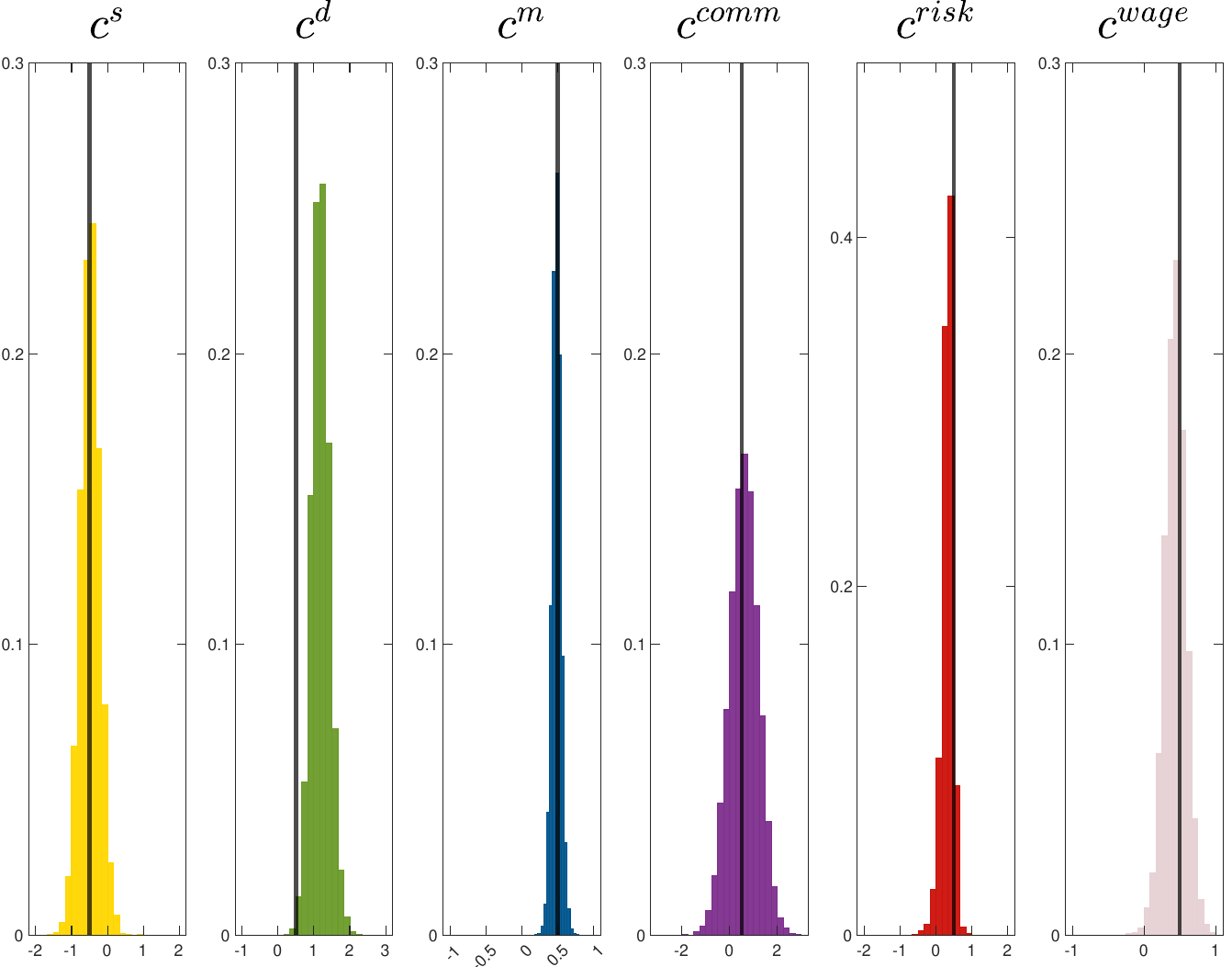}
  \label{fig:BH_6var_CONSTANTS}
\noindent \flushleft \footnotesize
Figure \ref{fig:BH_6var_CONSTANTS} plots the posterior distribution of the structural constants of the six variable VAR, for the case with additional priors over the structural constants, as described in section \ref{sec:Structural_priors}. The mean of the prior is depicted by a vertical black line in each case. 
\vfill      
\end{figure}

The extra variables in the VAR require additional identifying assumptions. Our approach is to maintain the intuition of the three variable  model described of section \ref{sec:BH3VAR}, whilst at the same time identifying three additional shocks using sign restrictions consistent with standard macroeconomic theory. We therefore impose priors over the structural constants for the supply, demand and monetary policy equations that $c^s$ is negative, and $c^d$ and $c^m$ are positive.  Using the notation of section \ref{sec:Structural_priors}, we set the prior mean for $c^s$ equal to $-0.5$ and for that for $c^d$ and $c^m$ equal to $0.5$. We also set the prior mean of the structural constants for the new variables equal to $0.5$.  We set $V_i$ equal to $0.1$ for all the structural constants  except for $c^s$ and $c^m$ where it is set to $0.01$. The posteriors from the structural constants are shown in Figure \ref{fig:BH_6var_CONSTANTS}.

For the sign restrictions on the enlarged VAR we follow Baumeister and Hamilton (2018) in assuming that a monetary policy shock has a negative effect on output, inflation and real wages and a positive effect on interest rates and the risk spread. We differ from Baumeister and Hamilton (2018) however, in assuming that it has no impact on commodity prices, which we model as being determined globally.  For the demand shock we assume that it is positively correlated with all domestic variables and for the supply shock we assume that real wages and output are positively correlated, and that these shocks are negatively correlated with inflation and the interest rate, which is consistent with its interpretation as a productivity shock. These sign restrictions also distinguish it from the real wage shock described below. 

The three new identified shocks are a commodity price shock, a risk shock and a cost push wage shock.  We regard commodity prices to be exogenous to the domestic economy  and so are not affected contemporaneously by domestic variables.\footnote{ \ \  Note that the commodity price process still depends on the lagged values of domestic variables. A stronger exogeneity assumption would impose a stronger prior on the non-own term autoregressive parameters then the Minnesota prior as in e.g. Mountford and Wadsworth (2025).} Thus the $S$ matrix has a row of zeros  in  row 4, for the commodity price inflation equation.  We also assume that commodity price shocks are positively correlated with inflation and the risk spread.  For the risk spread equation we assume that risk spread is mostly determined globally but we allow it to respond contemporaneously to domestic monetary policy, following Baumeister and Hamilton  (2018), and also to commodity price shocks. We also employ sign restrictions to distinguish risk spread shocks from monetary policy shocks which are that risk spread shocks are negatively correlated with all the other domestic variables.  Finally the real wage shock is assumed to be a `cost push' shock  where the real wage is positively correlated with inflation and the interest rate and negative correlated with output. These sign restrictions are described in equation (\ref{eq:SR6}) below 

\newpage

The estimated VAR therefore includes the following 6 variables: the output gap, $y_t$, inflation, $\pi_t$, the nominal interest rate, $r_t$, commodity price inflation, $\pi^{\Scale[0.65]{\mbox{comm}}}_t$, the corporate risk spread $\Scale[0.85]{\mbox{risk}}^{\Scale[0.65]{\mbox{spread}}}_t$,  and real wage growth, $w_t$. Given the restrictions described above, it can be written  
\begin{equation}
\begin{aligned}
\Scale[0.875]{
\underbrace{\addstackgap[6pt]{\left[
\begin{array}{cccccc}
1 & -a_{12}&0 &-s_1&-s_2&-s_3\\
1 & -a_{22} &-a_{23} &-s_4&-s_5&-s_6\\
-a_{31} & -a_{32} &1 &-s_7&-s_8&-s_9\\
0 &0&0&1&0&0\\
0&0&-s_{10} &-s_{11}&1&0\\
-s_{12} &-s_{13}&-s_{14} &-s_{15}& -s_{16}&1\\
\end{array}
\right]}}_{\addstackgap[6pt]{\Scale[1.01]{ S }}}
}
\Scale[0.875]{
\underbrace{\addstackgap[6pt]{\left[
\begin{array}{c}
y_t\\
\pi_t \\
r_t \\
\pi^{\Scale[0.65]{\mbox{comm}}}_t \\
\Scale[0.85]{\mbox{risk}}^{\Scale[0.65]{\mbox{spread}}}_t\\
w_t
\end{array}
\right]}}_{\addstackgap[6pt]{\Scale[1.1]{ \ y_t}}}
= 
\underbrace{\addstackgap[6pt]{\left[
\begin{array}{c}
c^s \\
c^{d} \\
c^r \\
c^{\Scale[0.65]{\mbox{comm}}} \\
c^{\Scale[0.65]{\mbox{risk}}} \\
c^{\Scale[0.65]{\mbox{wage}}} \\
\end{array}
\right]}}_{\addstackgap[6pt]{\Scale[1.01]{C}}} 
+ 
\underbrace{\addstackgap[6pt]{
\begin{array}{c}
\ \\
bx_{t-1}\\
\
\end{array}
}}_{\addstackgap[6pt]{\Scale[0.66]{(6\times 6p)(6p \times 1)}}}+ 
\underbrace{\addstackgap[6pt]{ \left[ 
\begin{array}{c}
u^s_{t} \\
u^d_{t} \\
u^m_{t} \\
u^{\Scale[0.65]{\mbox{comm}}}_t  \\
u^{\Scale[0.65]{\mbox{risk}}}_t\\
u^{\Scale[0.65]{\mbox{wage}}}_t
\end{array}
\right]}}_{\addstackgap[6pt]{\Scale[1.15]{u_t}}} 
} 
\label{eq:SVAR6}
\end{aligned}
\end{equation}
\normalsize

The identifying sign restrictions are summarized in equation (\ref{eq:SR6}) which uses the reduced form VAR.  This shows that each shock has a distinctive pattern. For the new shocks the commodity price shocks is identified as being the only shock that has a contemporaneous effect on commodity prices. The real wage shock is the only shock that has zero impact on commodity prices and a negative correlation between real wages and output. The risk shock is distinguished from the demand shock by the negative correlation between the risk spread and the interest rate. The relationship between monetary policy and risk is an important one and we explore this further in section \ref{sec:robust}.

\begin{equation}
\begin{aligned}
\Scale[0.95]{
\underbrace{\addstackgap[6pt]{\left[
\begin{array}{c}
y_t\\
\pi_t \\
r_t \\
\pi^{\Scale[0.65]{\mbox{comm}}}_t \\
\Scale[0.85]{\mbox{risk}}^{\Scale[0.65]{\mbox{spread}}}_t\\
w_t
\end{array}
\right]}}_{\addstackgap[6pt]{\Scale[1.1]{ \ y_t}}}
= 
\underbrace{\addstackgap[6pt]{\left[
\begin{array}{c}
\kappa^s \\
\kappa^{d} \\
\kappa^r \\
\kappa^{\Scale[0.65]{\mbox{comm}}} \\
\kappa^{\Scale[0.65]{\mbox{risk}}} \\
\kappa^{\Scale[0.65]{\mbox{wage}}} \\
\end{array}
\right]}}_{\addstackgap[6pt]{\Scale[1.01]{S^{-1}C}}} 
+ 
\underbrace{\addstackgap[6pt]{
\begin{array}{c}
\ \\
bx_{t-1}\\
\
\end{array}
}}_{\addstackgap[6pt]{\Scale[0.66]{(6\times 6p)(6p \times 1)}}}
+ 
\Scale[0.995]{
\underbrace{\addstackgap[6pt]{\left[
\begin{array}{cccccc}
\mathbf{+} & \mathbf{+} &\mathbf{-} &           & \mathbf{-}  &\mathbf{-}\\
\mathbf{-} & \mathbf{+} &\mathbf{-} &\mathbf{+} &  \mathbf{-} &\mathbf{+}\\
\mathbf{-} & \mathbf{+} &\mathbf{+} &           &  \mathbf{-} &\mathbf{-}\\
         0 &  0         &  0        &\mathbf{+} & 0            &    0\\
           & \mathbf{+} & \mathbf{+} &\mathbf{+} & \mathbf{+}  &     \\
\mathbf{+} &\mathbf{+}  &\mathbf{-} &            &  \mathbf{-} &\mathbf{+}\\
\end{array}
\right]}}_{\addstackgap[6pt]{\Scale[1.01]{ S^{-1} }}}
}
\underbrace{\addstackgap[6pt]{ \left[ 
\begin{array}{c}
u^s_{t} \\
u^d_{t} \\
u^m_{t} \\
u^{\Scale[0.65]{\mbox{comm}}}_t  \\
u^{\Scale[0.65]{\mbox{risk}}}_t\\
u^{\Scale[0.65]{\mbox{wage}}}_t
\end{array}
\right]}}_{\addstackgap[6pt]{\Scale[1.15]{u_t}}} 
} 
\label{eq:SR6}
\end{aligned}
\end{equation}
\normalsize

\begin{figure}[t]
\caption{Impulse responses from the six variable VAR}
\vspace{0.1in}   
\centering\includegraphics[width=0.995\linewidth, keepaspectratio]{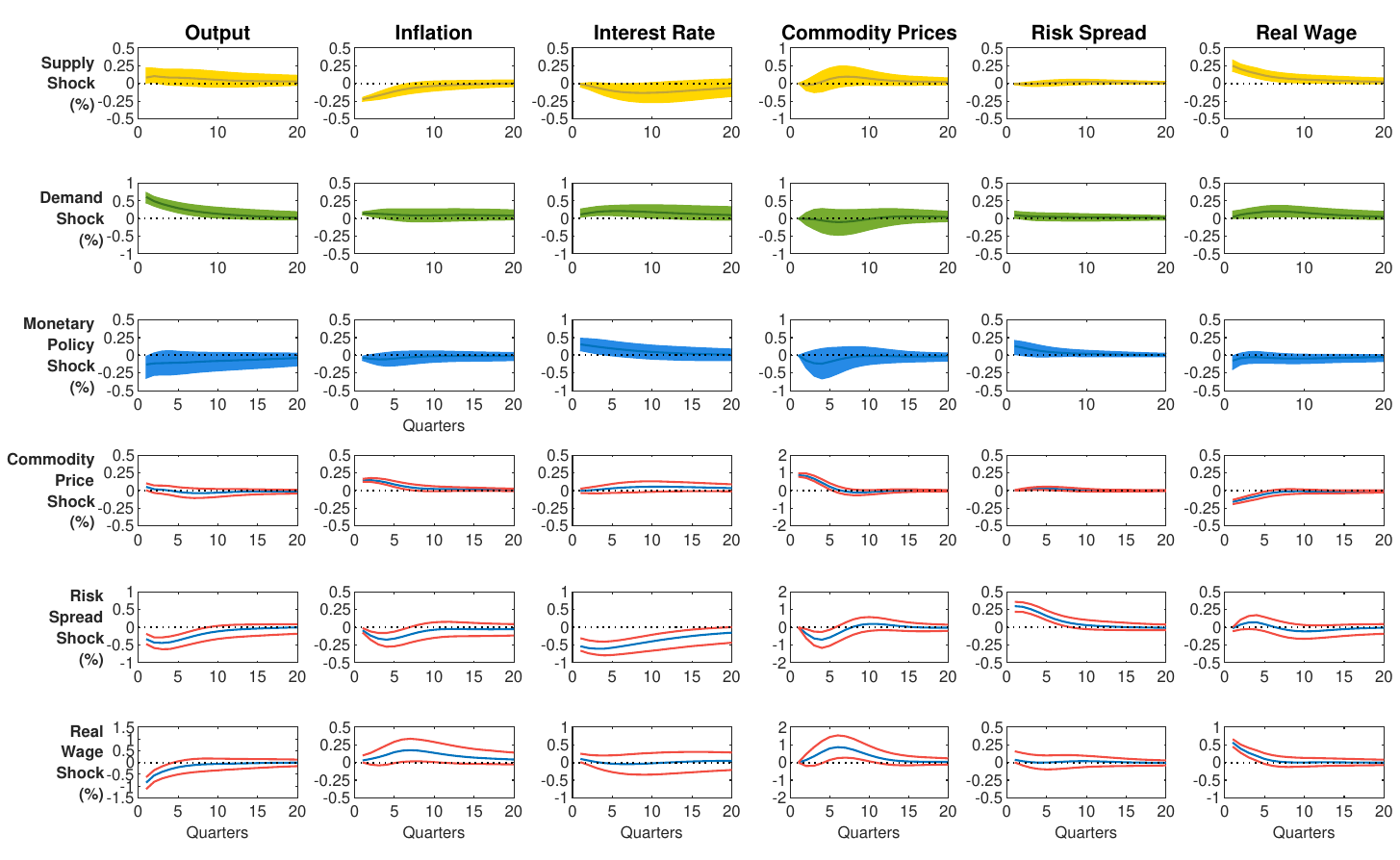}
\label{fig:IRF6_PRIOR}
\noindent \flushleft \footnotesize
Figure \ref{fig:IRF6_PRIOR} plots the impulse responses of the variables in the six variable VAR with the additional priors over the structural constants described in section \ref{sec:BH6_identify}.  The median impulses and 95\% confidence bands of the supply shocks are colored yellow, those of the demand shock colored green and those of the monetary policy shock colored blue, to relate them to the corresponding impulses in Figure \ref{fig:IRF3}. The impulse responses of the three additional shocks are shown in the last three rows where the median is colored blue and the upper and lower edges of the confidence bands are colored red. 
 \vfill
\end{figure}

These sign restrictions are intuitive in and of themselves,  but their plausibility is reinforced by the resulting impulse responses functions displayed Figure \ref{fig:IRF6_PRIOR} and the associated historical decompositions in Figures \ref{fig:BHHD_6var_1} and \ref{fig:BHHD_6var_2}, which we discuss below.

\newpage

\subsection{\Scale[0.95]{\mbox{Results from the six variable VAR}}}
\label{sec:BH6_results}

In this section we display results for the case where we impose priors on the structural constants of the VAR. The results for the case where there are no additional priors on the structural constants are given in the Appendix \ref{sec:appendix_NOPRIOR}. These are similar to the case of no additional prior in the three variable VAR in that the posterior distributions for  $c^{m}$, is close to zero and the monetary policy equation plays very little role in the historical decomposition of the interest rate.  In contrast for the case with priors on the structural constants, which we examine in this section,  the structural constants of the main structural equations take the sign that one would expect, with $c^m$ and $c^d$, positive and  $c^s$ negative for almost all draws. These are shown in Figure \ref{fig:BH_6var_CONSTANTS}.

The impulse responses of the six variable VAR are displayed in Figure \ref{fig:IRF6_PRIOR}. For the three original structural shocks of section \ref{sec:BH3VAR} we have used the same color scheme as in Figure \ref{fig:IRF3} so that the impulse responses to a monetary policy shock are colored blue, those of the supply shocks are colored yellow and those of the demand shock colored green.  The responses follow the sign restrictions of equation (\ref{eq:SR6}) and are mostly intuitive. The impulses of commodity prices which are restricted to be zero on impact, stay close to zero after impact for the demand and monetary policy shocks, although that for the supply shock has large positive section between 1 and 3 years after impact. 

The impulse responses of the three additional shocks are shown in the last three rows of Figure  \ref{fig:IRF6_PRIOR}. The  median response for these shocks are  colored blue and the upper and lower edges of the confidence bands are colored red. The impulses for the commodity price shock are intuitive, inflation rises and real wages correspondingly fall. The interest rate rises and output falls, although with a lag.  The risk spread shock is also intuitive and has large negative output, inflation and interest rate effects. As we show in the historical decomposition in Figure \ref{fig:BHHD_6var_2} below, these responses are  picking up a lot of the effects of the financial crisis of 2007-2008.  

The real wage shock is mostly intuitive in that it resembles a cost push shock with a positive inflation response and negative output response. However the response of commodity prices is noticeably significantly positive after a lag.  It would be strange if a cost push shock which reduces US output had a positive effect on world commodity prices. Therefore we think that  it is best to regard this shock as an anticipated commodity price increase shock, as in e.g. Mountford and Uhlig (2009). The anticipation of a future commodity price rise causes real wages to rise temporarily until the shock is realized. This dynamic is not picked up by the commodity price shock which causes real wages to fall. Thus in this interpretation the commodity price shock and the real wage shock should be seen together as different aspects - instantaneous and anticipated -  of a commodity price shock. We explore this interpretation of the real wage shock when we discuss the historical decompositions below. 

\subsubsection*{Historical Decompositions}

\begin{figure}[H]
\caption{Historical decomposition of output, inflation and interest rate in the six variable VAR}
  \begin{subfigure}{.99\textwidth}
\vspace{0.1in}   \caption{Output gap }
\centering\includegraphics[width=0.95\linewidth, height= 6.0cm]{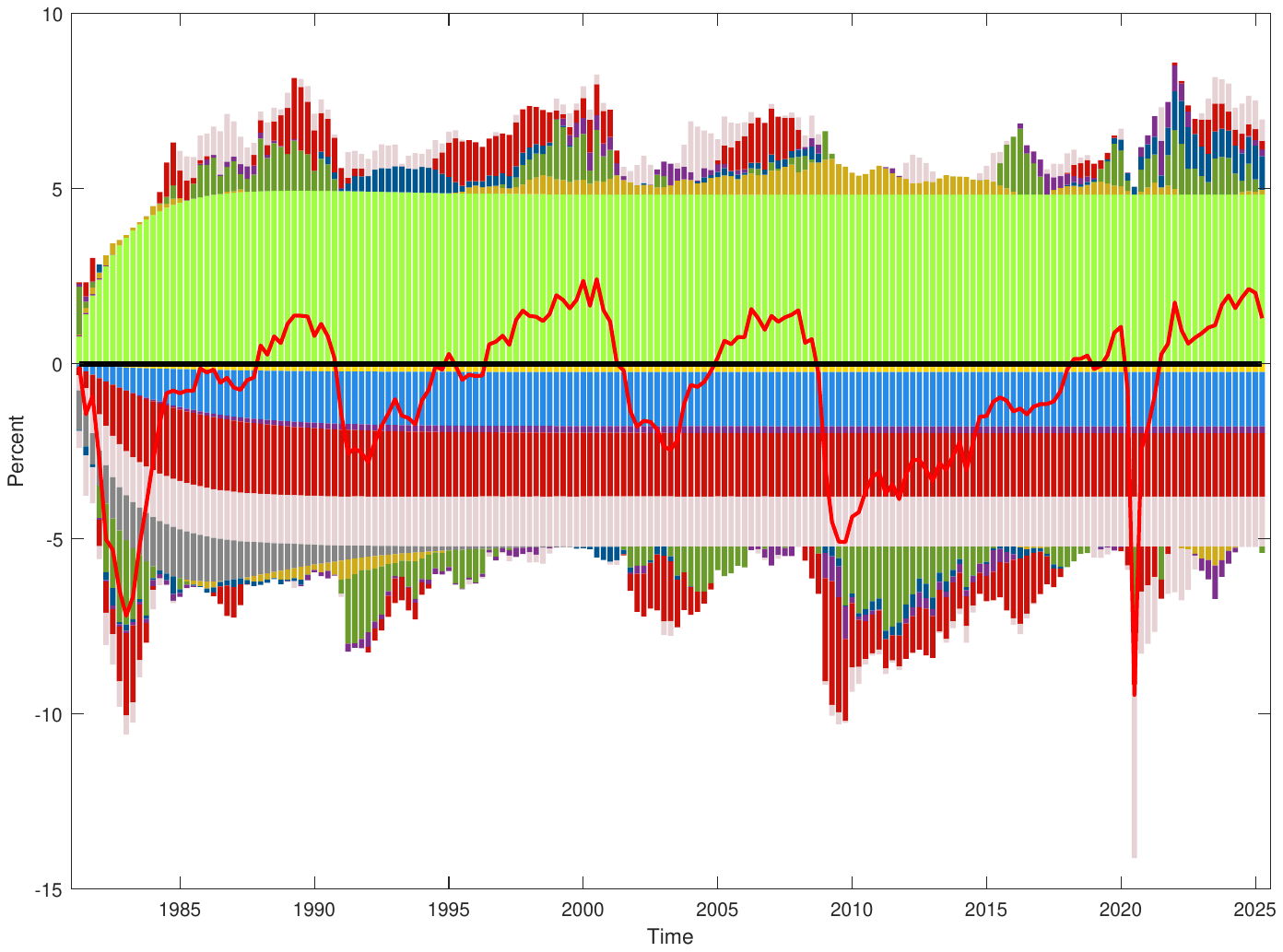}
\label{fig:OutputHD_6var}
  \end{subfigure}
  \begin{subfigure}{.99\textwidth}
\vspace{0.1in}  \caption{Inflation}
\centering\includegraphics[width=0.95\linewidth, height= 6.0cm]{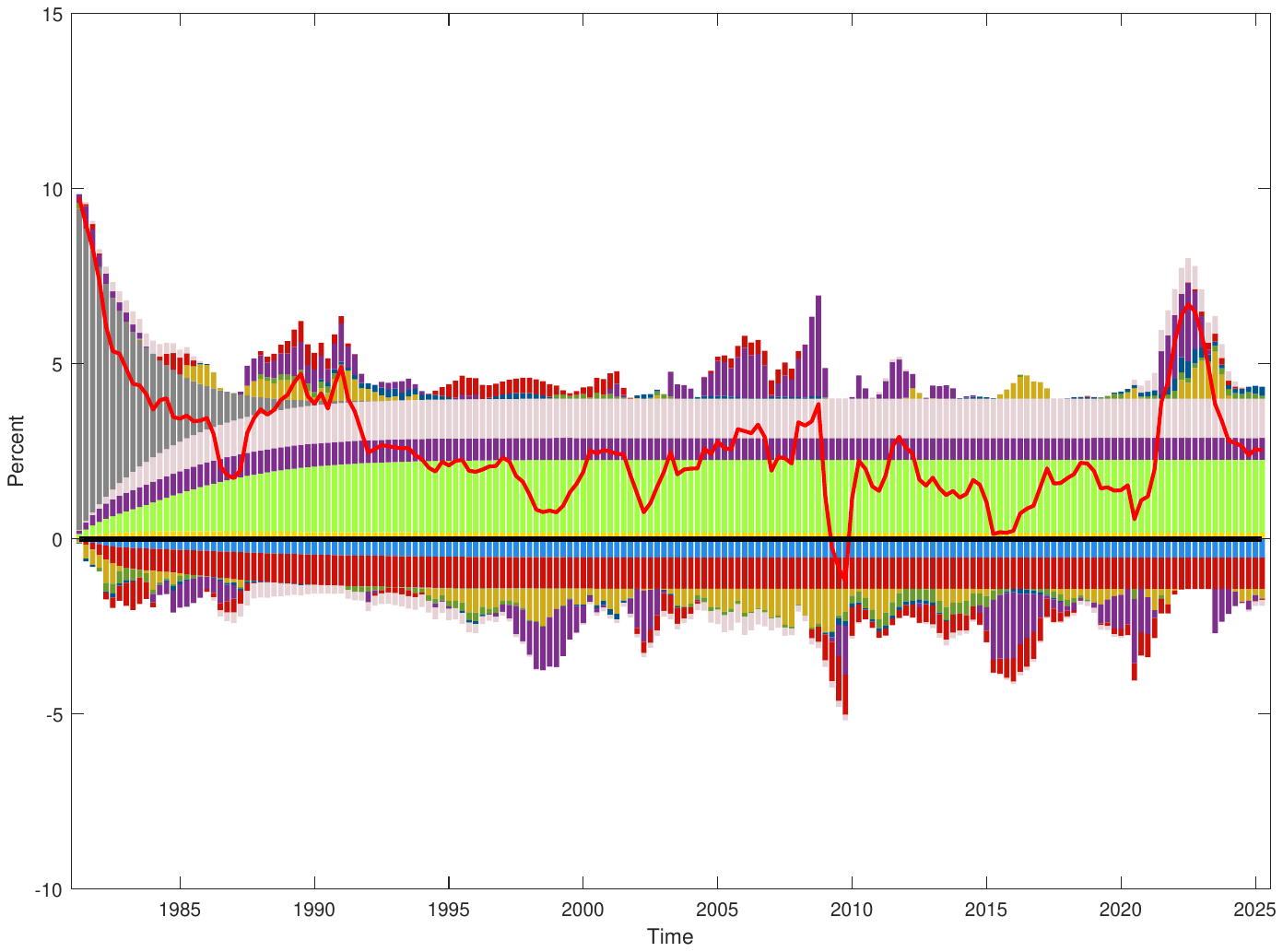}
 \label{fig:InflationHD_6var}
  \end{subfigure}
  \begin{subfigure}{.99\textwidth}
\vspace{0.1in}       \caption{Interest Rate}
\centering\includegraphics[width=0.95\linewidth, height= 7.5cm]{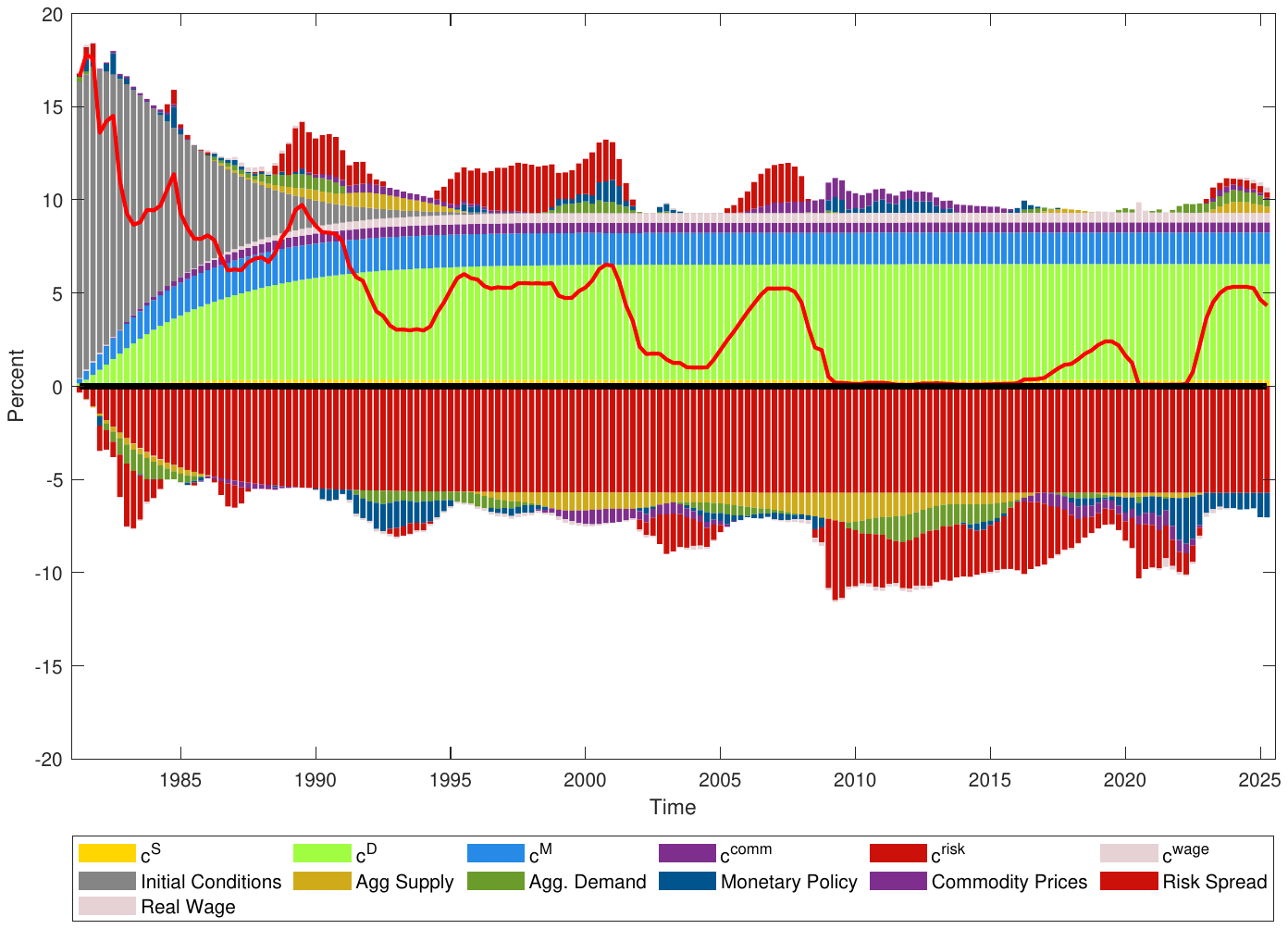}  
 \label{fig:InterestHD_6var}
  \end{subfigure}
   \label{fig:BHHD_6var_1}
\noindent \flushleft \footnotesize
Figure \ref{fig:BHHD_6var_1} plots the median estimates of the historical decomposition of the output gap, inflation, and the interest rate for the six variable VAR with additional priors over the structural constants as described in section \ref{sec:BH6_identify}. For the supply, demand, monetary policy equations and the initial conditions, the same color scheme is used as in Figure \ref{fig:IRF3}. For the new equations, the colors are  purple for the commodity price equation, red, for risk spread equation, and pink, for the real wage equation's contribution. We have used a slightly lighter shade for the structural constant terms.  In each panel the red line plots observed series.      
 \vfill
\end{figure}

\begin{figure}[H]
\caption{Historical decomposition of commodity inflation, corporate risk spread and real wage growth in the six variable VAR}
  \begin{subfigure}{.99\textwidth}
\vspace{0.1in}   \caption{Commodity price inflation}
\centering\includegraphics[width=0.95\linewidth, height= 6.0cm]{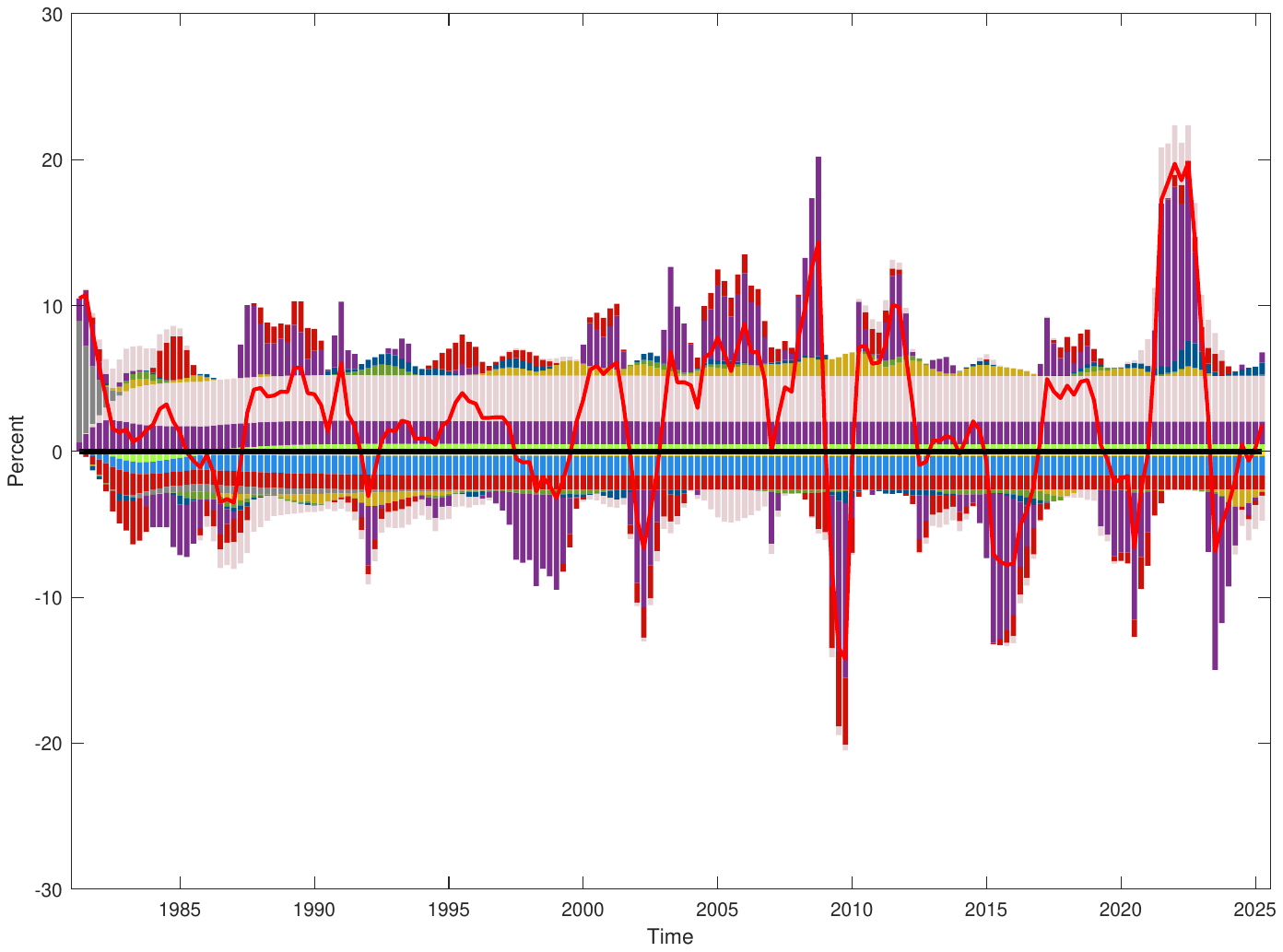}
\label{fig:COMMINFLHD_6var}
  \end{subfigure}
  \begin{subfigure}{.99\textwidth}
\vspace{0.1in}  \caption{Corporate risk spread}
\centering\includegraphics[width=0.95\linewidth, height= 6.0cm]{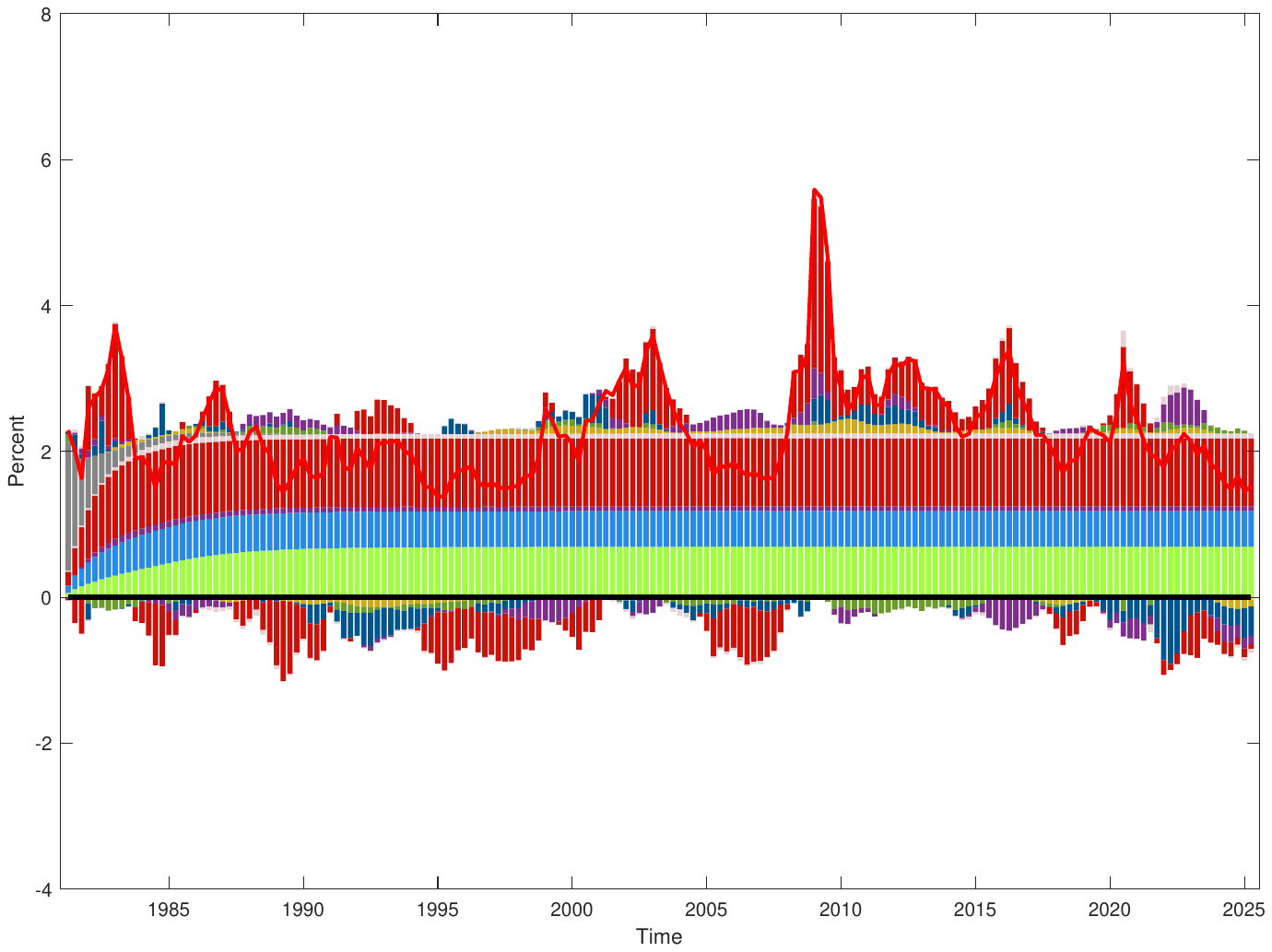}
 \label{fig:RiskSpreadHD_6var}
  \end{subfigure}
  \begin{subfigure}{.99\textwidth}
\vspace{0.1in}       \caption{Real wage growth}
\centering\includegraphics[width=0.95\linewidth, height= 7.5cm]{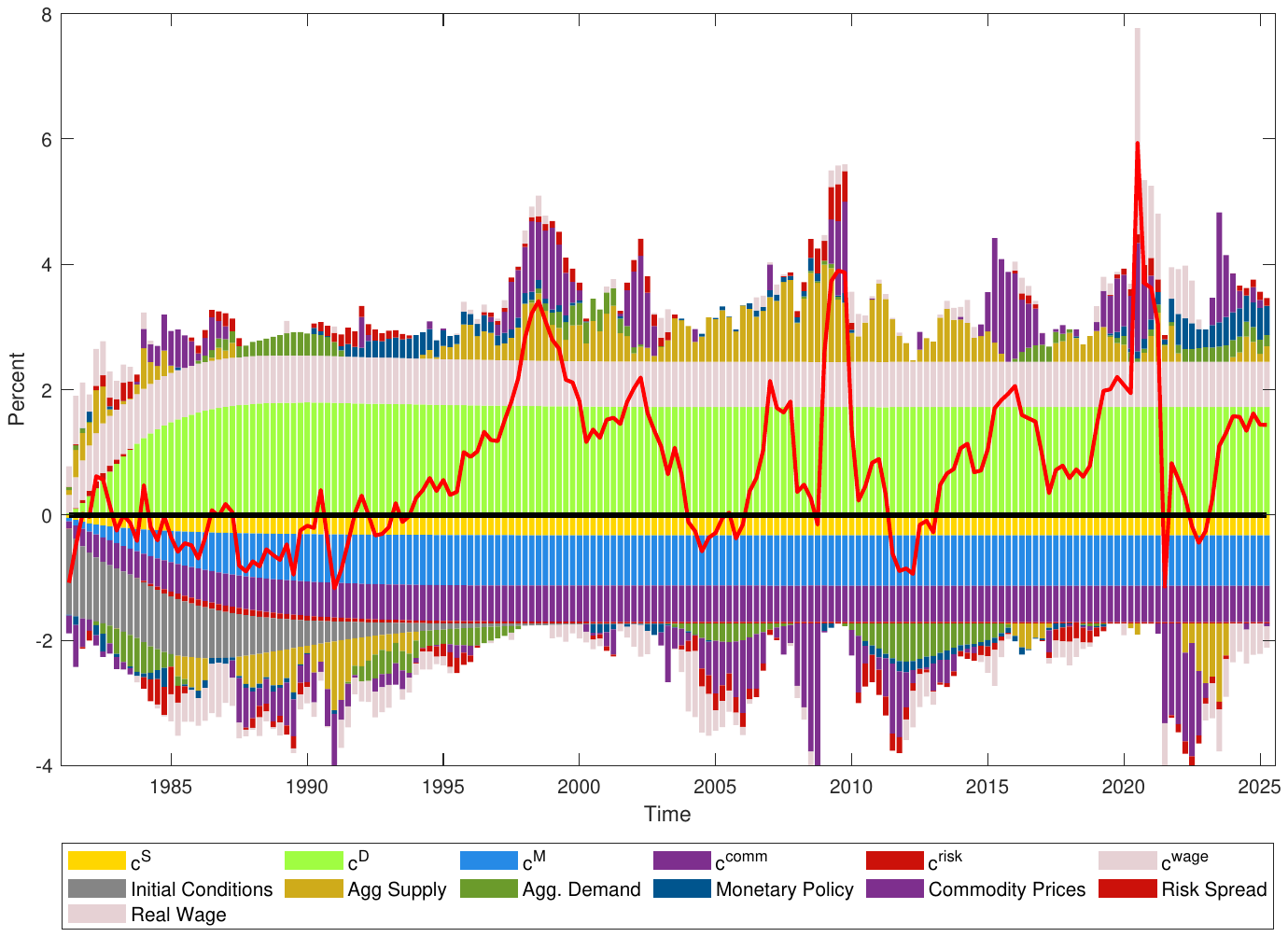}  
 \label{fig:RealWagetHD_6var}
  \end{subfigure}
   \label{fig:BHHD_6var_2}
\noindent \flushleft \footnotesize
Figure \ref{fig:BHHD_6var_2} plots the median estimates of the historical decomposition of commodity price inflation, the risk spread and real wages for the six variable VAR with additional priors over the structural constants as described in section \ref{sec:BH6_identify}. For the supply, demand, monetary policy equations and the initial conditions, the same color scheme is used as in Figure \ref{fig:IRF3}. For the new equations, the colors are  purple for the commodity price equation, red, for risk spread equation, and pink, for the real wage equation's contribution. We have used a slightly lighter shade for the structural constant terms.  In each panel the red line plots observed series.   
 \vfill
\end{figure}

Figures \ref{fig:BHHD_6var_1} and  \ref{fig:BHHD_6var_2} show the median historical decompositions for the six variable VAR  with the additional priors over the structural constants described in section \ref{sec:BH6_identify}. For the three original structural shocks we have used the same color scheme as in section \ref{sec:BH3VAR}. Thus the contribution of the demand equation to the evolution of the data is colored green, that of the supply equation colored yellow and that of the Taylor Rule colored blue. The new shocks' colors are purple, for the commodity price shock, red for the risk spread shock and pink for the real wage shock. As before we have used a slightly lighter shade for the structural constant terms to differentiate.  

Figure \ref{fig:BHHD_6var_1} shows the historical decompositions for the original three data series. For output, Panel a) tells a very intuitive story with long run output positively related to demand and negatively related to monetary policy, the risk spread and cost push wage equations. In terms of the variation the big decline in output after the 2007-2008 financial crisis is explained by the risk shock and demand shock which seems reasonable. The recovery of output after the covid epidemic was helped by positive demand and monetary policy effects. For inflation, Panel b) also tells a common sense story with long run inflation positively influenced by demand, commodity price and cost push effects which are offset by negative monetary policy and credit risk effects. For the variation in inflation, commodity prices are the most prominent shock especially for the larger deviations away from the mean.   Panel c) tells the same competing effects story for the interest rate, with the long run interest rate positively related to the Taylor rule and demand equations and negatively related to the risk spread. In terms of its variation, the risk spread shock provides most of the explanation for the low interest rates after the 2007-2008 financial crisis.

Figure \ref{fig:BHHD_6var_2} shows the historical decomposition for the three additional data series. For commodity price inflation panel a) shows that a very large proportion of the variation in the commodity price inflation is determined by the commodity price shock which is consistent with its identification as an exogenous shock. However the long run level of commodity price inflation is positively related to the real wage equation. This would appear counter-intuitive without reference to the discussion of the impulse responses of the real wage shock in Figure \ref{fig:IRF6_PRIOR} above. The discussion argued that this shock is best regarded as an anticipated future commodity price shock and that the commodity price shock and the real wage shock should be taken together as different elements of a commodity price shock. Thus a commodity price shock has an instantaneous element which reduces real wages, and an anticipated element where real wages rise temporarily in anticipation of future commodity price increases. This interpretation is supported in Figure \ref{fig:BHHD_6var_2} panel c)  which shows negative effects of commodity prices on the real wage, offset by positive real wage effects. This is true for both the level and variation in the real wages, for example during the large commodity price increases after 2020. 

The risk spread decomposition is shown in Panel b) of Figure \ref{fig:BHHD_6var_2}. This shows that for both its long run level and variation, risk spread is mostly determined by its own equation. However it is also positively related to interest rates and the level of demand. This is very intuitive and consistent with the view of the credit market of Stiglitz and Weiss (1981) and Mankiw (1986). The estimated VAR allowed for this channel through the sign restrictions in equation (\ref{eq:SR6}). We examine the robustness of this assumption below in section \ref{sec:robust}.

\newpage
\thispagestyle{empty}

\newpage
\thispagestyle{empty}

\normalsize

\section{\Scale[0.995]{\mbox{Robustness and further work}}}
\label{sec:robust} 

The sign restrictions in section \ref{sec:BH6VAR} follow Baumeister and Hamilton (2018) in assuming that increases in the interest rate increase the average level of risk in an economy. This is a reasonable assumption which is consistent with the seminal work on the credit market by Stiglitz and Weiss (1981) and Mankiw (1986). However it is interesting to explore how sensitive the results are to this assumption by instead assuming that the risk spread is not affected contemporaneously by domestic variables. This implies setting $s_{10}=0$ in equation (\ref{eq:SVAR6}). In this case risk spread and commodity prices form an exogenous block in the $S$ matrix with identification between them achieved via a Cholesky ordering. This assumption also means that the sign restriction on the impact of domestic variables on the risk spread in equation (\ref{eq:SR6}) are now zero restrictions.  Note that the estimated model still allows lagged values of the domestic variables to affect commodity prices and the risk spread. Greater exogeneity could be imposed by imposing a stronger prior that the non-own term autoregressive parameters are very small, as in Blake and Mumtaz (2017).  

\begin{figure}[H]
\caption{Historical decomposition of output, inflation and interest rate in the six variable VAR when both the risk and commodity shocks are exogenous to the domestic economy}
  \begin{subfigure}{.99\textwidth}
\vspace{0.1in}   \caption{Output gap }
\centering\includegraphics[width=0.95\linewidth, height= 6.0cm]{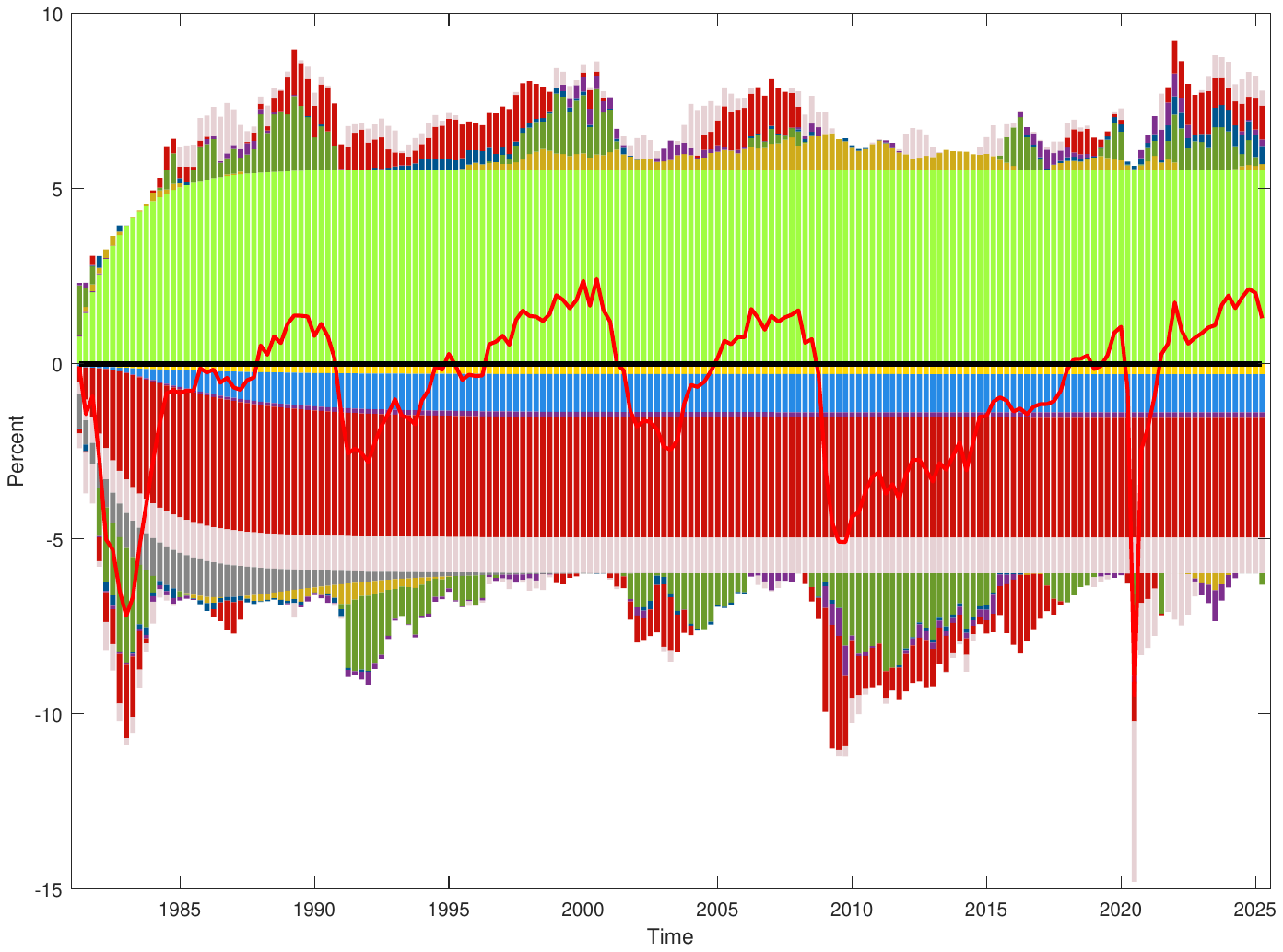}
\label{fig:OutputHD_6var}
  \end{subfigure}
  \begin{subfigure}{.99\textwidth}
\vspace{0.1in}  \caption{Inflation}
\centering\includegraphics[width=0.95\linewidth, height= 6.0cm]{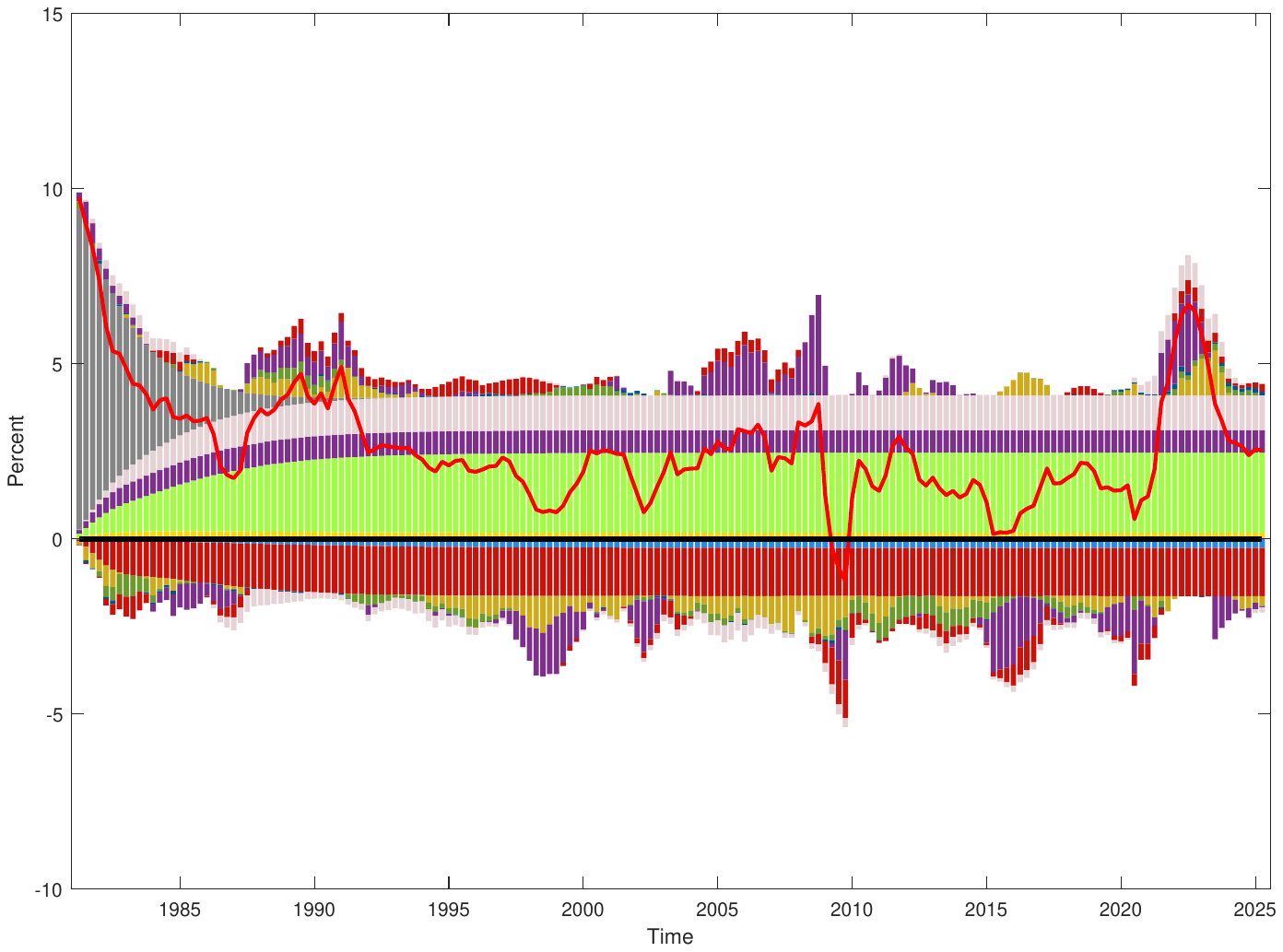}
 \label{fig:InflationHD_6var}
  \end{subfigure}
  \begin{subfigure}{.99\textwidth}
\vspace{0.1in}       \caption{Interest Rate}
\centering\includegraphics[width=0.95\linewidth, height= 7.5cm]{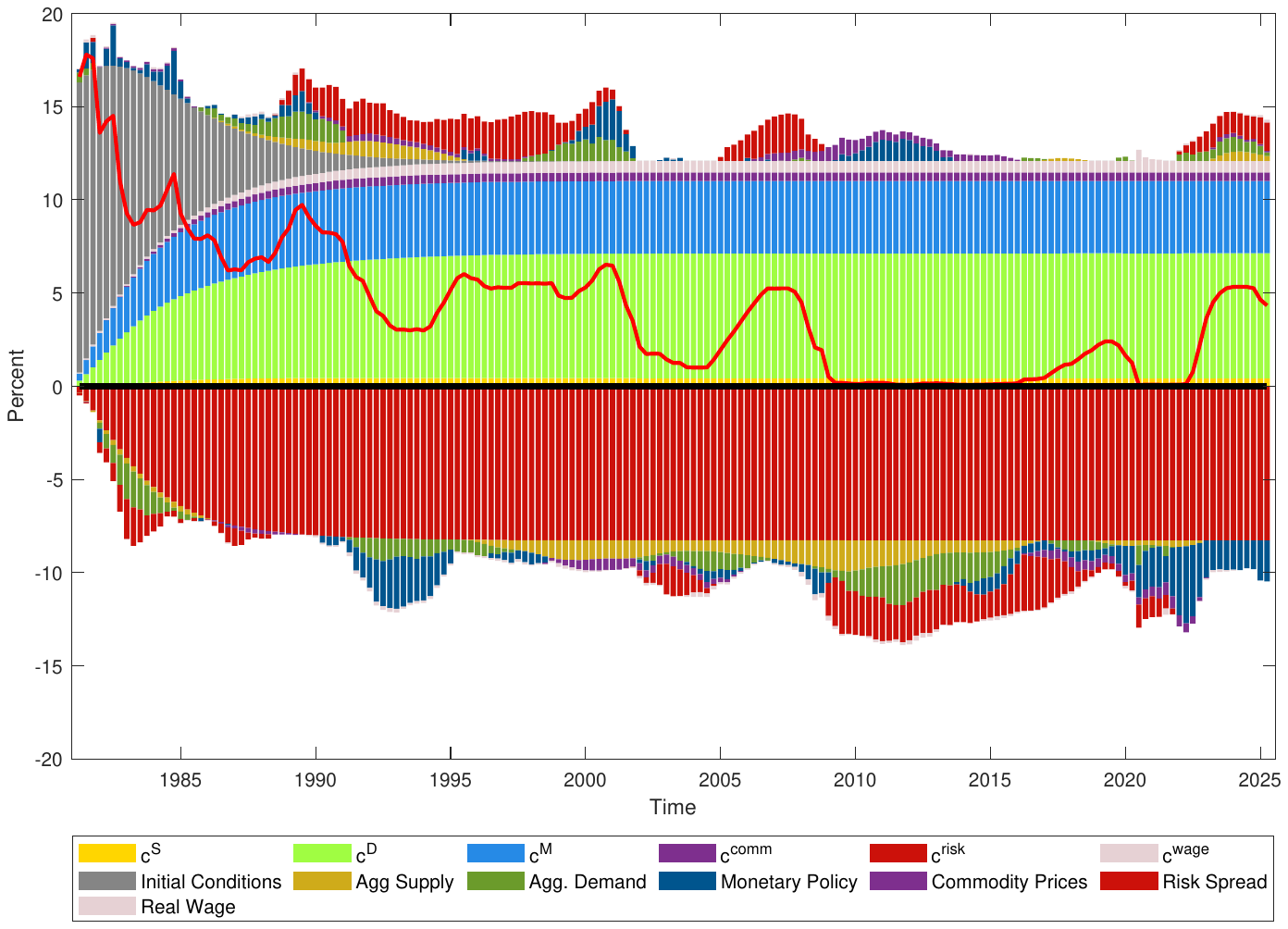}  
 \label{fig:InterestHD_6var}
  \end{subfigure}
   \label{fig:BHHD_6var_1_RISK}
\noindent \flushleft \footnotesize
Figure \ref{fig:BHHD_6var_1_RISK} plots the median estimates of the historical decomposition of the output gap, inflation, and the interest rate for the six variable VAR for the case where both the risk and commodity shocks are exogenous to the domestic economy as described in section \ref{sec:robust}. The color scheme is the same as in Figure \ref{fig:BHHD_6var_1}.      
 \vfill
\end{figure}

\begin{figure}[H]
\caption{Historical decomposition of commodity inflation, corporate risk spread and real wage growth in the six variable VAR when both the risk and commodity shocks are exogenous to the domestic economy}
  \begin{subfigure}{.99\textwidth}
\vspace{0.1in}   \caption{Commodity price inflation}
\centering\includegraphics[width=0.95\linewidth, height= 6.0cm]{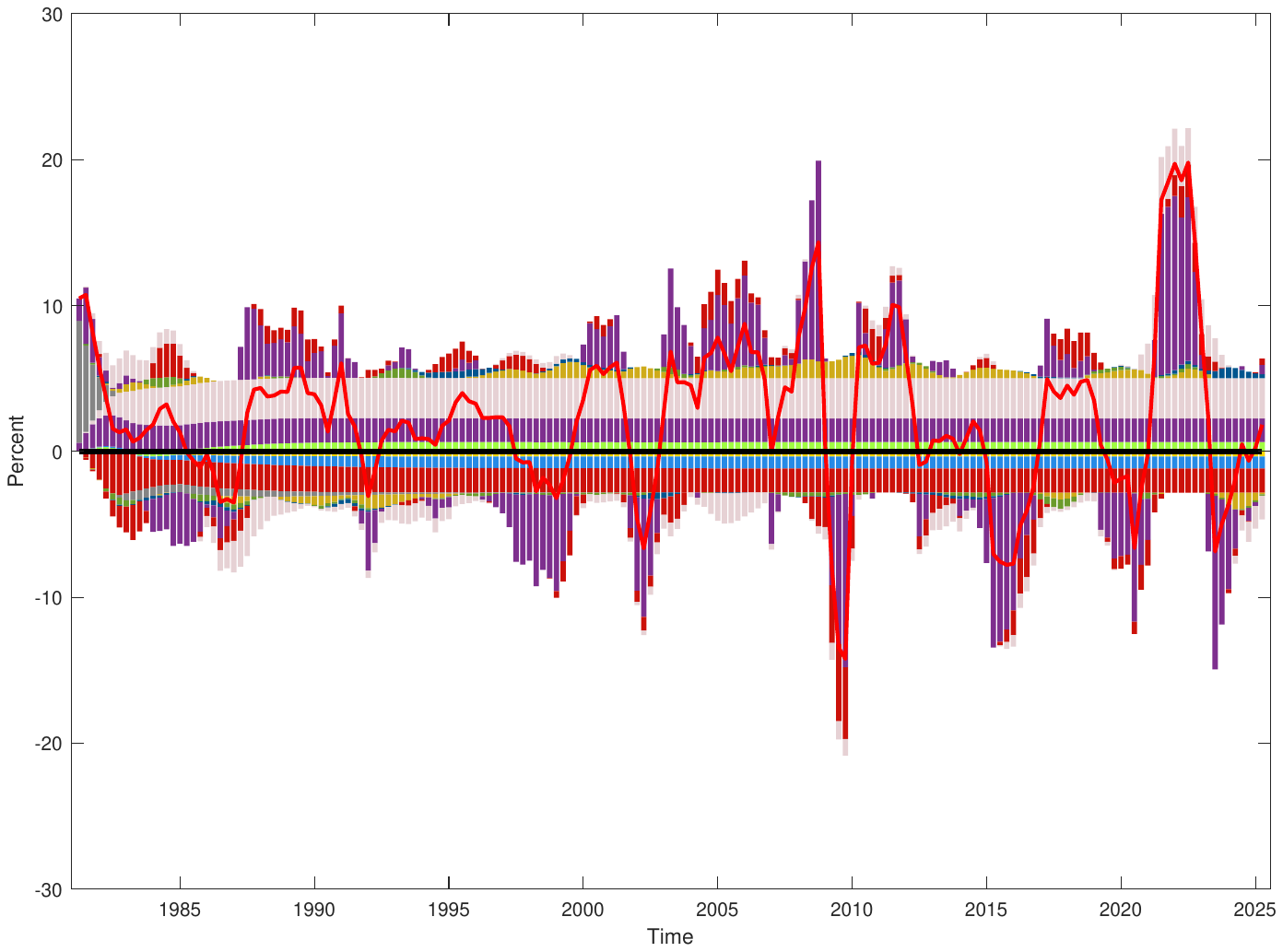}
\label{fig:COMMINFLHD_6var}
  \end{subfigure}
  \begin{subfigure}{.99\textwidth}
\vspace{0.1in}  \caption{Corporate risk spread}
\centering\includegraphics[width=0.95\linewidth, height= 6.0cm]{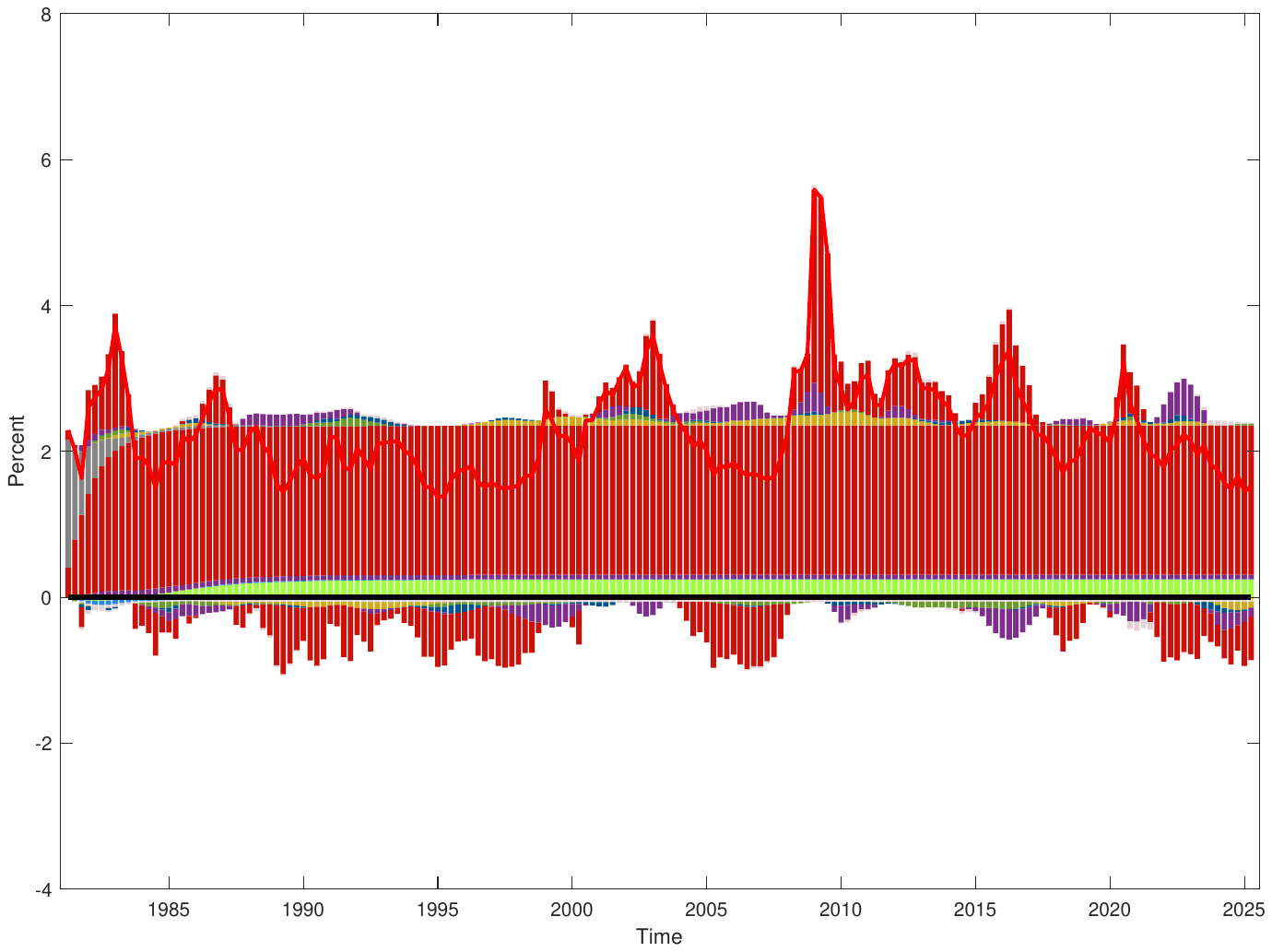}
 \label{fig:RiskSpreadHD_6var}
  \end{subfigure}
  \begin{subfigure}{.99\textwidth}
\vspace{0.1in}       \caption{Real wage growth}
\centering\includegraphics[width=0.95\linewidth, height= 7.5cm]{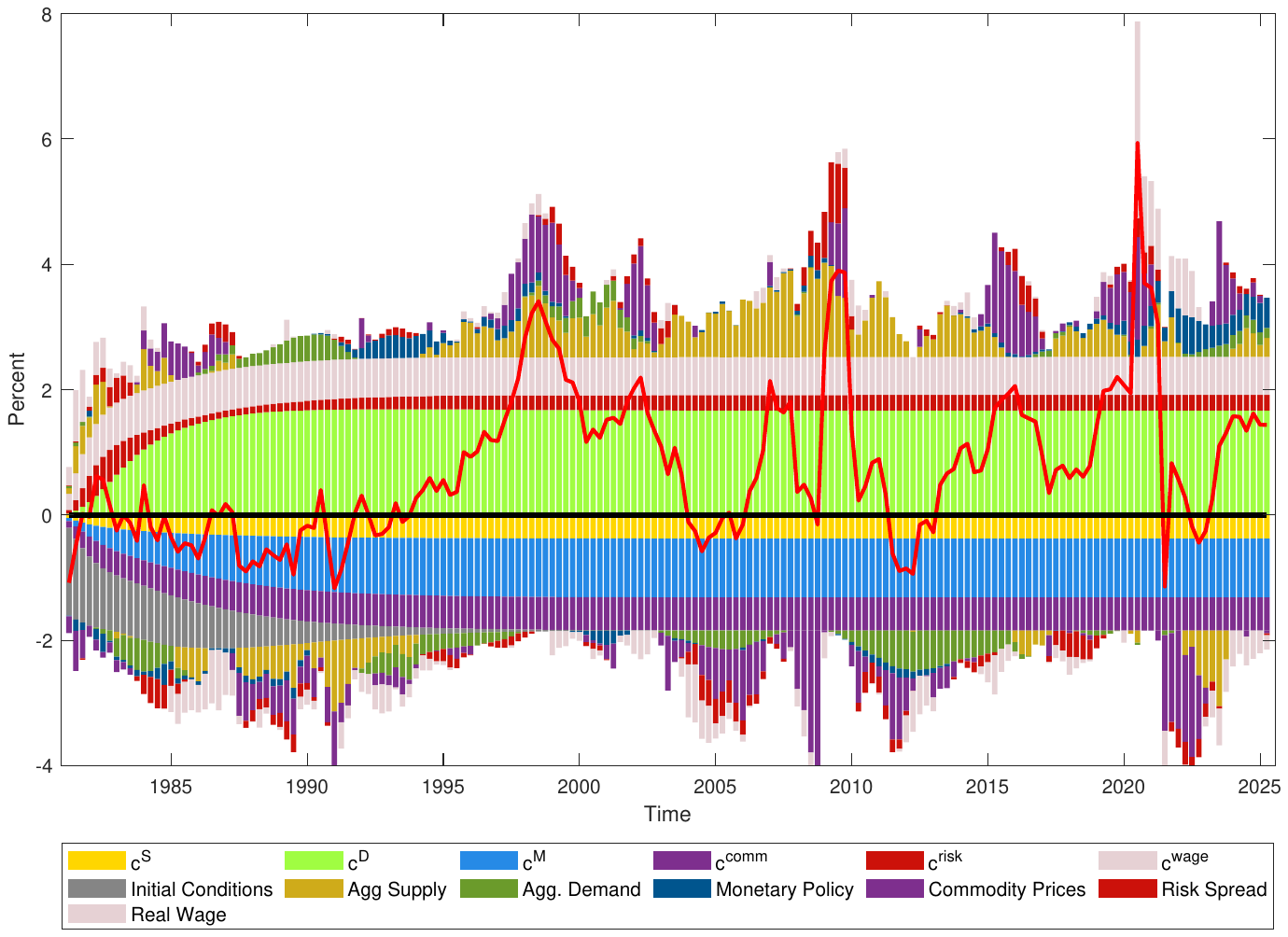}  
 \label{fig:RealWagetHD_6var}
  \end{subfigure}
   \label{fig:BHHD_6var_2_RISK}
\noindent \flushleft \footnotesize
Figure \ref{fig:BHHD_6var_2_RISK} plots the median estimates of the historical decomposition of commodity prices, risk spread, and the real wage for the six variable VAR for the case where both the risk and commodity shocks are exogenous to the domestic economy as described in section \ref{sec:robust}. The color scheme is the same as in Figure \ref{fig:BHHD_6var_2}.       
 \vfill
\end{figure}

The historical decompositions for this case are shown in Figures \ref{fig:BHHD_6var_1_RISK} and  \ref{fig:BHHD_6var_2_RISK}. Comparing Figures \ref{fig:BHHD_6var_1} and \ref{fig:BHHD_6var_1_RISK} one can see that the principle effect of setting  $s_{10}=0$ is to raise the importance of the risk spread equation at the expense of the Taylor Rule equation.  This is true for both the long run effects and the variation with a smaller role for monetary policy in the rebound of output after the covid epidemic and a smaller role for monetary policy in the low interest rates than their counterparts in Figure \ref{fig:BHHD_6var_1}. One exception being in the effect of monetary policy on the long run interest rate where the role of $c^{m}$ is increased due to the need to offset the greater negative effect of $c^{risk}$.  The same holds in the comparison of Figures \ref{fig:BHHD_6var_2} and \ref{fig:BHHD_6var_2_RISK}. The risk spread is now almost entirely accounted for by its own equation in panel b) of Figure \ref{fig:BHHD_6var_2_RISK}, which was the purpose of this exercize. The risk spread is also associated with a small increase in long run real wages as it is associated with lower inflation which in turn is associated with a higher real wage.   

The conclusion that we draw from this exercize is that identification is still important. This paper is not attempting to settle the debate about the identification of structural VAR models. Its purpose is to show how including discussion and analysis of the structural constants can help us in this process. For further refinement of the identifying assumptions used in this paper, the Baumeister and Hamilton approach describes a promising direction for future study. External information could be used to calibrate stronger priors over the structural parameters of the three additional variables. However this would be a significant task.

\newpage
\thispagestyle{empty}

\newpage
\thispagestyle{empty}

\section{\Scale[0.995]{\mbox{How large are monetary policy effects?}}}
\label{sec:howlarge} 

As we have noted, an important strand of the literature has only found weak and transient effects of monetary policy on the macroeconomy. In this paper we have shown that imposing priors on the structural constants can lead to an increased estimated role for  monetary policy in the evolution of the data. This is true for both the long run equilibrium levels and the variation in  the data. To depict this clearly we present a `counterfactual' analysis. This sets the Taylor rule equation's constant and estimated error terms to zero and then recalculates the historical decomposition using this counterfactual data. This 'counterfactual' analysis is just an alternative representation of the historical decomposition analysis of section \ref{sec:BH6_results}. It should not be interpreted as performing any kind of `pseudo experiment'.  

\begin{figure}[H]
\caption{Counterfactual with the monetary policy rule equation set to zero}
   \begin{subfigure}{.99\textwidth}
\vspace{0.1in}   \caption{Output gap }
\centering\includegraphics[width=0.85\linewidth, height= 4.5cm]{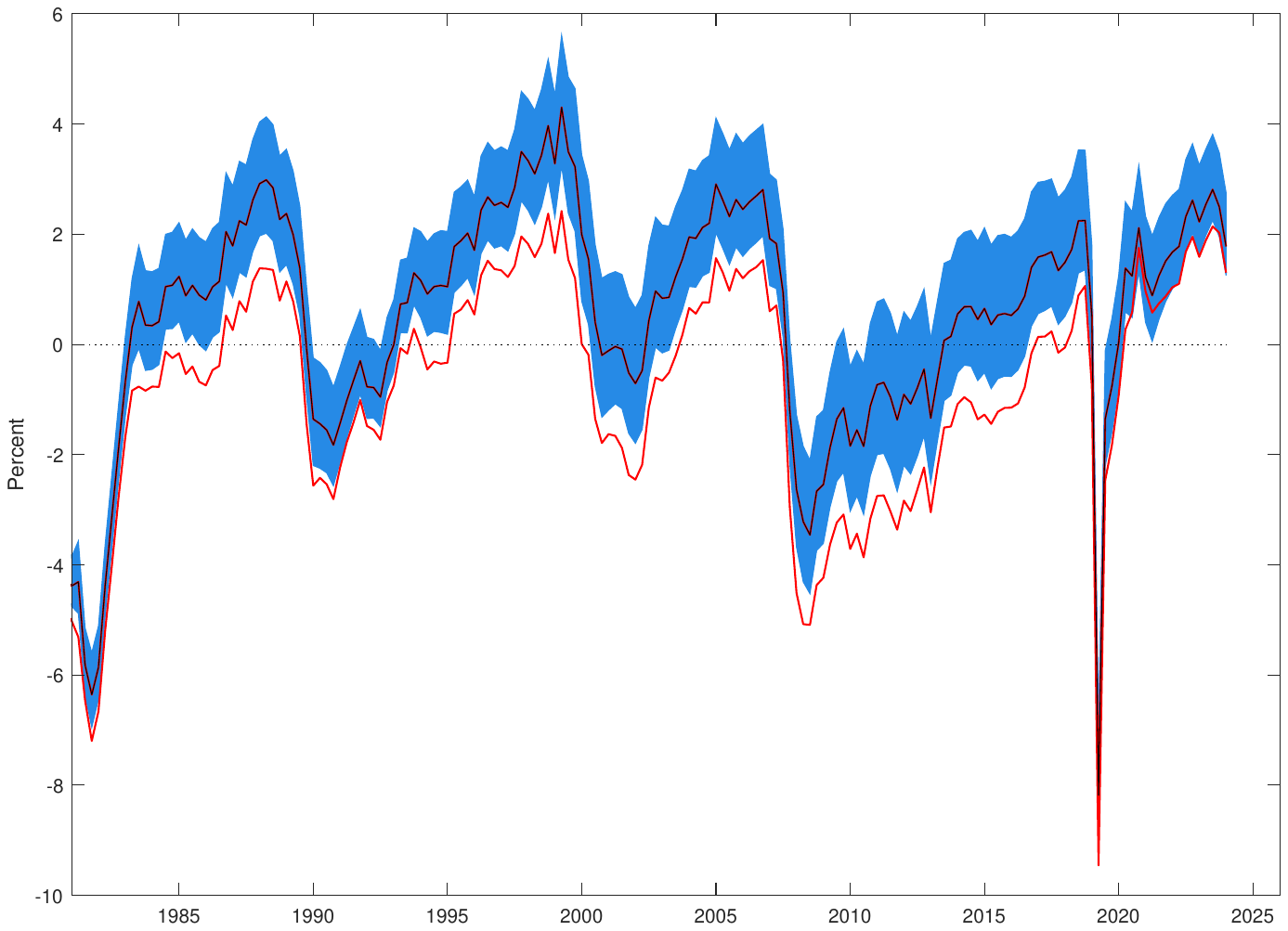}
\label{fig:Output_CF3_ORIG}
  \end{subfigure}
  \begin{subfigure}{.99\textwidth}
\vspace{0.1in}  \caption{Inflation}
\centering\includegraphics[width=0.85\linewidth, height= 4.5cm]{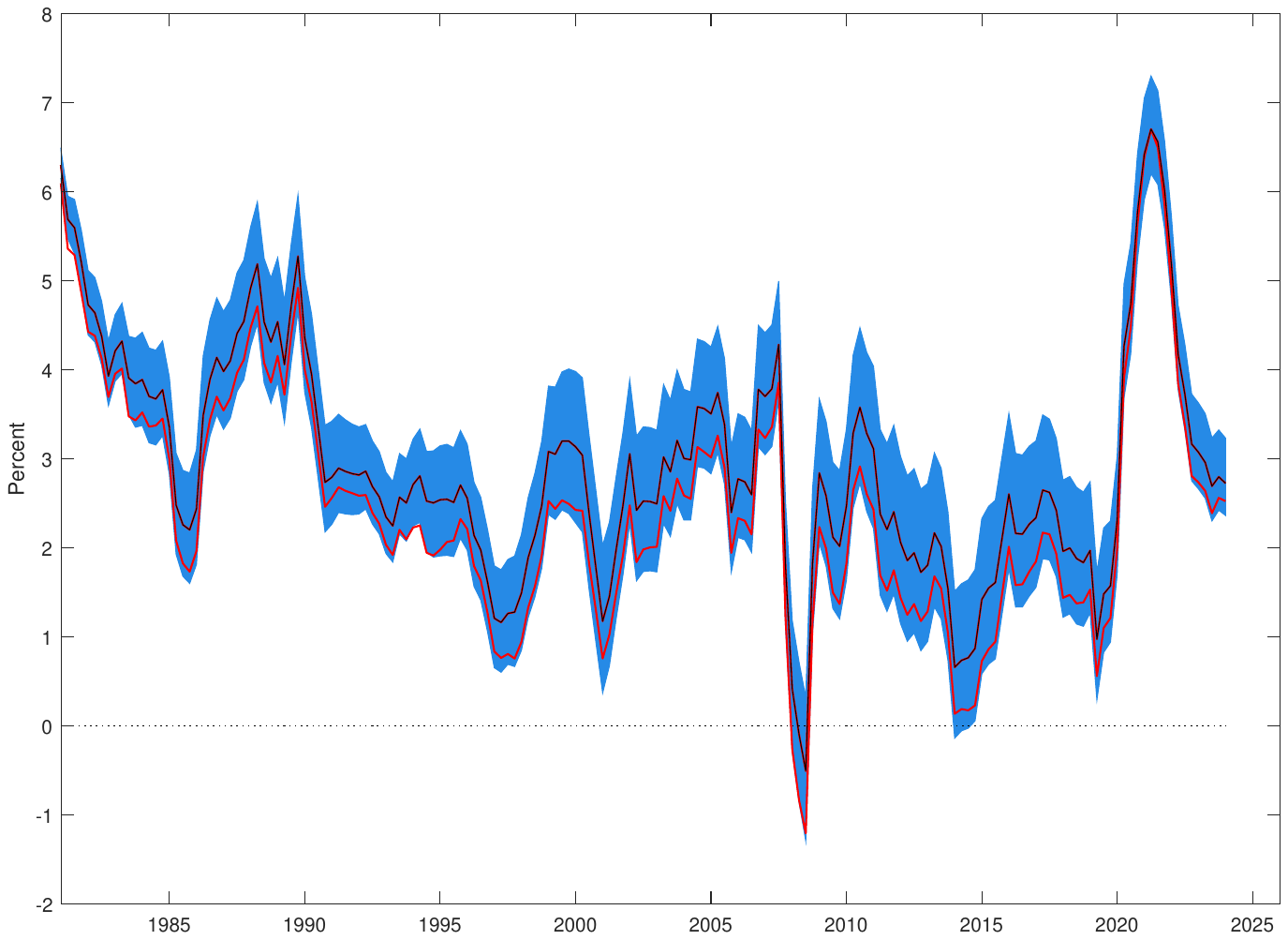}
\label{fig:Inflation_CF3_ORIG}
  \end{subfigure}
  \begin{subfigure}{.99\textwidth}
\vspace{0.1in}       \caption{Interest Rate}
\centering\includegraphics[width=0.85\linewidth, height= 4.5cm]{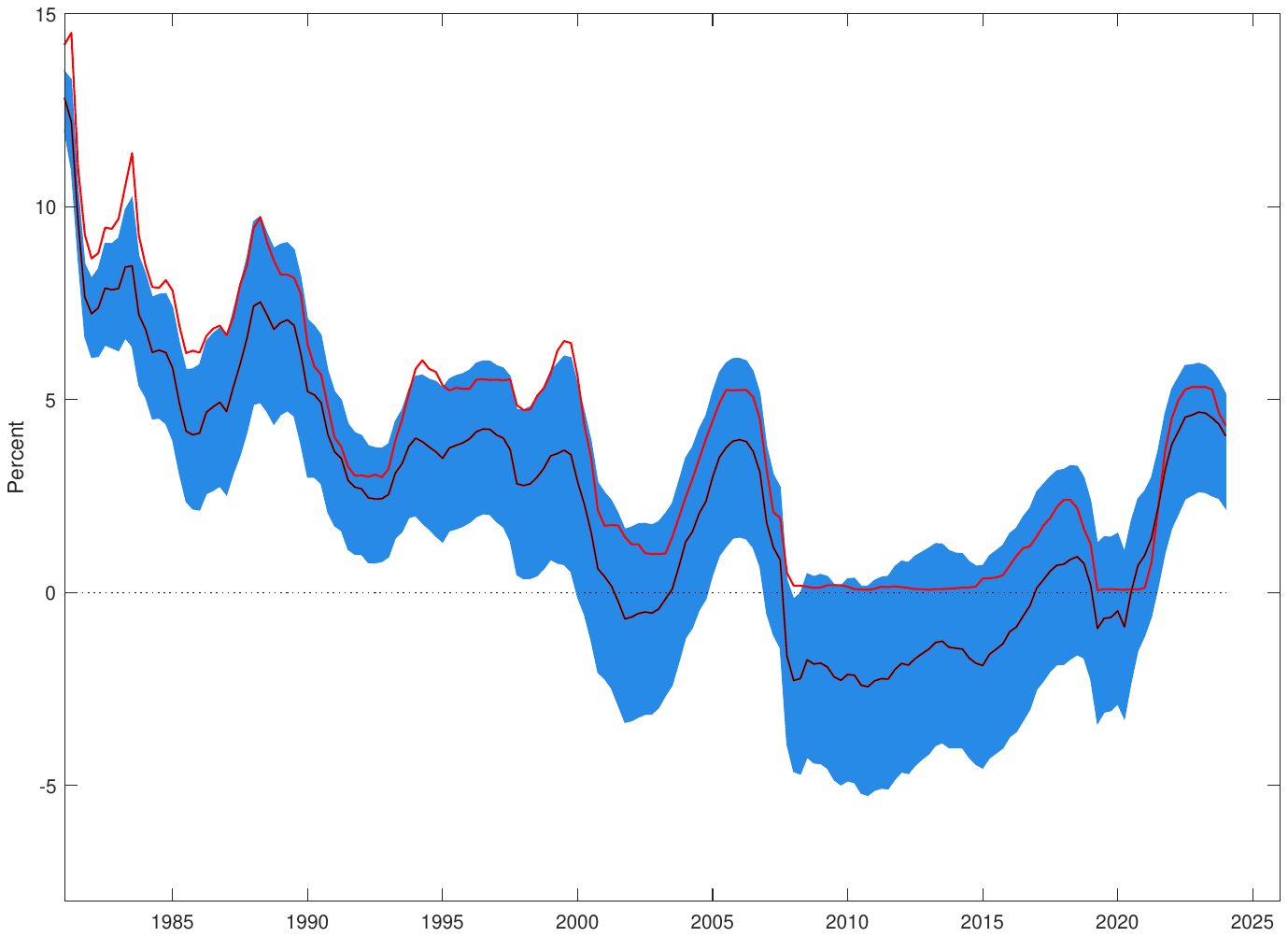}
\label{fig:Interest_CF3_ORIG}
  \end{subfigure}
   \label{fig:ORIG_CF3}
\noindent \vspace{0.5in}\flushleft \footnotesize
Figure \ref{fig:ORIG_CF3} plots  the counterfactual time series for the case where monetary policy rule equation is set to zero.  The red line plots the observed data series  and the shaded region shows the area between the 16\% and 84\% quantiles of the counterfactual posterior distribution with the median plotted in black.  The shading is blue, consistent with the color used for the monetary policy equation in the other Figures. The distribution is mostly below the data for the interest rate over the  sample period, illustrating that the monetary policy rule increases the interest rate above what it would otherwise be.   
 \vfill
\end{figure}

Figure \ref{fig:ORIG_CF3} plots this counterfactual exercize. This shows that the interest rate is below the observed path for most of the posterior distribution over the sample period. This results in inflation being higher by a small amount, less than 1\% on average, but output is significantly higher by over 1\% on average. Thus the cumulative effects on GDP will be very significant in the longer term. However the impact on inflation appears low. This may be because the model is estimated on a data sample without explosive wage price dynamics. Thus the effect of monetary policy in keeping the economy away from such an explosive regime will not be seen in the estimated model.

\newpage

\newpage 

\section{\Scale[0.95]{\mbox{Conclusion}}}

This paper has shown how the simple addition of priors over the structural constants in a VAR can significantly change its results and interpretation.   We have illustrated this using the analysis of monetary policy by Baumeister and Hamilton (2018). We have shown that imposing priors over the structural constant terms can lead to a more intuitive estimated monetary policy rule and a much larger role for monetary policy in describing the evolution of the data. By depicting the contribution of the structural constant terms in the historical decompositions, this paper has also provided a clear visual representation of the multiple contending macroeconomic forces acting on the economy, which together determine equilibrium  interest, inflation and income.

 \newpage


\newpage

\renewcommand{\thefigure}{A\arabic{figure}}
\setcounter{figure}{0}
\setcounter{equation}{0}
\setcounter{table}{0}
\renewcommand{\theequation}{A\arabic{equation}}
\appendix

\begin{center}
\textbf{\Large{Appendix}}
\end{center}

\section{Formulas used in calculation}
\label{sec:appendixformulas}

Rewriting equation (\ref{eq:decomp}) in the text shows  that the observation $\widehat{Y_{t+s}}$ in period $\tiny{t+s}$ can be decomposed into three contributions, the initial conditions, the constant terms and the innovations in the previous s periods.  
\begin{equation}
\Scale[0.975]{Y_{t+s} = \underbrace{F^s Y_{t}}_{\text{initial conditions}}  + \underbrace{F^{s-1}K + F^{s-2}K +\dots +K}_{\text{constant terms}}  + \underbrace{F^{s-1}E_{t+1} + F^{s-2}E_{t+2}+ \dots + E_{t+s}}_{\text{innovations}} }
\label{eq:decomp_Appendix}
\tag{4}
\end{equation}

This is the equations behind the calculations in the historical decomposition and counterfactual exercises where the contribution of the fundamental innovations use the formula $u_t=S^{-1} \epsilon_{t}$  following equation (\ref{eq:VAR_rf}). Iterating backwards into infinite history, $\widehat{Y_{t}}$ can be expressed as an $MA(\infty)$ process 
\begin{equation}
\begin{aligned}
\widehat{Y_{t}} =& K + FK+ F^2 K  + \dots + E_{t} + FE_{t-1}+ F^2 E_{t-2} + \dots + \\
\widehat{Y_{t}} =& \widehat{\mu} + \Psi(L)\widehat{u_{t}}
\end{aligned}
\label{eq:maintfy}
\end{equation}
\noindent Where $\mu=(I_n - \Phi_1 - \dots \Phi_p)^{-1} K$ and $\widehat{\mu}=[\mu^\prime \ 0_{1 \times n(p-1)}]^\prime$ and where $\Psi(L)$ is an $MA(\infty)$ process. These can be written 

\begin{equation}
y_t = \mu + u_{t} +\Psi_1 u_{t-1} + \Psi_2 u_{t-2}+ \Psi_3 u_{t-2}+ \dots 
\label{eq:impulseF}
\end{equation}

where $\Psi_j$ is the upper left $n \times n$  block of matrix $F^j$. Equation (\ref{eq:impulseF}) describes  the impulse responses functions following  Hamilton (1994).

\newpage 

\renewcommand{\thefigure}{B\arabic{figure}}
\setcounter{figure}{0}
\setcounter{equation}{0}
\renewcommand{\theequation}{B\arabic{equation}}
\renewcommand{\thetable}{B\arabic{table}}
\section{Additional Figures for 6 variable VAR with no additional prior on structural constants \\ (Not Intended For Publication)}
\label{sec:appendix_NOPRIOR}

\begin{figure}[h!]
\caption{Posterior distribution of structural constants of six variable VAR }
  \vspace{0.1in}      
\centering\includegraphics[width=6.5in, height=2.5in]{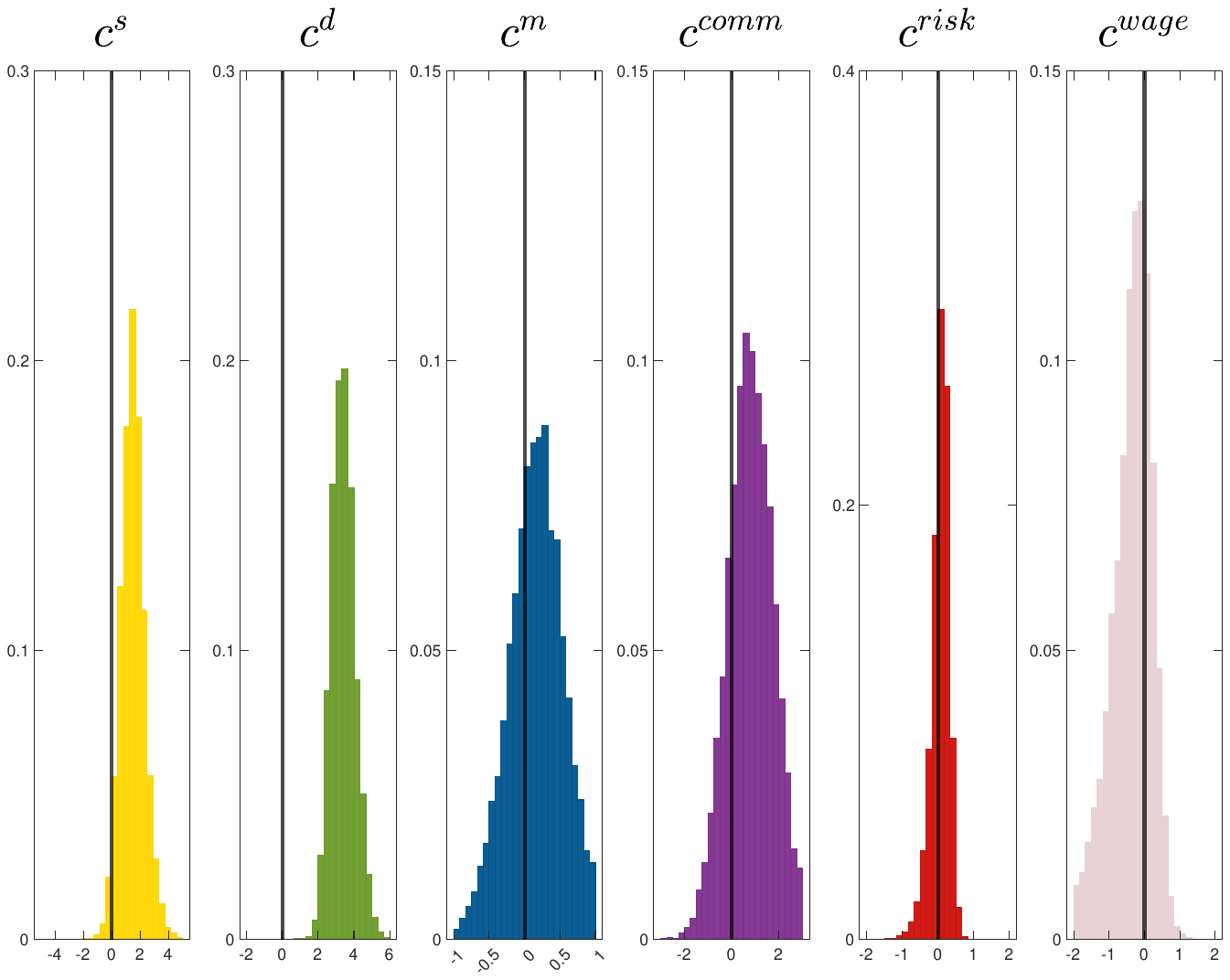}
  \label{fig:BH_6var_NOPRIOR_CONSTANTS}
\noindent \flushleft \footnotesize
Figure \ref{fig:BH_6var_NOPRIOR_CONSTANTS} plots the posterior distribution of the structural constants for the six variable VAR  without additional priors over the structural constant. The vertical black line is at zero in each case.       
 \vfill
\end{figure}

\begin{figure}[h!]
\caption{Historical decomposition of output, inflation and interest rate in six variable VAR with no additional priors over the structural constants}
  \begin{subfigure}{.99\textwidth}
\vspace{0.1in}   \caption{Output gap }
\centering\includegraphics[width=0.95\linewidth, height= 5.0cm]{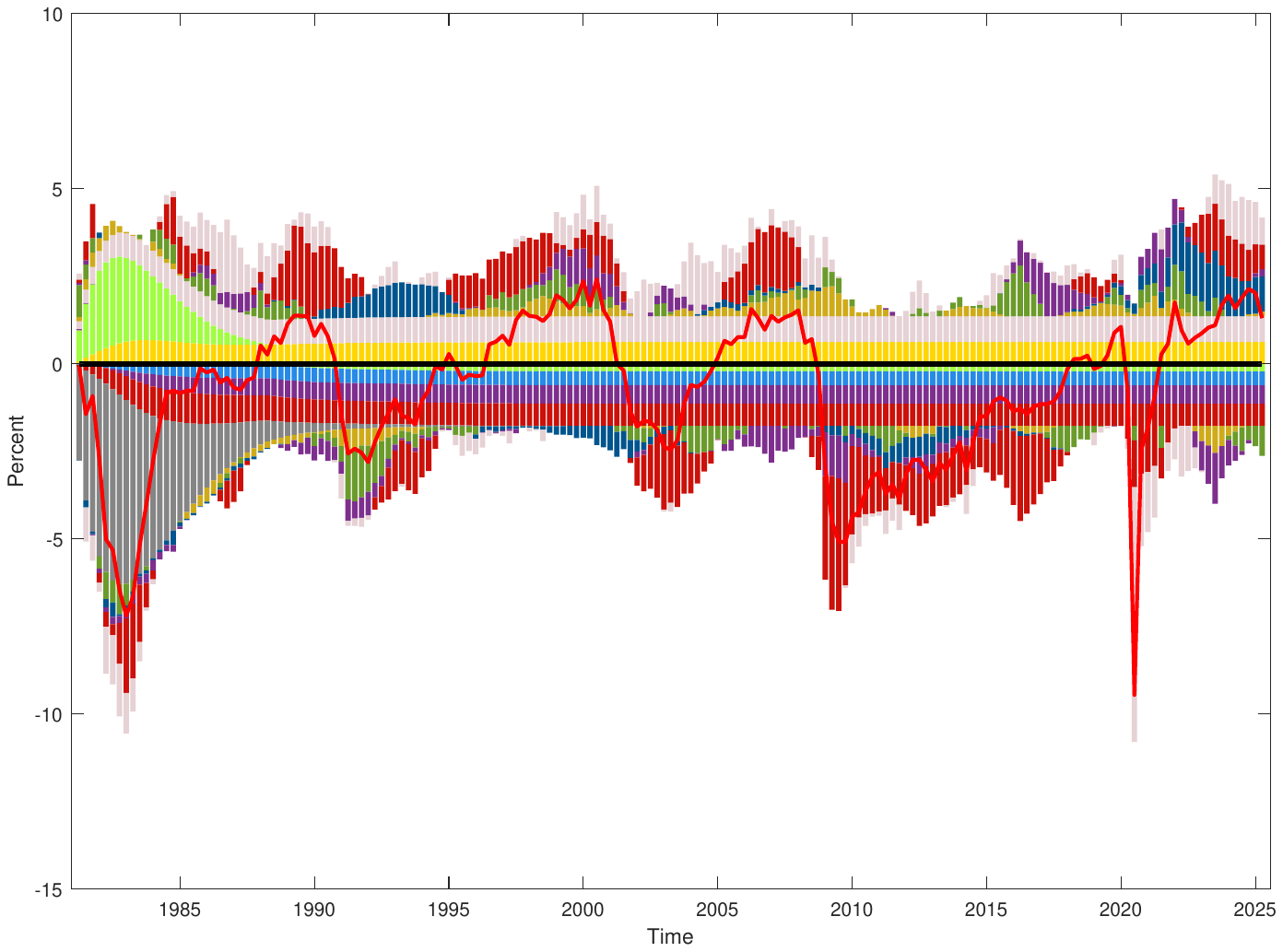}
\label{fig:OutputHD_6var}
  \end{subfigure}
  \begin{subfigure}{.99\textwidth}
\vspace{0.1in}  \caption{Inflation}
\centering\includegraphics[width=0.95\linewidth, height= 5.0cm]{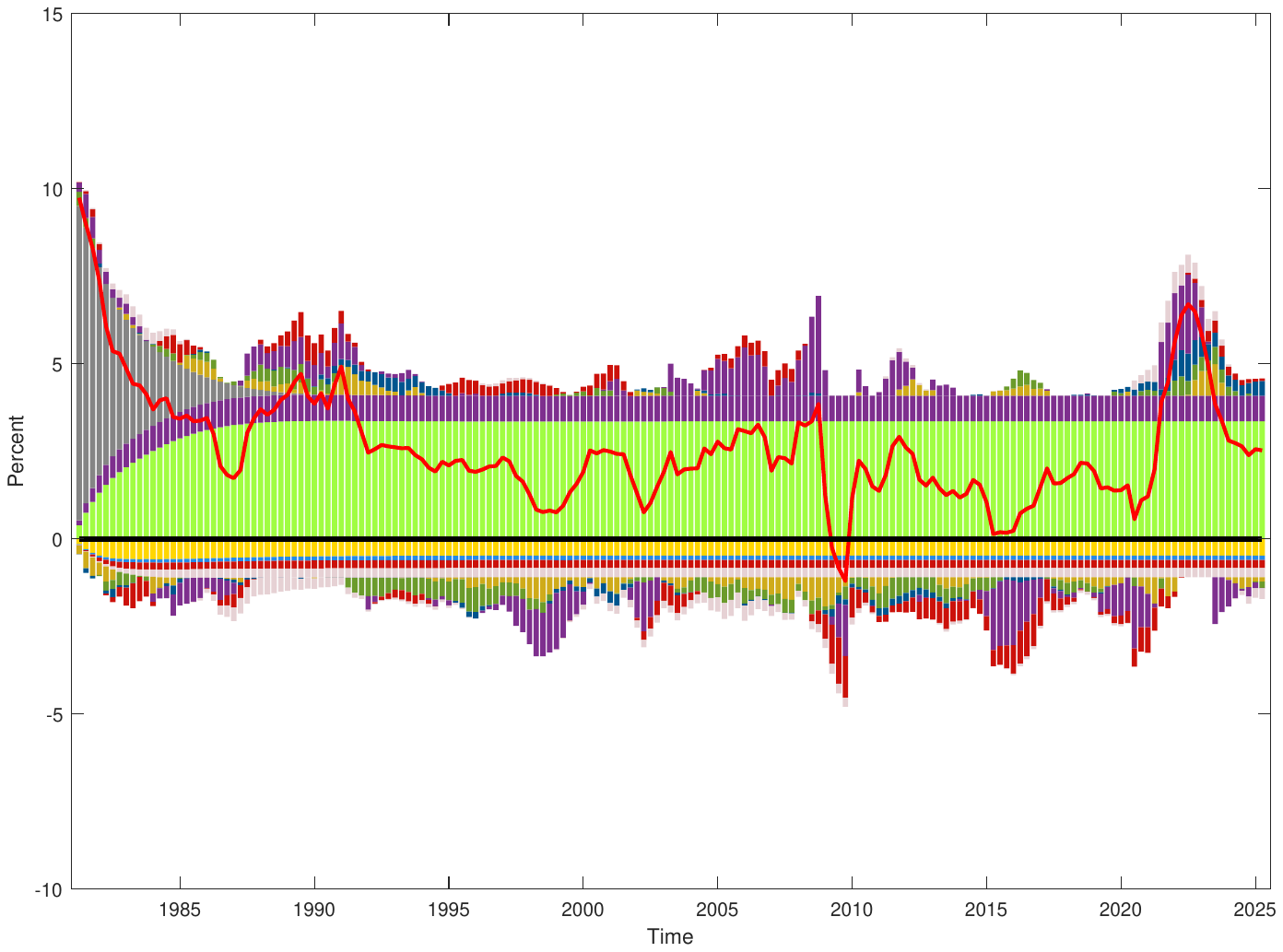}
 \label{fig:InflationHD_6var}
  \end{subfigure}
  \begin{subfigure}{.99\textwidth}
\vspace{0.1in}       \caption{Interest Rate}
\centering\includegraphics[width=0.95\linewidth, height= 6.5cm]{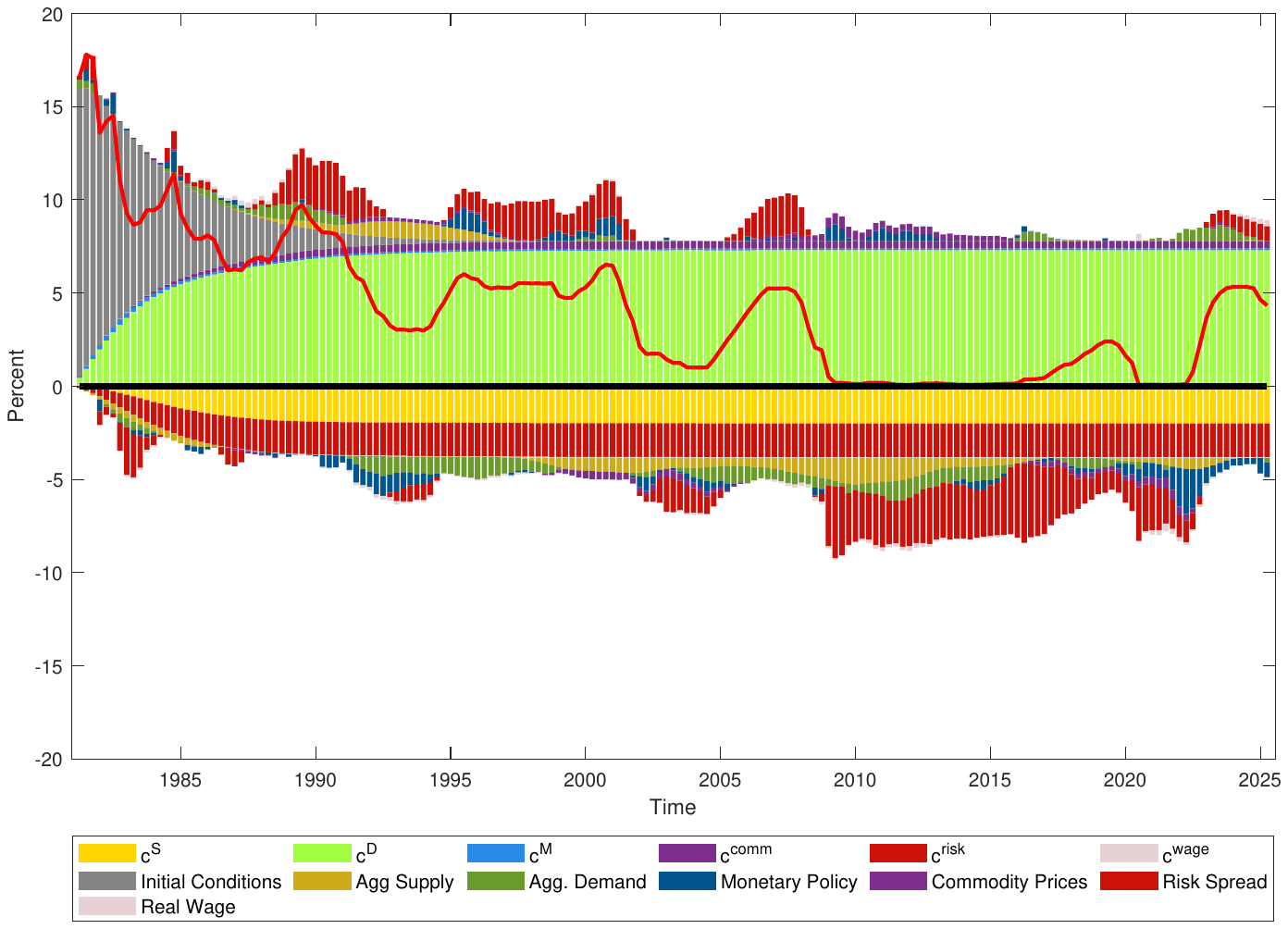}  
 \label{fig:InterestHD_6var}
  \end{subfigure}
   \label{fig:BHHD_6var_1_NOPRIOR}
\noindent \flushleft \footnotesize
Figure \ref{fig:BHHD_6var_1_NOPRIOR} plots the median estimates of the historical decomposition of the output gap, inflation, and the interest rate for the case where there is no additional prior for the structural constants. 
 \vfill
\end{figure}

\begin{figure}[h!]
\caption{Historical decomposition of commodity inflation, corporate risk spread and real wage growth in six variable VAR with no additional priors over the structural constants}
  \begin{subfigure}{.99\textwidth}
\vspace{0.1in}   \caption{Commodity inflation}
\centering\includegraphics[width=0.95\linewidth, height= 5.0cm]{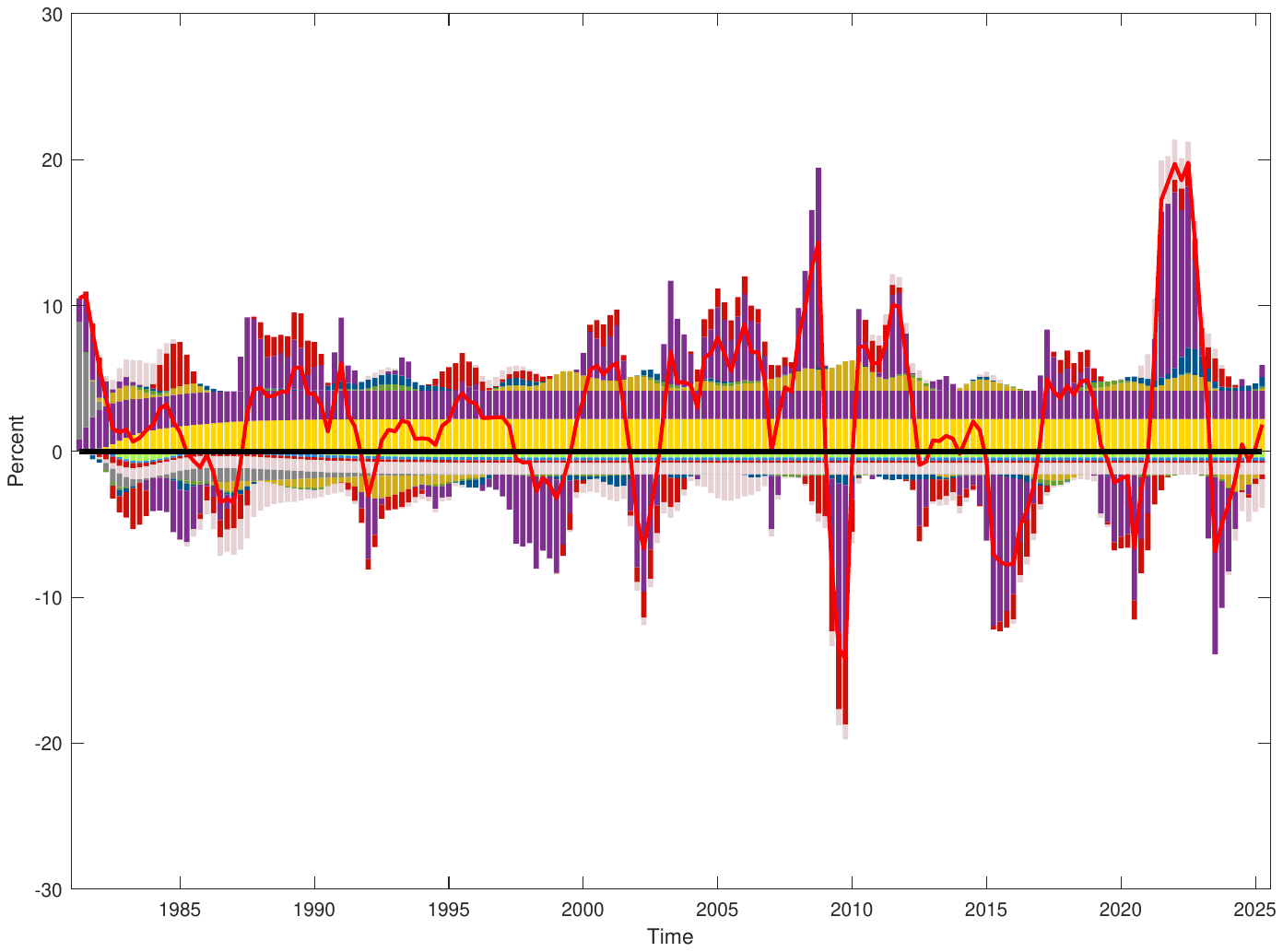}
\label{fig:COMMINFLHD_6var}
  \end{subfigure}
  \begin{subfigure}{.99\textwidth}
\vspace{0.1in}  \caption{Corporate risk spread}
\centering\includegraphics[width=0.95\linewidth, height= 5.0cm]{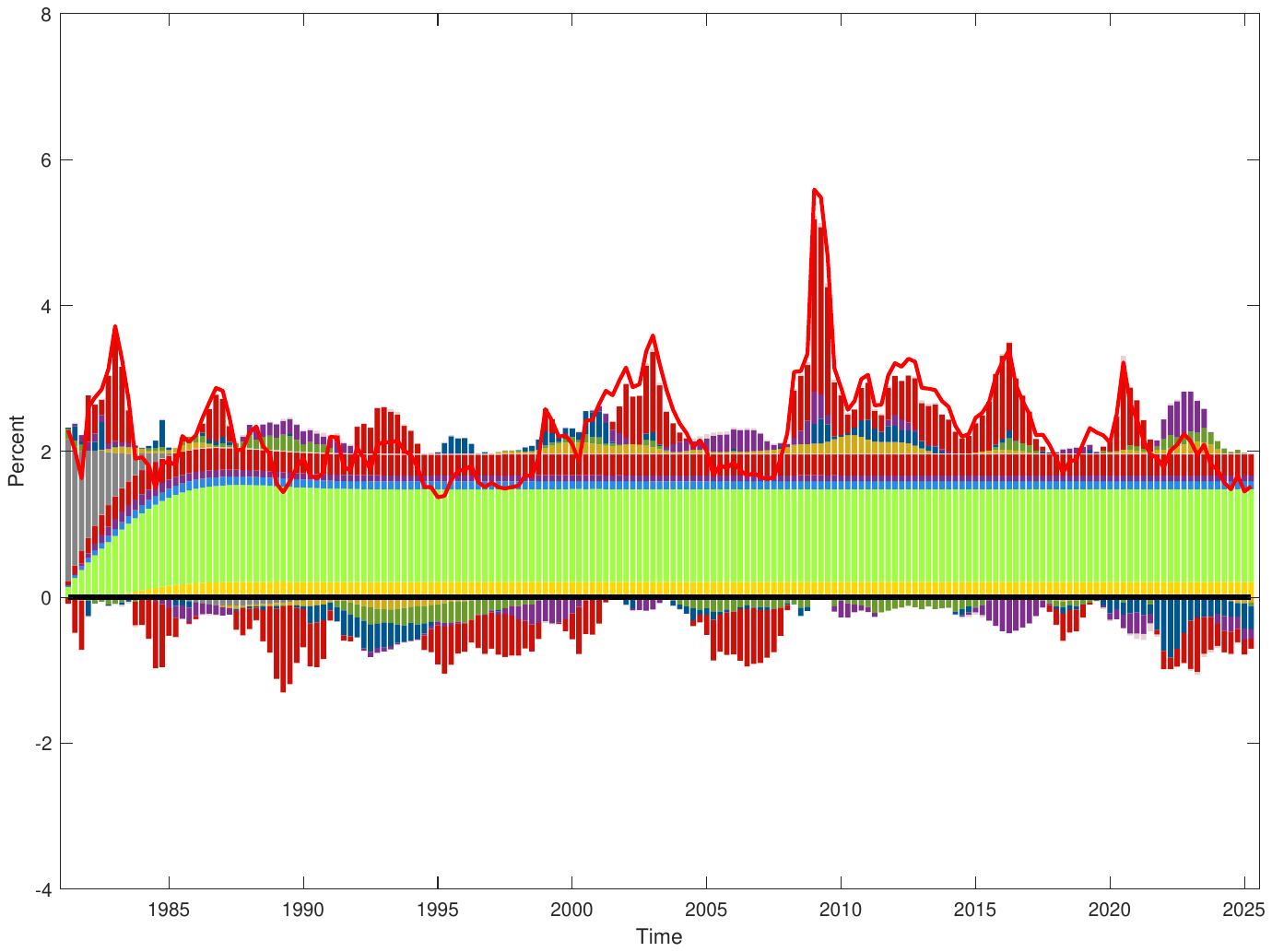}
 \label{fig:RiskSpreadHD_6var}
  \end{subfigure}
  \begin{subfigure}{.99\textwidth}
\vspace{0.1in}       \caption{Real wage growth}
\centering\includegraphics[width=0.95\linewidth, height= 6.5cm]{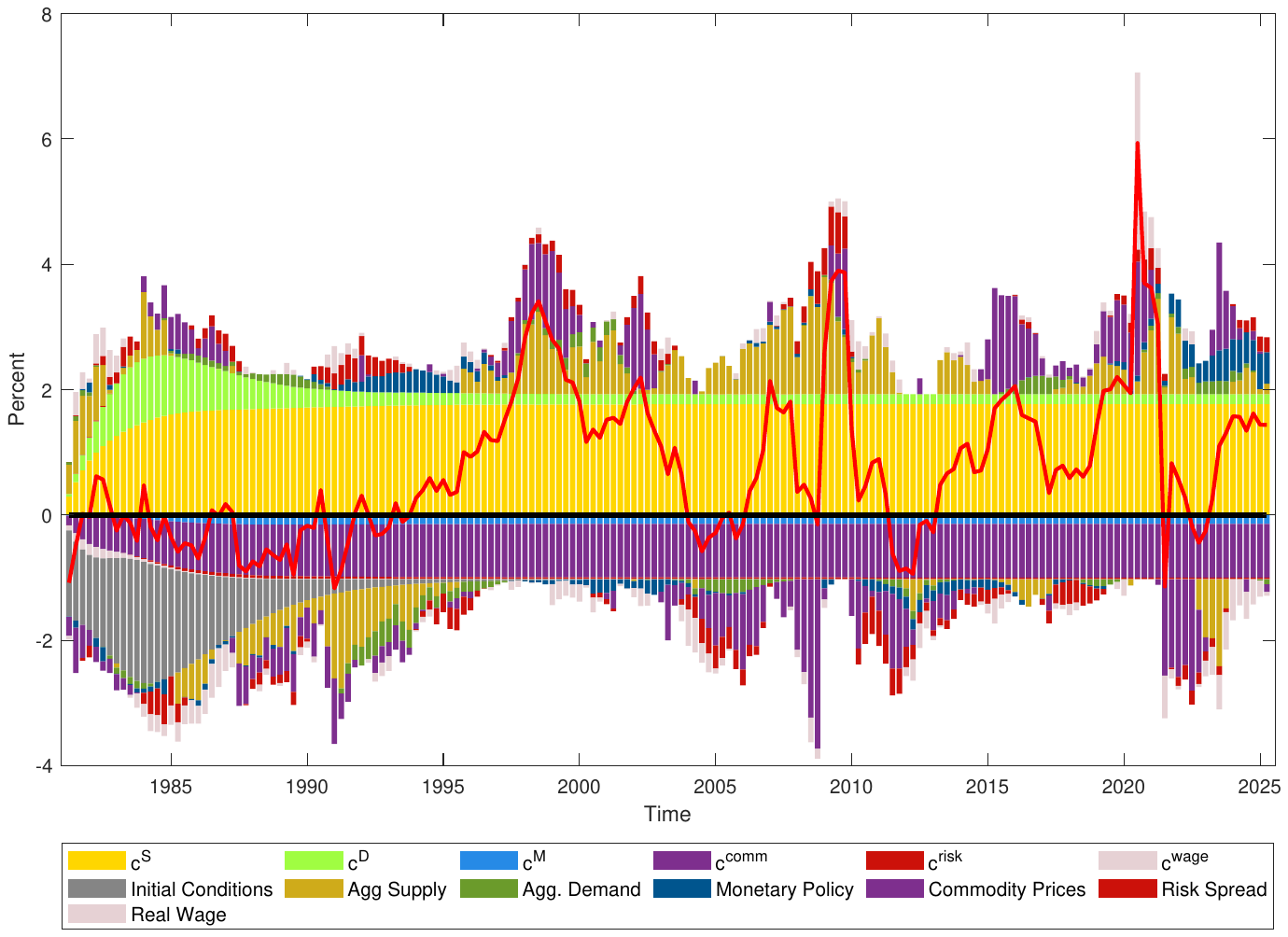}  
 \label{fig:RealWagetHD_6var}
  \end{subfigure}
   \label{fig:BHHD_6var_2_NOPRIOR}
\noindent \flushleft \footnotesize
Figure \ref{fig:BHHD_6var_2_NOPRIOR} plots the median estimates of the historical decomposition of commodity inflation, corporate risk spread and real wage growth for the case where there is no additional prior for the structural constants      
 \vfill
\end{figure}

In this section we present the Figures for the six variable VAR without additional priors on the structural constants.  The posterior distributions of the structural constants for this case are shown in Figure \ref{fig:BH_6var_NOPRIOR_CONSTANTS}. As in the three variable VAR the posterior for the constant in the Taylor Rule has large parts which are negative which is very unintuitive. The constants for supply shock are also positive which doesn't accord with the Phillips curve intuition of the structural model. If the supply equation does not represent a Phillips Curve and the monetary policy equation does not represent a Taylor rule then the resultant historical decompositions have no clear interpretation. We nevertheless present the historical decomposition for this case in Figures  \ref{fig:BHHD_6var_1_NOPRIOR} and \ref{fig:BHHD_6var_2_NOPRIOR}. These Figures serve to emphasize the importance of the structural model. It is only when the intuition of the structural model about its constant terms is able to influence the estimation, via the additional priors on these structural constants, that the results come into focus and provide an intuitive decomposition of the data.  

\newpage 

\newpage

\newpage

\section{Figures for 6 variable VAR with tighter priors on structural constants \\ (Not Intended For Publication)}

\begin{figure}[h!]
\caption{Posterior distribution of structural constants of six variable VAR }
  \vspace{0.1in}      
\centering\includegraphics[width=6.5in, height=2.5in]{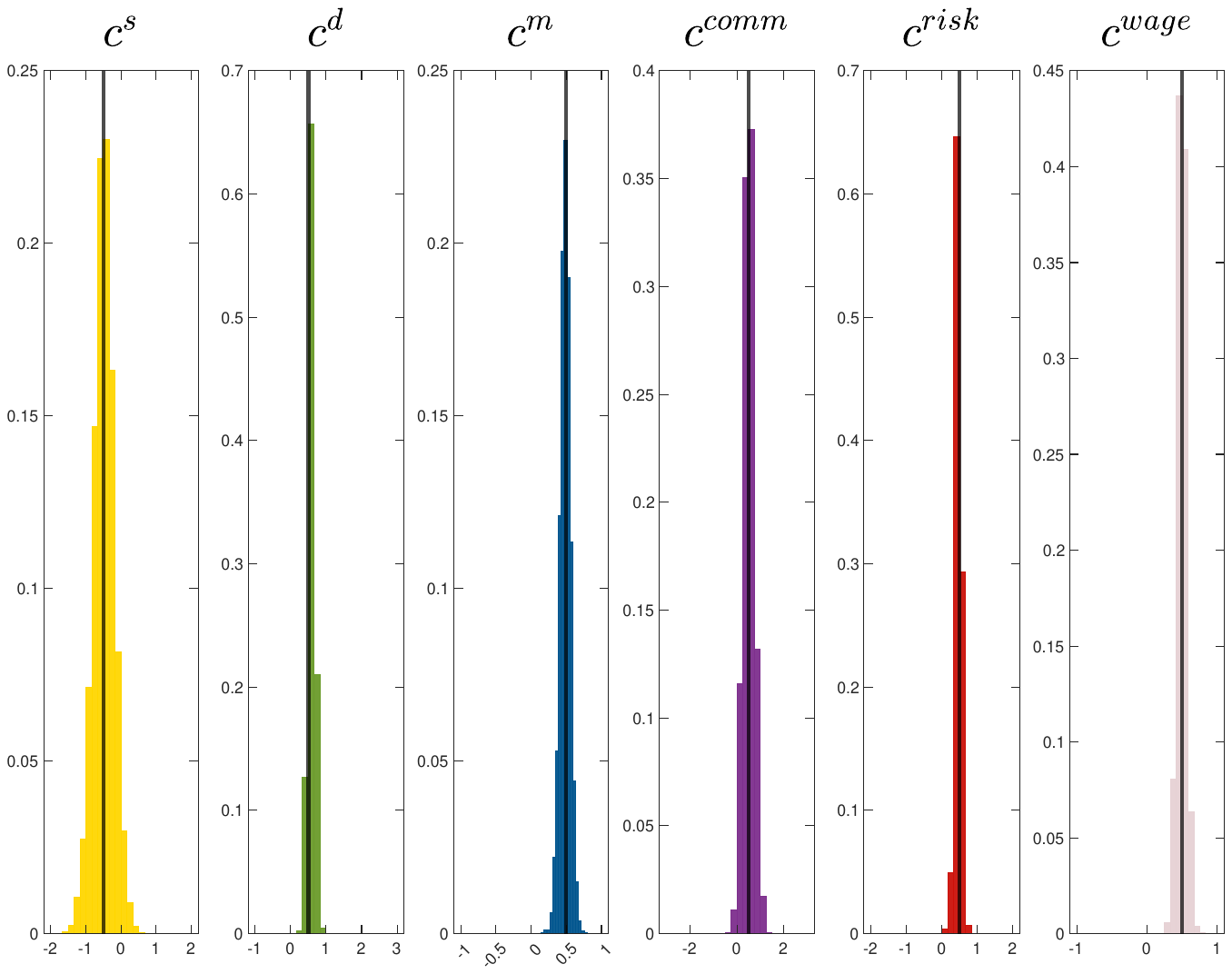}
  \label{fig:BH_6var_TIGHT_CONSTANTS}
\noindent \flushleft \footnotesize
Figure \ref{fig:BH_6var_TIGHT_CONSTANTS} plots the posterior distribution of the structural constants for the six variable VAR  with tighter additional priors over the structural constant. The vertical black line is at the prior median in each case.       
 \vfill
\end{figure}

\begin{figure}[h!]
\caption{Historical decomposition of output, inflation and interest rate in six variable VAR with tighter additional priors over the structural constants}
  \begin{subfigure}{.99\textwidth}
\vspace{0.1in}   \caption{Output gap }
\centering\includegraphics[width=0.95\linewidth, height= 5.0cm]{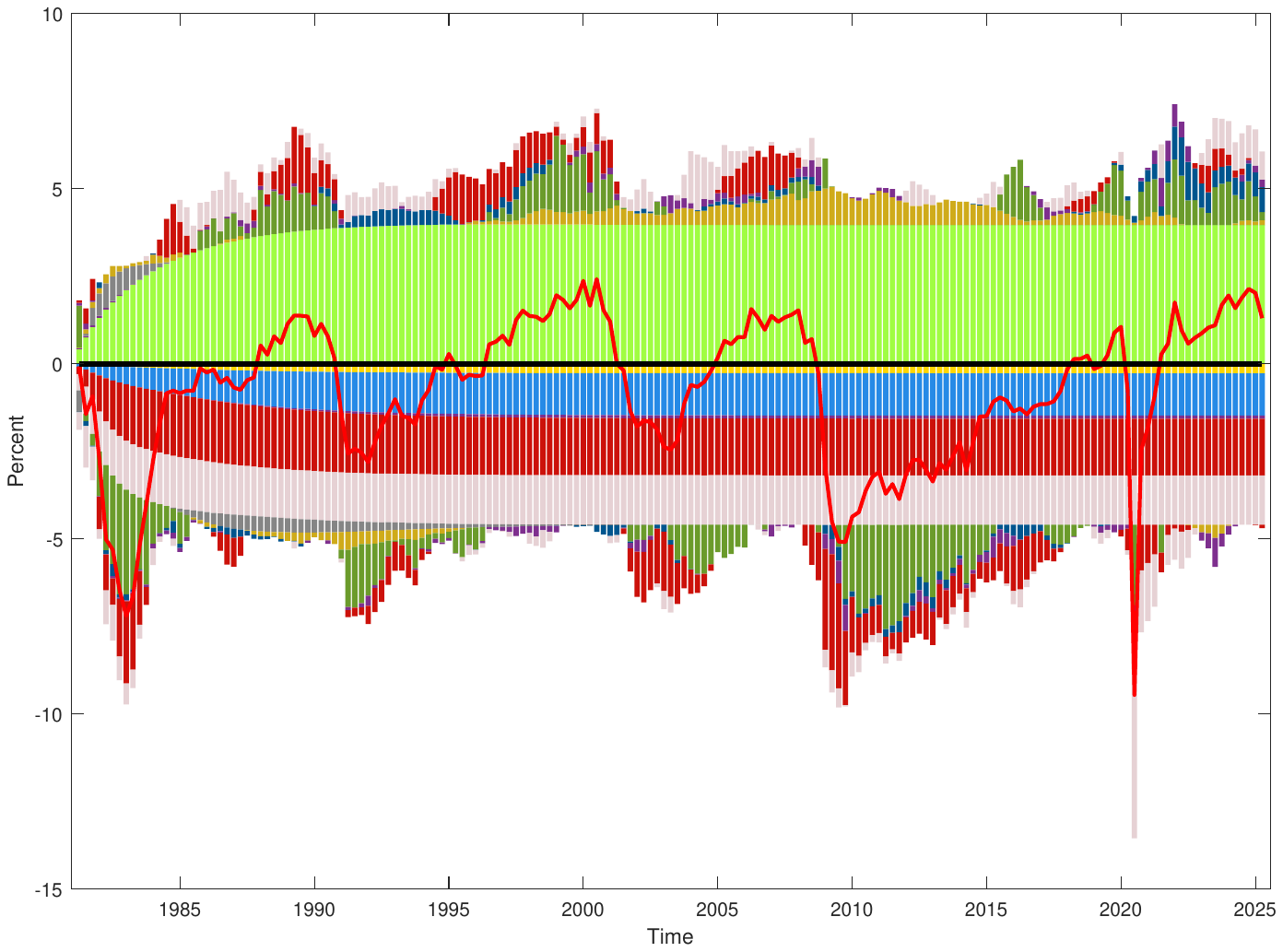}
\label{fig:OutputHD_6var}
  \end{subfigure}
  \begin{subfigure}{.99\textwidth}
\vspace{0.1in}  \caption{Inflation}
\centering\includegraphics[width=0.95\linewidth, height= 5.0cm]{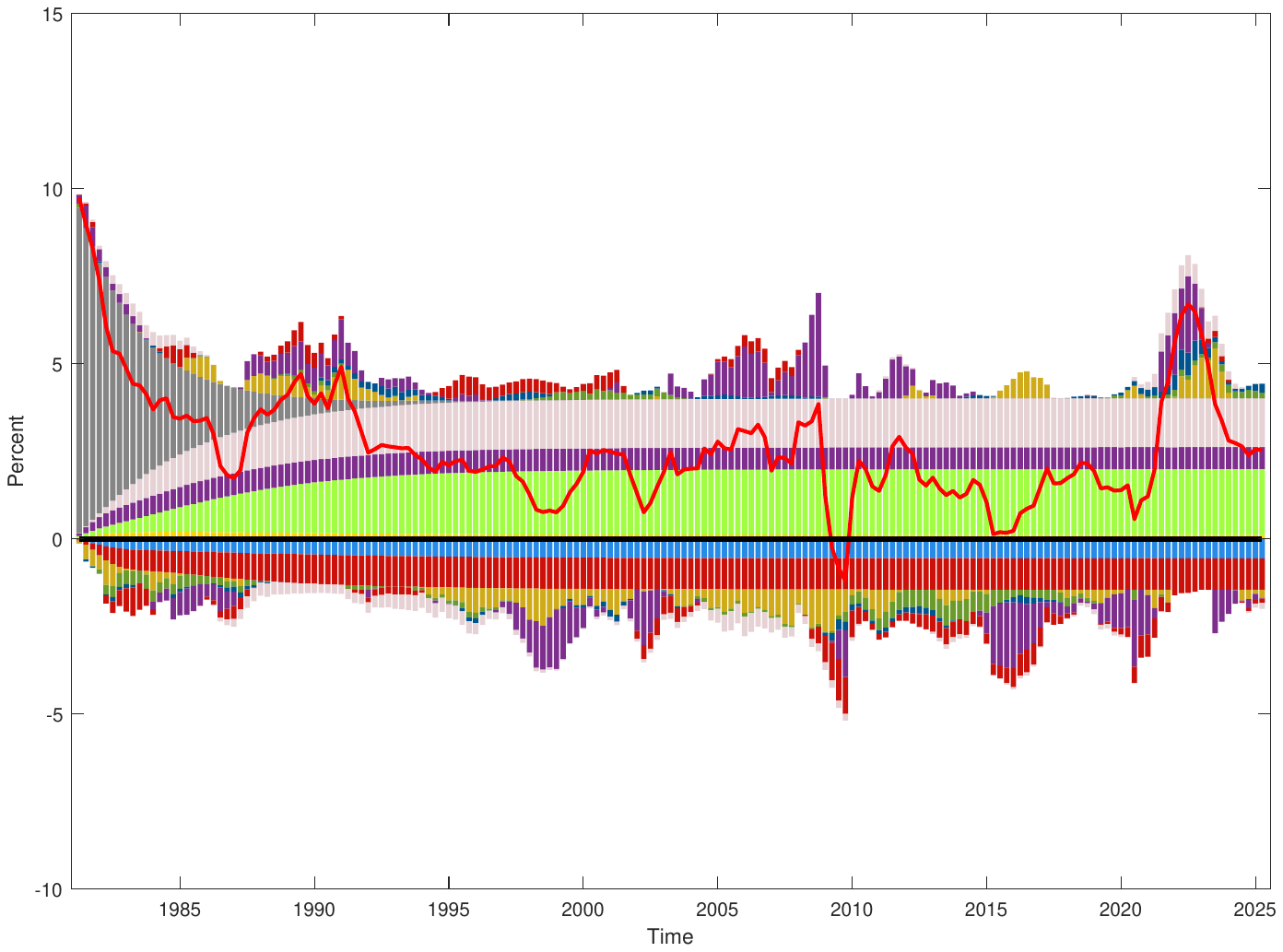}
 \label{fig:InflationHD_6var}
  \end{subfigure}
  \begin{subfigure}{.99\textwidth}
\vspace{0.1in}       \caption{Interest Rate}
\centering\includegraphics[width=0.95\linewidth, height= 6.5cm]{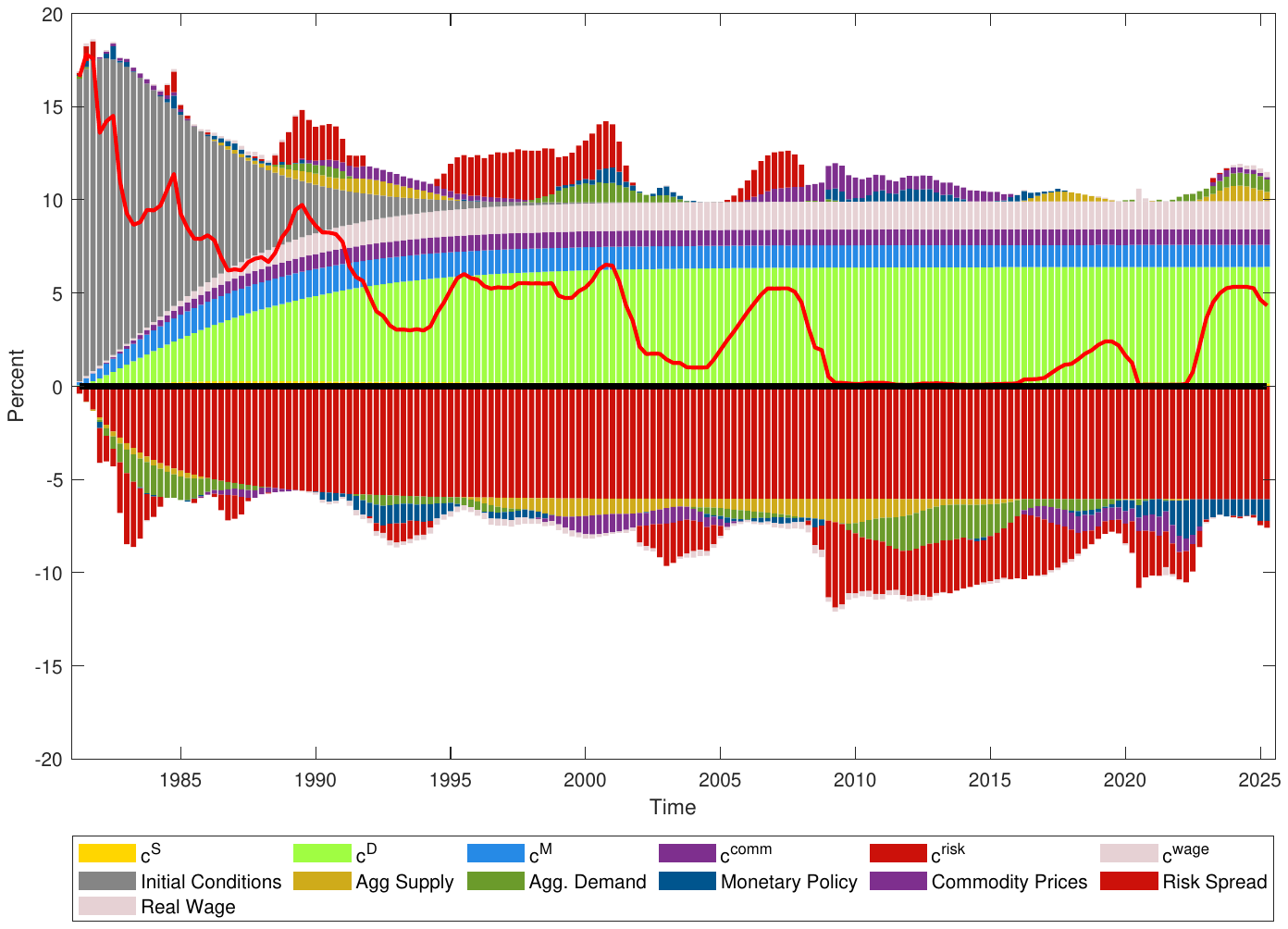}  
 \label{fig:InterestHD_6var}
  \end{subfigure}
   \label{fig:BHHD_6var_1_TIGHT}
\noindent \flushleft \footnotesize
Figure \ref{fig:BHHD_6var_1_TIGHT} plots the median estimates of the historical decomposition of the output gap, inflation, and the interest rate for the case with tighter additional priors over the structural constant. 
 \vfill
\end{figure}

\begin{figure}[h!]
\caption{Historical decomposition of commodity inflation, corporate risk spread and real wage growth in six variable VAR with tighter additional priors over the structural constants}
  \begin{subfigure}{.99\textwidth}
\vspace{0.1in}   \caption{Commodity inflation}
\centering\includegraphics[width=0.95\linewidth, height= 5.0cm]{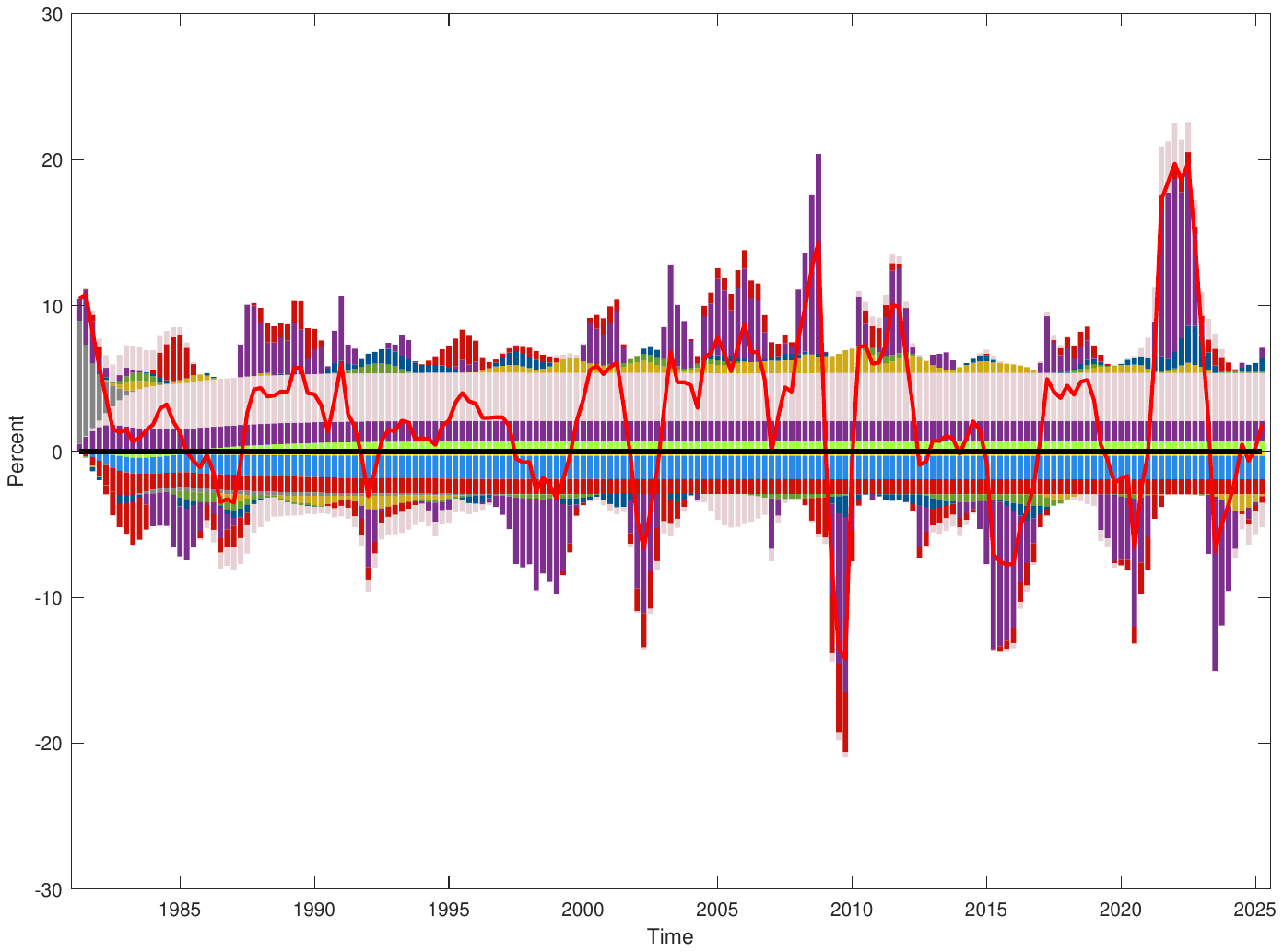}
\label{fig:COMMINFLHD_6var}
  \end{subfigure}
  \begin{subfigure}{.99\textwidth}
\vspace{0.1in}  \caption{Corporate risk spread}
\centering\includegraphics[width=0.95\linewidth, height= 5.0cm]{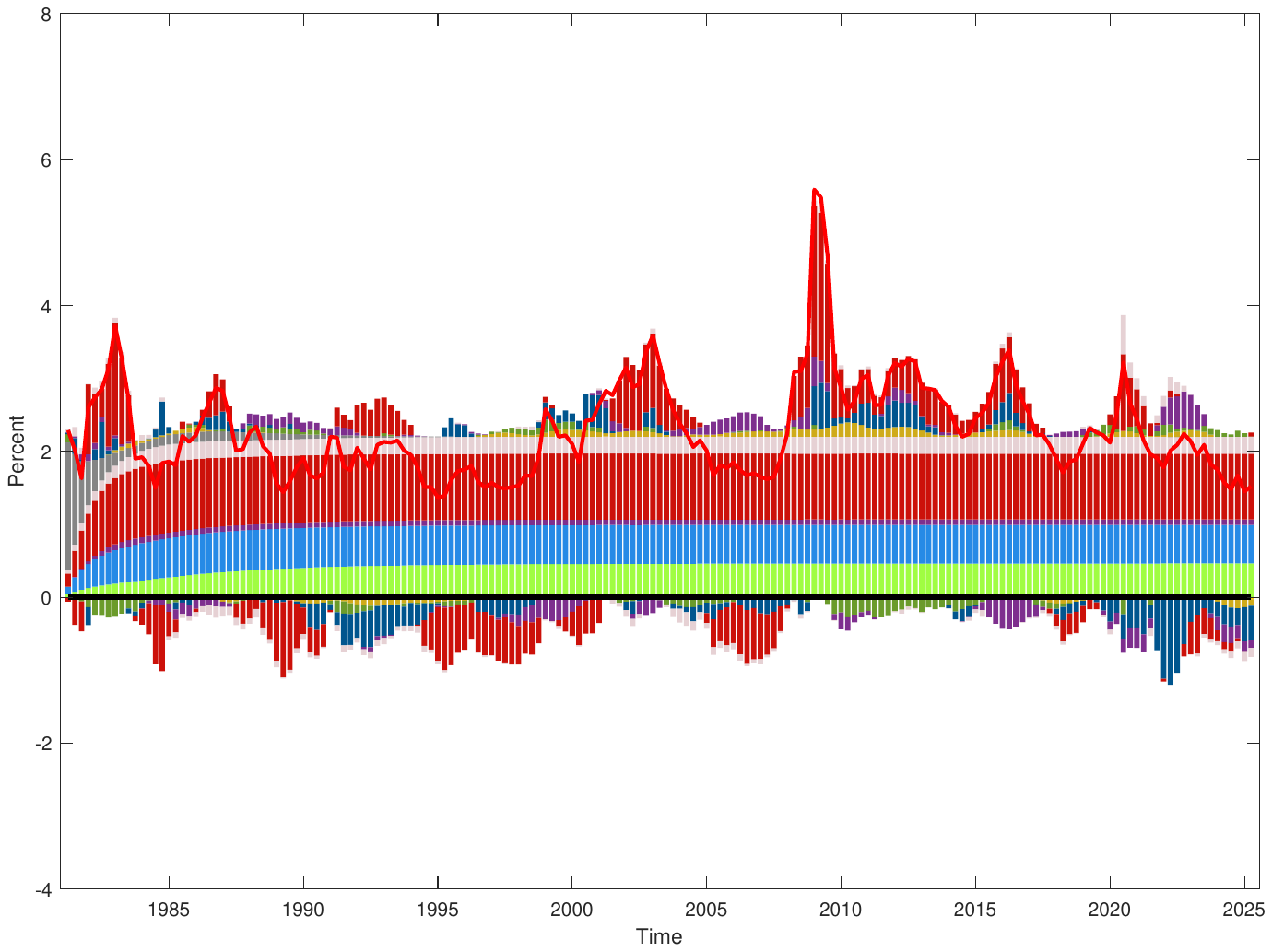}
 \label{fig:RiskSpreadHD_6var}
  \end{subfigure}
  \begin{subfigure}{.99\textwidth}
\vspace{0.1in}       \caption{Real wage growth}
\centering\includegraphics[width=0.95\linewidth, height= 6.5cm]{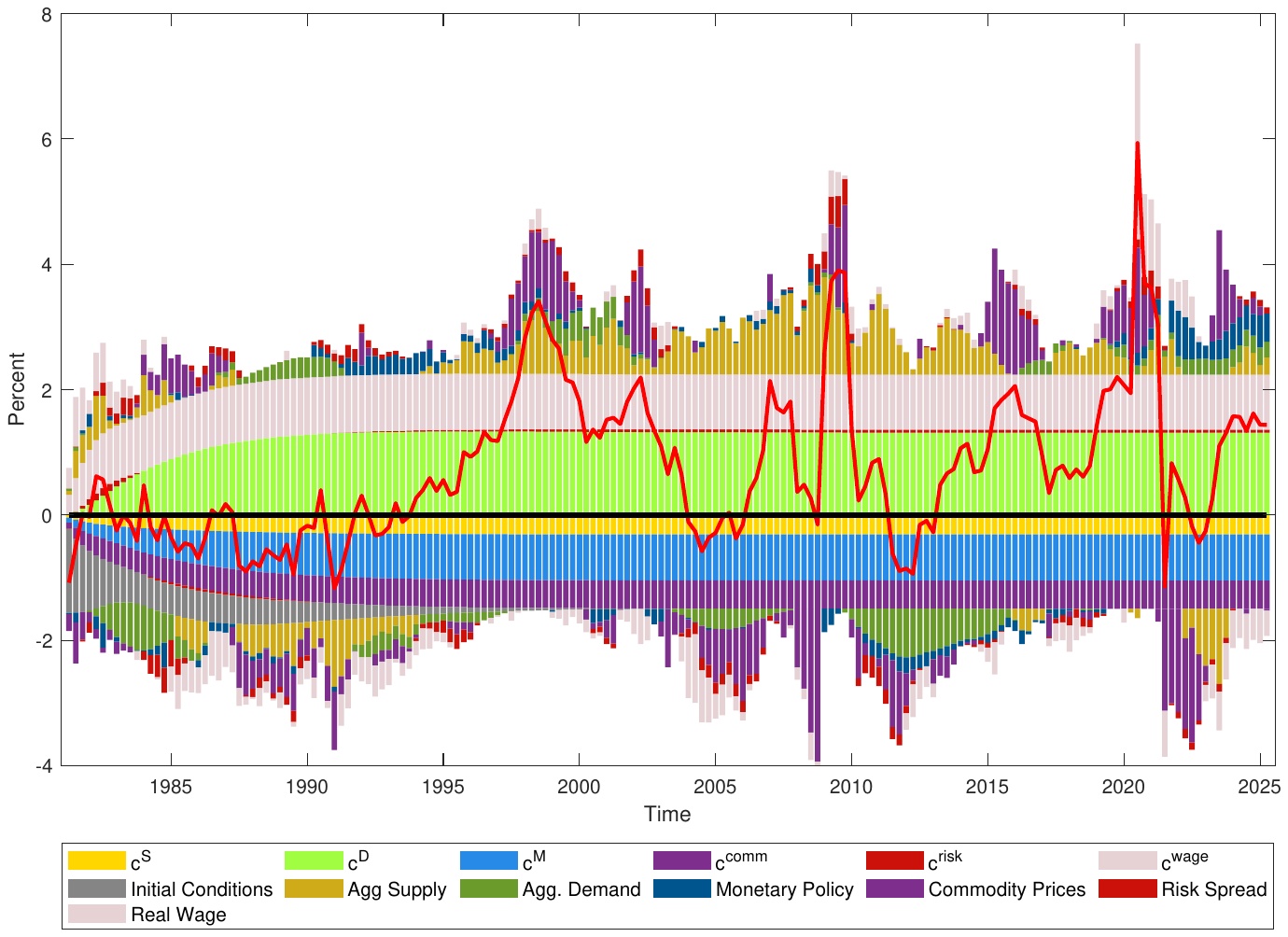}  
 \label{fig:RealWagetHD_6var}
  \end{subfigure}
   \label{fig:BHHD_6var_2_TIGHT}
\noindent \flushleft \footnotesize
Figure \ref{fig:BHHD_6var_2_TIGHT} plots the median estimates of the historical decomposition of the output gap, inflation, and the interest rate for the case with tighter additional priors over the structural constant. 
 \vfill
\end{figure}

We present the Figures for the six variable VAR with tighter priors on the structural constants.  The priors for the structural constants are the same as those described in section \ref{sec:BH6_identify} except that $V_i=0.01$ for all the structural constants.  The posteriors of the structural constants for this case are shown in Figure \ref{fig:BH_6var_TIGHT_CONSTANTS} and are much tighter than in Figure \ref{fig:BH_6var_CONSTANTS}.  However this does not have a great effect on the historical decomposition. Those for this case are shown in Figures  \ref{fig:BHHD_6var_1_TIGHT} and \ref{fig:BHHD_6var_2_TIGHT} which are very similar to Figures \ref{fig:BHHD_6var_1} and \ref{fig:BHHD_6var_2}. 

\newpage 

\newpage

\begin{thebibliography}{9}

\bibitem{BG} Barro, Robert and David B. Gordon, 1983, A Positive Theory of Monetary Policy in a Natural-Rate Model,  Journal of Political Economy. 91 (4): 589–610. 

\bibitem{BH1}  Baumeister, Christiane and James Hamilton, 2015, Sign Restrictions, Structural Vector Autoregressions, and Useful Prior Information,  Econometrica, 83(5):  1963-1999

\bibitem{BH2}  Baumeister, Christiane and James Hamilton, 2018, Inference in Structural Vector Autoregressions When the Identifying Assumptions are Not Fully Believed: Re-evaluating the Role of Monetary Policy in Economic Fluctuations (with James D. Hamilton), Journal of Monetary Economics, 100: 48-65.

\bibitem{BH3}  Baumeister, Christiane and James Hamilton, 2019, Structural Interpretation of Vector Autoregressions with Incomplete Identification: Revisiting the Role of Oil Supply and Demand Shocks, American Economic Review, 109(5): 1873-1910.



\bibitem{BH5}  Baumeister, Christiane, and James D. Hamilton, 2024, Advances in Using Vector Autoregressions to Estimate Structural Magnitudes, Econometric Theory, 40(3), 472-510.

\bibitem{Furlanetto2} Bergholt, Drago, Fabio Canova, Francesco Furlanetto, Nicol\`o Maffei-Faccioli and P\.al Ulvedal, 2025, What Drives the Recent Surge in Inflation? : The Historical Decomposition Rollercoaster, American Economic Journal: Macroeconomics (Forthcoming).



\bibitem{BH4}  Blake, Andrew, and Haroon Mumtaz, 2017, Applied Bayesian econometrics for central bankers, Centre for Central Banking Studies, Bank of England

\bibitem{BofE} Brignone D.  and M. Piffer, 2025, A structural VAR model for the UK economy, Bank of England Macro Technical Paper, No. 3 July 2025

\bibitem{BofE} Brignone D.  and M. Piffer, 2026, Structural forecast analysis,  Bank of England Staff Working Paper No. 1165  January 2026



\bibitem{Canova3} Canova, Fabio, 2007, Methods for Applied Macroeconomic Research, Princeton University Press.

\bibitem{Canova1} Canova, Fabio and Gianni De Nicol\`o, 2002, Monetary disturbances matter for business fluctuations in the G-7, Journal of Monetary Economics,
49(6), 1131-1159.

\bibitem{Canova2} Canova, Fabio, and Filippo Ferroni, 2022, Mind the Gap! Stylized Dynamic Facts and Structural Models, American Economic Journal: Macroeconomics, 14(4):104–35








\bibitem{Furlanetto} Furlanetto, Francesco and Orjan Robstad, 2019, Immigration and the macroeconomy: some new empirical evidence, Review of Economic Dynamics, Elsevier for the Society for Economic Dynamics, vol. 34, 1-19. 


\bibitem{Hamilton} Hamilton, James, 1994, Time Series Analysis, Princeton University Press.


\bibitem{LSZ} Leeper, E. M. , Sims, C.A. and T. Zha, 1996,  What does monetary policy do?,  Brookings Papers on Economic Activity, 1996(2):1-78

\bibitem{Lubik} Lubik, T. A., and F.Schorfheide, 2004, Testing for Indeterminacy: An Application to U.S. Monetary Policy, American Economic Review, 94(1):190–217




\bibitem{Mankiw} Mankiw, N. Gregory, 1986, The Allocation of Credit and Financial Collapse, Quarterly Journal of Economics, Vol. 101(3):455-470.


\bibitem{C} Mountford, Andrew and Jonathan Wadsworth, 2025, Immigration, demand, supply and sectoral heterogeneity in the UK labour market,  CEP discussion paper No. 2089

\bibitem{AMUlhlig} Mountford, Andrew and Harald Uhlig, 2009,  What are the effects of fiscal policy shocks?, Journal of Applied Econometrics, 24(6), 960-992

\bibitem{Mumtaz3} Mumtaz, Haroon and Paolo Surico, 2009, The Transmission of International Shocks: A Factor‐Augmented VAR Approach, Journal of Money, Credit and Banking, vol. 41(s1), pages 71-100, February. 





\bibitem{Sims1} Sims, C A, 1992, Interpreting the macroeconomic time series facts: the effects of monetary
policy, European Economic Review, Vol. 36(5), 975-1000

\bibitem{Sims2} Sims, C, 2003, Comments on Smets and Wouters, available at  http://sims.princeton.edu/yftp/Ottawa/SWcommentSlides.pdf.

\bibitem{Stiglitz} Stiglitz, Joseph E. and Andrew Weiss, 1981,
Credit Rationing in Markets with Imperfect Information, American Economic Review, Vol. 71(3), 393–410


\bibitem{Taylor} Taylor, C, 1993, Discretion versus policy rules in practice, Carnegie-Rochester Conference Series on Public Policy, 3:195-214.

\bibitem{Theil} Theil, Henri, 1971, Principles of Econometrics, John Wiley \& Sons Inc. 

\bibitem{Uhlig1994} Uhlig, Harald,  1994, What Macroeconomists should know about unit roots: a Bayesian Perspective, Econometric Theory, 10 645-671.

\bibitem{UhligJME} Uhlig, Harald, 2005, What are the effects of monetary policy on output? Results from an agnostic identification procedure, Journal of Monetary Economics, 52 381-419.

\bibitem{Uhlig2017} Uhlig, Harald, 2017, Shocks, sign restrictions, and identification, in Advances in economics and econometrics, edited by Bo Honoré, Ariel Pakes, Monika Piazzesi, and Larry Samuelson, 2 95-127, Cambridge University Press
\end{thebibliography}
\end{document}